\documentclass[aps,twocolumn,superscriptaddress,english,10pt,nofootinbib]{revtex4}

\usepackage{float}

\usepackage{amssymb, amsmath, bm, dcolumn, epsf, graphicx, latexsym, slashed, simplewick}
\usepackage[utf8]{inputenc}
\usepackage[normalem]{ulem}

\usepackage{color}

\def\be{\begin{equation}}
\def\ee{\end{equation}}
\def\bea{\begin{eqnarray}}
\def\eea{\end{eqnarray}}

\usepackage{hyperref}
\usepackage{comment}

\usepackage[normalem]{ulem}

\begin{document}

\title{Early Dark Energy Does Not Restore Cosmological Concordance}

\author{J.~Colin Hill}
\affiliation{Department of Physics, Columbia University, New York, NY, USA 10027}
\affiliation{Center for Computational Astrophysics, Flatiron Institute, New York, NY, USA 10010}

\author{Evan McDonough}
\affiliation{Brown Theoretical Physics Center and Department of Physics, Brown University, Providence, RI, USA 02912}
\affiliation{Center for Theoretical Physics, Massachusetts Institute of Technology, Cambridge, MA 02139, USA}

\author{Michael W.~Toomey}
\affiliation{Brown Theoretical Physics Center and Department of Physics, Brown University, Providence, RI, USA 02912}

\author{Stephon Alexander}
\affiliation{Brown Theoretical Physics Center and Department of Physics, Brown University, Providence, RI, USA 02912}

\begin{abstract}
Current cosmological data exhibit a tension between inferences of the Hubble constant, $H_0$, derived from early- and late-universe measurements.  One proposed solution is to introduce a new component in the early universe, which initially acts as ``early dark energy'' (EDE), thus decreasing the physical size of the sound horizon imprinted in the cosmic microwave background (CMB) and increasing the inferred $H_0$.  Previous EDE analyses have shown this model can relax the $H_0$ tension, but the CMB-preferred value of the density fluctuation amplitude, $\sigma_8$, increases in EDE as compared to $\Lambda$CDM, increasing tension with large-scale structure (LSS) data. We show that the EDE model fit to CMB and SH0ES data yields scale-dependent changes in the matter power spectrum compared to $\Lambda$CDM, including $10$\% more power at $k = 1 \, h$/Mpc. Motivated by this observation, we reanalyze the EDE scenario, considering LSS data in detail. We also update previous analyses by including \emph{Planck} 2018 CMB likelihoods, and perform the first search for EDE in \emph{Planck} data alone, which yields no evidence for EDE. We consider several data set combinations involving  the primary CMB, CMB lensing, supernovae, baryon acoustic oscillations, redshift-space distortions, weak lensing, galaxy clustering, and local distance-ladder data (SH0ES). While the EDE component is weakly detected (3$\sigma$) when including the SH0ES data and excluding most LSS data, this drops below 2$\sigma$ when further LSS data are included. Further, this result is in tension with strong constraints imposed on EDE by CMB and LSS data without SH0ES, which show no evidence for this model.  We also show that physical priors on the fundamental scalar field parameters further weaken evidence for EDE.  We conclude that the EDE scenario is, at best, no more likely to be concordant with all current cosmological data sets than $\Lambda$CDM, and appears unlikely to resolve the $H_0$ tension.
\end{abstract}

\maketitle

\section{Introduction}

The value of the Hubble constant $H_0$, the present-day expansion rate of the Universe, is crucial to cosmology.  All cosmological quantities are connected to $H_0$, which effectively sets the scale of the Universe.  In recent years, the value of $H_0$ inferred from probes of the early universe has been in persistent disagreement with that measured from probes of the late universe, a discrepancy that has now reached $\approx 4-6\sigma$ significance~(e.g.,~\cite{Verde:2019ivm}). Assuming that systematic errors in one or more measurements are not responsible for the disagreement, this ``Hubble tension'' may be a first sign of physics beyond the standard $\Lambda$ Cold Dark Matter ($\Lambda$CDM) cosmological model.

The cosmic microwave background (CMB) allows for a precise, albeit indirect, inference of the Hubble constant in the context of a cosmological model \cite{2013ApJS..208...19H,Ade:2013zuv,Ade:2015xua,Aghanim:2018eyx,Louis:2016ahn,Bianchini:2019vxp}. The angular size of the sound horizon, combined with constraints on the energy density in each component of the $\Lambda$CDM model derived from the CMB temperature, polarization, and lensing power spectra, allow for a determination $H_0 = 67.36 \pm 0.54\, {\rm km/s /Mpc}$ using the \emph{Planck} 2018 data alone \cite{Aghanim:2018eyx}. The same approach can be taken \emph{without} CMB anisotropy data, instead using an early-universe measurement of the baryon density, namely, that inferred from Big Bang nucleosynthesis \cite{2016ApJ...830..148C}, and late-universe measurements of the matter density to calibrate the sound horizon measured in baryon acoustic oscillation (BAO) experiments.\footnote{Note that the early-universe methods also require knowledge of the radiation density, as inferred from the CMB monopole temperature \cite{1996ApJ...473..576F}.} Applied to Dark Energy Survey (DES) data combined with Baryon Oscillation Spectroscopic Survey (BOSS) BAO data, this methodology leads to $H_0 = 67.4 ^{+1.1} _{-1.2}\, {\rm km/s /Mpc}$ \cite{Abbott:2017smn}, in near-perfect agreement with the CMB constraints, albeit with error bars doubled in size.  Recent analyses have further refined this cosmological approach to constrain $H_0$ using not only sound horizon information, but also information in the shape of the matter power spectrum, as measured from the redshift-space galaxy power spectrum~ \cite{Ivanov:2019hqk,DAmico:2019fhj,2020A&A...633L..10T}.  The results are consistent with those from the \emph{Planck} CMB analysis, again albeit with somewhat larger error bars.

The opposite approach is to constrain $H_0$ directly via late-universe measurements, without assuming a cosmological model. Historically, these constraints have been obtained via the classical distance ladder (e.g.,~\cite{2006ApJ...653..843S}). In this procedure, parallax measurements are used to calibrate the period-luminosity relation of Cepheid variable stars, which are then used to calibrate the luminosity of nearby Type Ia supernovae (SN), allowing distant SNIa to be used as proxies for the Hubble flow. The SH0ES collaboration has applied this method in recent years, most recently obtaining $H_0=74.03 \pm 1.42\,{\rm km/s /Mpc}$ \cite{Riess:2019cxk}. The Cepheid calibration step can alternatively be swapped out for a calibration using the ``tip of the red giant branch'' in the Hertzsprung-Russell diagram.  The most recent analysis with this method yields $H_0 = 69.6 \pm 1.9 \,{\rm km/s /Mpc}$ \cite{Freedman:2020dne}.  Another alternative approach replaces the Cepheids with Miras, variable red giant stars, leading to $H_0=73.3\pm3.9\,{\rm km/s /Mpc}$ \cite{Huang:2019yhh}. Very recently, late-universe $H_0$ probes have emerged that are independent of, and statistically competitive with, the distance ladder. In particular, the H0LiCOW collaboration has constrained $H_0$ by measuring time delays in strongly lensed quasar systems, obtaining $H_0= 73.3^{+1.7} _{-1.8}\, {\rm km/s /Mpc}$ \cite{Wong:2019kwg}, although recently concerns have been raised regarding the sensitivity to details of the lens modeling \cite{Kochanek:2019ruu,Blum:2020mgu}. The Megamaser Cosmology Project finds $H_0=73.9\pm3.0\, {\rm km/s /Mpc}$ \cite{Pesce:2020xfe} from very-long-baseline interferometry observations of water masers orbiting supermassive black holes. A futuristic, but already fruitful, alternative is to directly infer $H_0$ using not the brightness of standard candles, but the ``volume'' of standard \emph{sirens}, i.e., gravitational waves from merging binary neutron stars \cite{Abbott:2017xzu, Soares-Santos:2019irc}.  In this work, we focus on the $H_0$ constraint from \cite{Riess:2019cxk}, as this has been the most widely analyzed late-universe measurement to date.

 There exist varied theoretical proposals to explain or ameliorate the $H_0$ discrepancy, ranging from new physics in the very early to late universe.  It has been argued \cite{Knox:2019rjx} that the proposal ``least unlikely to be successful'' is an increase in the expansion rate just prior to recombination, which acts to shrink the sound horizon at last scattering. There is now a growing body of  work to realize this in concrete theoretical models. A popular subclass of these models has been termed ``Early Dark Energy'' (EDE) \cite{Poulin:2018cxd}, and many EDE-like models have been proposed, both in the context of the $H_0$ tension~ \cite{Poulin:2018cxd,Smith:2019ihp,Agrawal:2019lmo,Alexander:2019rsc,Lin:2019qug,Sakstein:2019fmf,Niedermann:2019olb,Kaloper:2019lpl,Berghaus:2019cls} and other areas of cosmological phenomenology~(e.g.,~\cite{Kamionkowski:2014zda,Poulin:2018a,Hill2018}).

In the EDE implementation that we will focus on, an ultra-light scalar field, significantly lighter than canonical ultra-light axion or fuzzy dark matter, is displaced from the minimum of its potential at early times, and, held up by Hubble friction, effectively acts as an additional contribution to dark energy. When the Hubble parameter becomes less than the mass of the field, it rolls down its potential and begins to oscillate about the minimum. If the potential about the minimum is steeper than quadratic, the EDE field quickly becomes a subdominant component of the universe; hence the name ``early dark energy". The model can be parameterized by the field's maximal fractional contribution to the energy density of the universe, $f_{\rm EDE}\equiv {\rm max}(\rho_{\rm EDE}/3 M_{pl}^2 H^2)$, and the critical redshift $z_c$ at which this maximum is reached, which roughly corresponds to the moment before the onset of oscillations. This evolution is encoded in the Hubble parameter as an enhancement (compared to $\Lambda$CDM) localized in redshift-space in the epoch before the onset of oscillations. The consequent decrease in the sound horizon $r_s$ increases the inferred $H_0$ value from the early-universe approach described above, in principle bringing it into agreement with late-universe measurements.

  However, the apparent success of the EDE scenario in resolving the Hubble tension comes at a cost: in order to preserve the fit to CMB data, some of the standard $\Lambda$CDM parameters shift.  In particular, the physical CDM density $\Omega_c h^2$ increases substantially, as does the spectral index $n_s$ and to a lesser extent the physical baryon density $\Omega_b h^2$~ \cite{Smith:2019ihp}.  Primarily due to the increase in $\Omega_c h^2$, the late-time amplitude of density fluctuations, $\sigma_8$, increases as well.  This increase exacerbates the current mild tension between CMB and large-scale structure (LSS) inferences of this parameter.  Thus, one expects that the fit to LSS data will be degraded in EDE models that fit the CMB and the distance-ladder $H_0$ data.  We confirm this expectation in this paper.
  
  The physical origin of these parameter shifts is fairly straightforward, and likely applies to \emph{any} scenario in which the sound horizon is decreased through the introduction of a new dark-energy-like component in the pre-recombination universe (so as to increase the expansion rate during this epoch over that in $\Lambda$CDM). This new component (e.g., EDE) acts to slightly suppress the growth of perturbations during the period in which it contributes non-negligibly to the cosmic energy density.  Thus, in order to preserve the fit to the CMB data, the CDM density must be increased to compensate for this loss in the efficiency of perturbation growth.  Since the EDE field is only relevant for a short period of time, the suppression is scale-dependent, and thus a small change in $n_s$ is also required to preserve the CMB fit.  While we carefully quantify these effects in the EDE scenario here, the basic physics indicates that similar considerations would afflict any $H_0$-tension-resolving scenario that works in a similar manner.
  
  \begin{figure*}[t]
\centering
\includegraphics[width=0.76\textwidth]{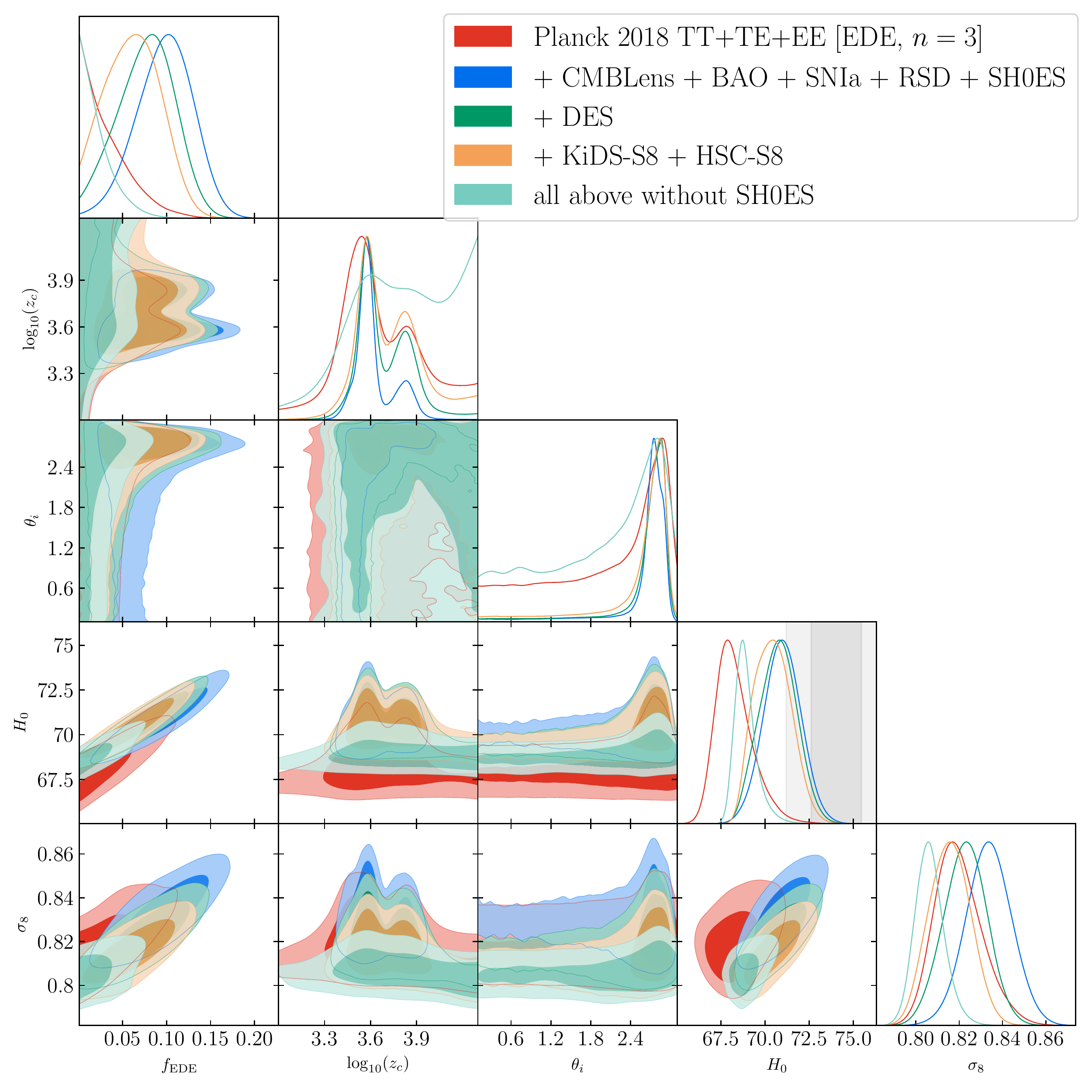}
    \caption{Constraints on the EDE scenario from \emph{Planck} 2018 primary CMB data (TT+TE+EE); \emph{Planck} 2018 CMB lensing data; BAO data from 6dF, SDSS DR7, and SDSS DR12; Pantheon SNIa data; the latest SH0ES $H_0$ constraint; SDSS DR12 RSD data; and the DES-Y1 3x2pt data. Here we present a subset of the parameters: the EDE energy-density fraction, timing, and initial condition, denoted $f_{\rm EDE}$, $\log_{10}(z_c)$, and $\theta_i$, respectively (note $\theta_i$ is distinct from $\theta_s$, the latter being the angular size of the sound horizon), along with $H_0$ [km/s/Mpc] and $\sigma_8$ . The contours show $1\sigma$ and $2\sigma$ posteriors for various data set combinations, computed with \texttt{GetDist} \cite{GetDist}.  The red contours show results for \emph{Planck} primary CMB data alone; the blue contours additionally include \emph{Planck} CMB lensing data, BAO data, SNIa data, SH0ES, and RSD data (matching the data sets used in~\cite{Poulin:2018cxd} and~\cite{Smith:2019ihp}, but with \emph{Planck} 2018 replacing 2015); and the dark green contours further include the DES-Y1 3x2pt likelihood.  The orange contours add priors on $S_8$ derived from KiDS and HSC data (as an approximation to the full likelihoods from these data sets).  The \emph{Planck} primary CMB data already place relatively strong constraints on the EDE scenario. Inclusion of the DES, KiDS, and HSC data significantly weakens the moderate evidence for EDE seen when analyzing the data sets used in~\cite{Smith:2019ihp}.  The $H_0$ increase found in the EDE model fit in~\cite{Smith:2019ihp} (blue contours here) is noticeably reduced by the inclusion of LSS data, and the tension with SH0ES (shown in the gray bands) is no longer reconciled. The light green contours include all data sets except SH0ES, yielding a stringent upper bound $f_{\rm EDE} < 0.060$ at 95\% CL, and a value for $H_0$ consistent with the fit to the primary CMB alone.  Fig.~\ref{fig:no-SH0ES-logf-logm} in Appendix~\ref{app:mfconstraints} shows these constraints in terms of fundamental physics parameters.}
    \label{fig:mainEDEconstraints}
\end{figure*}

  \begin{table*}[ht!]
Constraints on EDE ($n=3$) for varying data sets  \\
  \centering
  \begin{tabular}{|l|c|c|c|c|c|}
    \hline\hline Parameter & \hspace{0cm} \begin{tabular}[t]{@{}c@{}}\emph{Planck}  2018 \\ TT+TE+EE\end{tabular} & \hspace{0cm}\begin{tabular}[t]{@{}c@{}}\emph{Planck} 2018\\ TT+TE+EE, \\  CMB lensing,  BAO,\\ RSD, SNIa,\\ and SH0ES\end{tabular} & \hspace{0cm}\begin{tabular}[t]{@{}c@{}}\emph{Planck} 2018\\ TT+TE+EE, \\  CMB lensing,  BAO,\\ RSD, SNIa,\\ SH0ES, \\ and DES-Y1\end{tabular} &
    \hspace{0cm}\begin{tabular}[t]{@{}c@{}}\emph{Planck} 2018 \\ TT+TE+EE, \\  CMB lensing,  BAO,\\ RSD, SNIa,\\ SH0ES,\\  DES-Y1, \\ and HSC, KiDS ($S_8$) \end{tabular} & 
    \hspace{0cm}\begin{tabular}[t]{@{}c@{}}\emph{Planck} 2018 \\ TT+TE+EE, \\  CMB lensing,  BAO,\\ RSD, SNIa,\\  DES-Y1, \\ and HSC, KiDS ($S_8$) \\ (no SH0ES)  \end{tabular}
    \\\hline \hline

    {\boldmath$f_\mathrm{EDE} $} & $< 0.087 $ & $ 0.098  \pm 0.032 $ & $ 0.077 \, ^{+0.032}_{-0.034} $&$ 0.062 \, ^{+0.032}_{-0.033} $&$ < 0.060$\\

    {\boldmath$\mathrm{log}_{10}(z_c)$} & $3.66 \, ^{+0.28}_{-0.24}$ &$3.63 \,^{+0.17}_{-0.10}$ &$3.69 \, ^{+0.18}_{-0.15}$ &$3.73 \, ^{+0.20}_{-0.19}$&$>3.28$ \\

    {\boldmath$\theta_i$} & $> 0.36  $ &  $ 2.58 \, ^{+0.29}_{-0.09} $ &  $ 2.58 \, ^{+0.32}_{-0.15} $ &  $ 2.49 \, ^{+0.40}_{-0.38} $&  $ > 0.35  $\\

    \hline

    $H_0 \, [\mathrm{km/s/Mpc}]$ & $68.29 \,^{+1.02}_{-1.00}$ & $70.98 \,  \pm 1.05$ & $70.75 \, ^{+1.05}_{-1.09}$ & $70.45 \,^{+1.05}_{-1.08}$ & $68.92 \,^{+0.57}_{-0.59}$\\

    $\sigma_8$ & $0.8198 \, ^{+0.0109}_{-0.0107}$ & $0.8337 \,  \pm 0.0105$ & $0.8228 \,  ^{+0.0099} _{-0.0101}$ & $0.8157 \, \pm 0.0096$&  $0.8064 \, \pm 0.0065 $\\
    
    \hline
  \end{tabular} 
  \caption{The mean $\pm1\sigma$ constraints on cosmological parameters in the EDE scenario (with index $n=3$, c.f. Eq.~\eqref{eq.EDEdef}) from \emph{Planck} 2018 primary CMB data (TT+TE+EE); \emph{Planck} 2018 CMB lensing data; BAO data from 6dF, SDSS DR7, and SDSS DR12 (BOSS); Pantheon SNIa data; the latest SH0ES $H_0$ constraint; SDSS DR12 RSD data; and the DES-Y1 3x2pt data. Parameters in bold are sampled parameters. The two furthest-right columns add priors on $S_8$ derived from KiDS and HSC data (as an approximation to the use of full likelihoods from these data sets). In the furthest right column we include all data sets except the SH0ES measurement. Upper and lower limits are quoted at 95\% CL; the one-sided $f_{\rm EDE}$ upper bounds for the +DES-Y1 (fourth column) and +DES-Y1+HSC+KiDS (fifth column) are  $f_{\rm EDE}<0.127$ and $f_{\rm EDE}<0.112$ at $95\%$ CL, respectively.  The best-fit parameter values for most of these analyses can be found in Sec.~\ref{sec:constraints}.  The data set combination that yields the strongest evidence for EDE is shown in the third column (analogous to that used in~\cite{Smith:2019ihp}); the preferred EDE model in that analysis is in tension with the constraints on EDE imposed in the final column by the combination of all data sets without SH0ES, indicating discordance between SH0ES and the other data sets, even in the broadened EDE parameter space.}
  \label{table:EDE-params-full}
\end{table*}
  
  A simple way to quantify the parameter shifts in the EDE scenario and the associated CMB-LSS tension is by examining $S_8 \equiv \sigma_8 (\Omega_m/0.3)^{0.5}$. Fitting the EDE model (with index $n=3$, see Eq.~\eqref{eq.PoulinEDE} below) to CMB and cosmological distance data, \cite{Smith:2019ihp} finds $S_8=0.842 \pm 0.014$, which is in $2.3\sigma$ tension with the DES-Y1 constraint $S_8 = 0.773^{+0.026} _{-0.020}$ \cite{Abbott:2017wau}. Moreover, $S_8$ is only a single number, while LSS data constrain the matter power spectrum $P(k)$ across a decade in $k$-space. As we show, EDE models that fit the CMB and SH0ES data produce significant effects on $P(k)$ beyond an overall amplitude change (as $S_8$ primarily captures), thereby suggesting the possibility of tightly constraining these models using LSS data. We note that late-universe constraints on early-universe resolutions have also been discussed in, e.g., \cite{Dutta:2019pio,Krishnan:2020obg}.
  
In this work, by ``large-scale structure data'' we refer to data sets that constrain not only the expansion history of the universe, e.g., via the BAO feature, but also the growth history, e.g., via weak gravitational lensing (including CMB lensing), photometric and/or spectroscopic galaxy clustering, galaxy cluster counts, etc.  In recent years, LSS data sets have delivered precise cosmological constraints, and any extension of the standard cosmological model must also satisfy these bounds.  Recent LSS breakthroughs have come from BOSS \cite{Alam:2016hwk}, a component of the Sloan Digital Sky Survey (SDSS), DES \cite{Abbott:2017wau}, the Kilo-Degree Survey (KiDS) \cite{Hildebrandt:2016iqg,2020A&A...633A..69H}, and the Subaru Hyper Suprime-Cam (HSC) survey \cite{Hikage:2018qbn}, amongst many others. Parallel to these new data sets, advances in the effective field theory of large-scale structure \cite{Carrasco:2012cv,Baumann:2010tm} have allowed $\Lambda$CDM parameters to be precisely measured with the redshift-space galaxy power spectrum \cite{Ivanov:2019hqk,Colas:2019ret,DAmico:2019fhj}, in particular, a CMB-independent $1.6\%$ measurement of the Hubble constant $H_0=68.6 \pm 1.1\, {\rm km/s /Mpc}$ \cite{Philcox:2020vvt}.  Powerful upcoming data sets from the Dark Energy Spectroscopic Instrument \cite{Levi:2019ggs}, Vera Rubin Observatory \cite{Ivezic:2008fe} (formerly LSST), Euclid \cite{Amendola:2016saw}, and WFIRST \cite{2019arXiv190205569A} are furthermore anticipated to significantly strengthen cosmological constraints.

In this work we reanalyze the EDE scenario taking into account \emph{Planck} 2018 and DES-Y1 data in detail (in addition to many other data sets), as well as approximate LSS constraints from KiDS and HSC. The DES-Y1 measurements are the most statistically powerful LSS data with publicly available likelihoods. We consider in detail the impact of EDE on the matter power spectrum and growth of structure, and the resulting constraints from LSS probes, in combination with CMB and cosmological distance information that has been used in previous EDE analyses.

The main results of this work are summarized in Fig.~\ref{fig:mainEDEconstraints} and Table~\ref{table:EDE-params-full}. We find no evidence for EDE in \emph{Planck} 2018 primary CMB data (TT+TE+EE) alone, but instead find an upper bound $f_{\rm EDE}<0.087$ at 95\% CL.  In contrast, when considering the same data set combination as used in~\cite{Poulin:2018cxd} and~\cite{Smith:2019ihp} (with CMB likelihoods updated to \emph{Planck} 2018), consisting of the primary CMB supplemented by \emph{Planck} 2018 CMB lensing data; BAO data from 6dF, SDSS DR7, and SDSS DR12; Pantheon SNIa data; the latest SH0ES $H_0$ constraint; and SDSS DR12 RSD data; we find $3.1 \sigma$ evidence for EDE, consistent with past claims in the literature. 

However, the inclusion of additional LSS data yields a downward trend in this result. The DES-Y1 3x2pt data bring the evidence for EDE down to 2.3$\sigma$.  Interestingly, we find that the results of this computationally expensive analysis are extremely well approximated by a simple Gaussian prior on $S_8$ (see Sec.~\ref{sec:constraintsKV450HSC}). Guided by this, we approximate HSC and KiDS data via priors on $S_8$, and find a further degradation of the evidence for EDE, $f_{\rm EDE}=0.062 ^{+0.032} _{-0.033}$, consistent with null at below $2\sigma$ (orange contours in Fig.~\ref{fig:mainEDEconstraints}). The one-sided upper bound is $f_{\rm EDE}<0.112$ at 95\% CL, and we constrain $H_0 = 70.45 \,^{+1.05}_{-1.08}$ km/s/Mpc. This constraint is in 2$\sigma$ tension with the SH0ES result on its own, shown by the gray bands in Fig.~\ref{fig:mainEDEconstraints}, indicating discordance.

To further assess the concordance of these varied data sets, we consider the fit to the combined data set including all likelihoods \emph{except} SH0ES. The results, shown as light green contours in Fig.~\ref{fig:mainEDEconstraints}, are statistically consistent with the fit to \emph{Planck} 2018 primary CMB data (TT+TE+EE) alone, and clearly inconsistent with SH0ES. This analysis yields an even tighter upper bound on EDE, $f_{\rm EDE}<0.060$ at 95\% CL, with $H_0 = 68.92 \, ^{+0.57} _{-0.59}$ km/s/Mpc, in strong tension with SH0ES.

Finally, we examine the choice of priors and the role of the axion decay constant. For computational efficiency, we limit ourselves to \emph{Planck} 2018 primary CMB data (TT+TE+EE) alone.  We find that uniform priors imposed directly on the particle physics parameters $f$ and $\log_{10}(m)$ (see Eq.~\eqref{eq.EDEdef}) strongly downweight large $f_{\rm EDE}$ values, in comparison to uniform priors placed on the effective EDE parameters $f_{\rm EDE}$ and $\log_{10}(z_c)$. This is reflected in the posterior distributions, and in particular, that for $H_0$, which is a near identical match to that in $\Lambda$CDM (see Fig.~\ref{fig:P18onlyconstraints}).

The outline of this paper is as follows: in Sec.~\ref{sec:EDE} we review the physics of the EDE proposal and its variants. In Sec.~\ref{sec:numimpl} we describe our numerical implementation of the EDE model in a publicly available code, \texttt{CLASS\_EDE} \cite{CLASS-EDE}.  In Sec.~\ref{sec:LSS} we study in detail the impact on LSS, particularly the matter power spectrum, and in Sec.~\ref{sec:Data}, we discuss the data sets used in our analysis.  We present our main results in Sec.~\ref{sec:constraints}, followed by an examination of physical priors in Sec.~\ref{sec:priors}, and we conclude in Sec.~\ref{sec:Discussion}. Additional figures can be found in the Appendices.

\section{The Early Dark Energy Proposal}
\label{sec:EDE}

The goal of the EDE proposal is to allow for larger values of $H_0$ than obtained in $\Lambda$CDM when analyzing CMB power spectrum data, while not degrading the overall quality of the fit. This goal is achieved by demanding that the angular acoustic scale, namely the ratio of the sound horizon at last scattering to the comoving angular diameter distance to last scattering (at redshift $z_* \approx 1100$),
\begin{equation}
    \theta_s = \frac{r_s (z_*)}{D_A(z_*)},
\end{equation}
be unchanged by the new physics introduced to solve the tension. The acoustic scale is the best-measured quantity in CMB data: it is constrained to 0.03\% precision in the \emph{Planck} 2018 analysis, $100 \theta_s = 1.0411 \pm 0.0003$ \cite{Aghanim:2018eyx}.  Upcoming CMB polarization data from Simons Observatory~ \cite{2019JCAP...02..056A} and CMB-S4~ \cite{2019arXiv190704473A} will independently constrain $\theta_s$ to this level (or better), providing a useful cross-check on the current CMB-temperature-dominated constraints.

The evolution of the Hubble parameter is encoded in $\theta_s$ via the integral expressions for $r_s$ and $D_A$ (here $c=1$), 
\begin{equation}
    r_s = \int _{z_*} ^\infty \frac{{\rm d} z}{H(z)} c_s(z) \;\; , \;\;    D_A = \int _0 ^{z_*} \frac{{\rm d} z}{H(z)} .
\end{equation}
The former depends sensitively on $H(z)$ in the two decades of scale factor evolution prior to recombination, while the latter depends directly on $H_0$ (and low-redshift cosmic evolution). It follows that a $\approx 10\%$ increase in $H_0$, i.e., of order the early-universe discrepancy with late-universe measurements, can be compensated for in $\theta_s$ by a $\approx 10\%$ increase in $H(z)$ just prior to recombination.

A simple mechanism to realize this effect, while not disrupting the rest of CMB physics and the ensuing cosmological evolution, is to introduce an additional contribution to the cosmic energy budget, which constitutes $\approx 10\%$ of the total energy density for a brief period just prior to recombination, and which rapidly decays away after achieving the required decrease in $r_s$.  Thus, the new component acts as dark energy at early times, and then rapidly becomes irrelevant after a critical redshift where it decays.  This early-time contribution to the cosmological constant is necessarily orders of magnitude greater than the present-day cosmological constant, $\rho_{\Lambda} ^{1/4}\simeq$ meV.  This hypothesized additional contribution is known as ``early dark energy''.

The simplest example of an effective cosmological constant which dynamically decays is that of a light scalar field. From the equation of motion of a scalar field $\phi$ with mass $m$ and potential $V(\phi) = m^2 \phi^2/2$,
\be
\ddot{\phi} + 3 H \dot{\phi} + m^2 \phi =0,
\ee
one can see that if initially $m \ll H$, then Hubble friction will freeze $\phi$ at its initial value $\phi_i$, contributing a vacuum energy $m^2 \phi_i ^2 / 2$ to the cosmological constant. Once the Hubble parameter drops below the mass, $m \gg H$, the field begins to oscillate, $\phi(t) = \phi_i a^{-3/2} \cos(m t)$, and the vacuum energy redshifts away as matter ($\propto a^{-3}$).

To utilize such a field to resolve the Hubble tension, the field must begin to oscillate in the rough ballpark of $z_{\rm CMB}$, at which point the Hubble parameter $H \sim T^2 / M_{pl}$ is roughly $10^{-27} \, {\rm eV}$.\footnote{We denote the reduced Planck mass as $M_{pl}=2.435 \times 10^{18} \, {\rm GeV}$ here and throughout.}  Thus the scalar field in question must be \emph{extremely} light. From a particle physics standpoint, the only known example of such a field is the axion \cite{Peccei:1977hh,Wilczek:1977pj,Weinberg:1977ma}.

The axion is defined as a (pseudo)-scalar endowed with a global $U(1)$ shift symmetry, broken by non-perturbative effects, namely instantons, that generate a potential (see, e.g., \cite{Montero:2015ofa}),
\be
\label{eq:instexp}
V(\phi) = \sum_{n} c_n e^{- S_n} \cos(n \phi/f)  \simeq V_0 \cos \frac{\phi}{f} + ... 
\ee
breaking the continuous shift symmetry to a discrete shift symmetry. This shift symmetry protects the axion mass from radiative corrections, allowing for extremely small values of the axion mass. The $...$ in the above equation indicates subdominant instantons, exponentially suppressed by the charge-$n$ instanton action $S_n$. Gravitational instantons scale as $S_n \simeq n M_{pl}/f$ \cite{Montero:2015ofa,Rudelius:2015xta}. If $f > M_{pl}$, the instanton expansion breaks down and the potential cannot be approximated by the leading term 
\cite{Banks:2003sx}, \cite{Rudelius:2014wla}.

To resolve the Hubble tension, the leading-order instanton will not suffice. The EDE field must rapidly decay away so as to leave low-redshift cosmic evolution unchanged, while the axion redshifts only as matter.  Thus, its effects would spoil late-time cosmology.  The proposal of \cite{Poulin:2018cxd} is then to consider potentials of the form (e.g.,~\cite{Kamionkowski:2014zda}),
\be
V = V_0 \left( 1 - \cos (\phi/f)\right)^n \;\; , \;\; V_0 \equiv m^2 f^2 \, ,
\label{eq.PoulinEDE}
\ee
corresponding to a careful fine-tuning of the hierarchy of instanton actions. For integer values of $n$, this fine-tuning is limited to the first $n$ terms in the expansion in Eq.~\eqref{eq:instexp}. For arbitrary real-valued $n$, one must instead fine-tune an \emph{infinite} tower of terms. For this reason, we will restrict our analysis to integer values of $n$ (primarily $n=3$).

The minimum of the potential \eqref{eq.PoulinEDE} is locally $V\sim \phi^{2n}$, in which case the oscillations of $\phi$ correspond to an equation of state \cite{PhysRevD.28.1243},
\be
w_{\phi} = \frac{n-1}{n+1} .
\ee
 For $n=2$, the initial energy stored in the field (i.e., the EDE) redshifts away during the oscillatory phase as radiation ($\propto a^{-4}$), and for $n\rightarrow \infty$ it redshifts as kinetic energy ($\propto a^{-6}$).

  These dynamics allow the model to be succinctly described in terms of an initial field displacement $\theta_i \equiv \phi_i/f$, and two effective parameters, $z_c$ and $f_{\rm EDE}$, which are defined by the redshift $z_c$ at which the EDE makes its maximal fractional contribution $f_{\rm EDE}$ to the total energy density of the universe, $f_{\rm EDE}(z_c) \equiv (\rho_{\rm EDE}/3M_{pl}^2 H^2) |_{z_c}$.\footnote{For notational simplicity, $f_{\rm EDE}(z_c)$ will be denoted as $f_{\rm EDE}$ in most contexts.} The dependence of $f_{\rm EDE}$ and $z_c$ on the particle physics parameters $m$ and $f$ and the initial angle $\theta_i$ is highly non-linear, a fact that we will return to in Sec.~\ref{sec:priors}.

\begin{figure}
\begin{center}
\includegraphics[width=\linewidth]{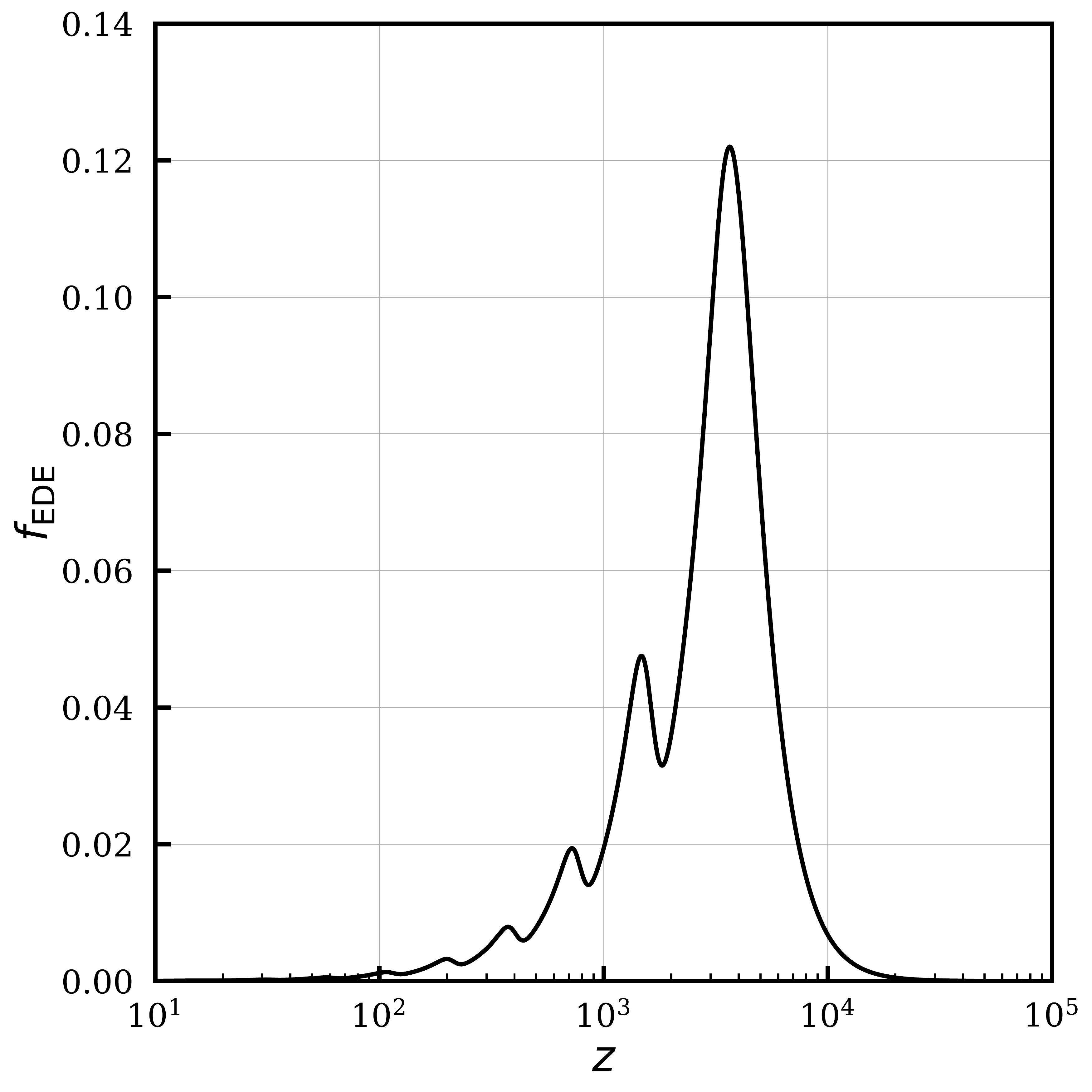}
\caption{Fraction of the cosmic energy density in the EDE field $\phi$ as a function of redshift, for the parameters in Eq.~\eqref{smithparams}. }
\label{EDEn3}
\end{center}
\end{figure}

 As a numerical example, we consider $n=3$, $f_{\rm EDE}=0.122$,  $\log_{10}(z_c)=3.562$, and $\theta_i = 2.83$. This is the best-fit integer-$n$ model reported in the \cite{Smith:2019ihp} fit to CMB, CMB lensing, BAO, redshift-space distortion (RSD), SNIa, and SH0ES data. We will refer to this example throughout; the full set of parameters is given by
 \begin{align}
     \label{smithparams}
     H_0&=72.19 \, {\rm km/s/Mpc} ,&100\omega_b=2.253 \\
     \omega_{\rm cdm}&=0.1306 ,&10^9 A_s = 2.215, \nonumber \\  n_s&=0.9889 ,&\tau_{\rm reio}=0.072 \nonumber \\
      f_{\rm EDE}&=0.122 \;\;,\;\;\log_{10}(z_c)=&3.562 \;\;,\;\;\theta_i = 2.83. \nonumber
 \end{align}
 The resulting evolution of the EDE component is shown in Fig.~\ref{EDEn3}. At its peak (near $z \approx 3650$), the EDE field comprises 12\% of the energy density of the universe. This is then rapidly dissipated as the field starts to roll and oscillate, and at $z= 10^3$ its contribution is less than 2\% of the energy density of the universe.

A minimal alternative to power-law cosine potentials is to consider instead the only aspect seemingly relevant to the Hubble tension, namely the shape of the potential at the minimum. This is the approach of \cite{Agrawal:2019lmo}, who studied
\be
V = V_0 \left( \frac{\phi}{M_{pl}} \right)^{2n} .
\ee
This coincides with the earlier models for small initial field values $\phi_i /f \ll 1$. The most recent statistical analysis \cite{Smith:2019ihp} found such monomial potentials are disfavoured relative to the cosine potential with a large initial field displacement.

There are now many realizations of the EDE scenario that have been proposed. Unstable dark energy, a.k.a ``Axion-Dilaton Destabilization'' \cite{Alexander:2019rsc}, is a realization of EDE \emph{without} higher-order instantons. This is done with a two-field model:
\be
\label{V}
V(\phi,\chi) = \frac{1}{2}m_{\chi}^2 f^2 e^{\beta(\phi)}  (1+\cos \frac{\chi}{f}) + V_0 e^{- \lambda \phi/M_{pl}} ,
\ee
where $\beta(0)>0$. The axion $\chi$ rolls down its potential at a time $z_{c}$, triggering the destabilization of a second scalar field $\phi$ with a steep potential, $\lambda \gg 1$. The two-field model \cite{Alexander:2019rsc} can be qualitatively understood by considering a single-field truncation, with a piece-wise-defined potential for the EDE field, in a manner resembling the best-fit ``Acoustic Dark Energy'' of \cite{Lin:2019qug}. Other EDE-like possibilities include a ``kick'' from neutrino freeze-out \cite{Sakstein:2019fmf}, a first-order phase transition \cite{Niedermann:2019olb}, parametric resonance \cite{Kaloper:2019lpl}, and dissipation into gauge fields \cite{Berghaus:2019cls}. For this work, we will concentrate on the cosine potentials as proposed in \cite{Poulin:2018cxd}, which have been shown to fit cosmological data well and serve as a canonical example of the EDE scenario.

\begin{figure*}
    \centering
    \includegraphics[width=\linewidth]{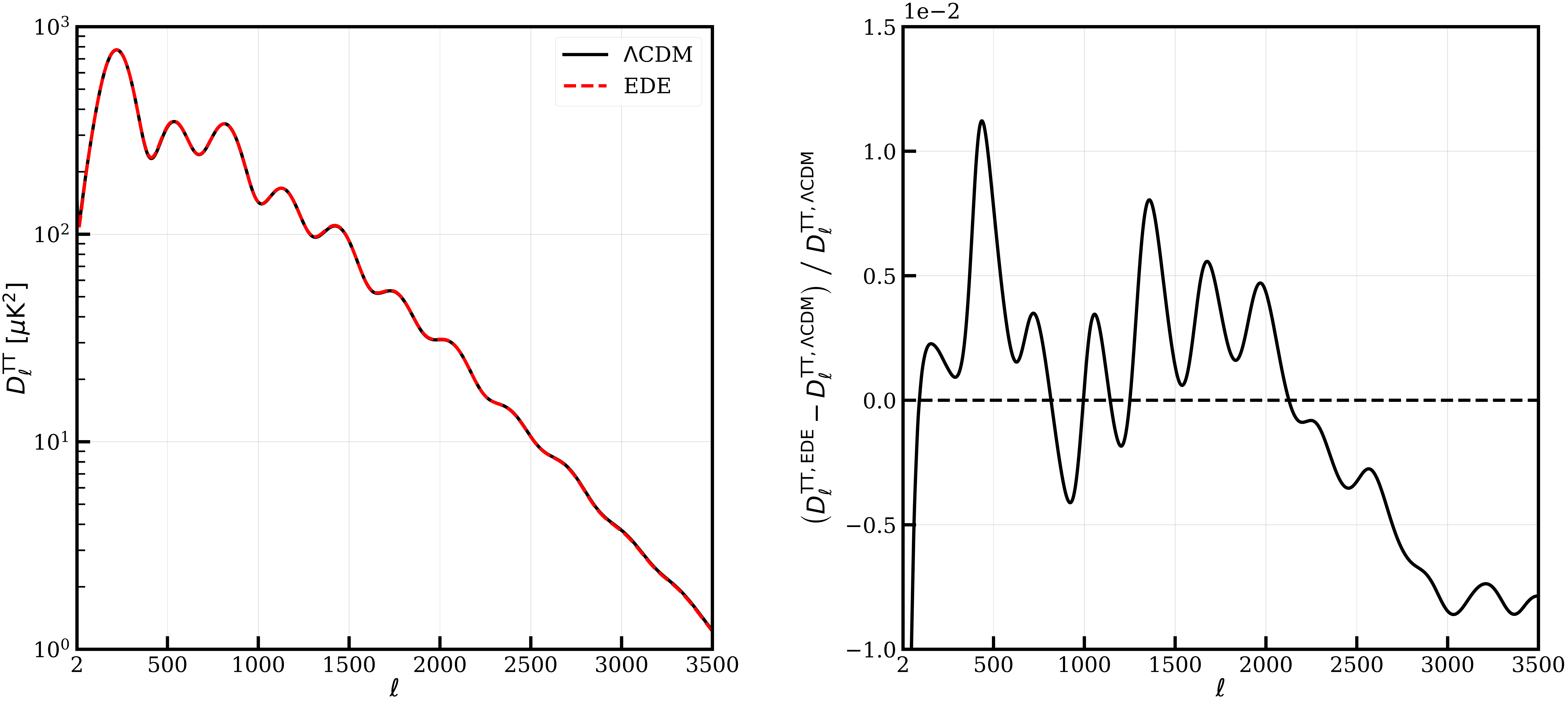}
    \caption{
    CMB temperature anisotropy power spectra (left panel) and residuals (right panel) for $\Lambda $CDM (black, solid) and EDE (red, dashed) models, with $H_0 = 68.21$ km/s/Mpc and $H_0 = 72.19$ km/s/Mpc, respectively. The curves are nearly indistinguishable in the left panel. The model parameters are given by Eqs.~\eqref{smithparams} and~\eqref{smithparamsLCDM} for EDE and $\Lambda$CDM, respectively, corresponding to the \cite{Smith:2019ihp} best-fit models to primary CMB, CMB lensing, BAO, RSD, SNIa, and SH0ES data.}
    \label{fig:CMB_TT}
\end{figure*}

The hallmark success of this proposal lies in preserving the fit to the \emph{Planck} CMB temperature power spectrum. The best-fit EDE model analyzed in \cite{Smith:2019ihp} has $H_0 \approx 72$ km/s/Mpc, while leaving no visible imprint on the CMB compared to a $\Lambda$CDM model with significantly lower $H_0$. To illustrate this, we show the CMB temperature power spectra in $\Lambda$CDM with $H_0 \approx 68$ km/s/Mpc and in a fiducial EDE model with $H_0 \approx 72$ km/s/Mpc, in Fig.~\ref{fig:CMB_TT}. Analogous figures for the CMB polarization and lensing power spectra (including the fractional change with respect to $\Lambda$CDM) can be found in Appendix~\ref{app:EDEvsLCDMfigs} (see Figs.~\ref{CMBTEgg} and \ref{CMBTTEE}).  The model parameters for these figures are chosen as the best-fit values reported in \cite{Smith:2019ihp}: EDE with parameters as in Eq.~\eqref{smithparams}, and $\Lambda$CDM with parameters given by
\begin{align}
     \label{smithparamsLCDM}
     H_0&=68.21 \, {\rm km/s/Mpc} &   100\omega_b&=2.253, \\
     \omega_{\rm cdm}&=0.1177 &10^9 A_s &= 2.216, \nonumber \\  n_s&=0.9686 &    \tau_{\rm reio}&=0.085. \nonumber 
 \end{align}
Figs.~\ref{fig:CMB_TT} and~\ref{CMBTTEE} show that the primary CMB power spectra are nearly indistinguishable in these two models, despite the EDE model having a significantly larger $H_0$ than the $\Lambda$CDM model.  This suggests that EDE can provide a simple early-universe solution to the Hubble tension.

\section{Numerical Implementation}
\label{sec:numimpl}

We implement the EDE scenario as a modification to the publicly available Einstein-Boltzmann code \texttt{CLASS} \cite{2011arXiv1104.2932L,2011JCAP...07..034B}.\footnote{\url{http://class-code.net}} Our modified version, \texttt{CLASS\_EDE}, is publicly available \cite{CLASS-EDE}.\footnote{\url{https://github.com/mwt5345/class_ede}} \texttt{CLASS\_EDE} solves for the evolution of the scalar field perturbations directly using the perturbed Klein--Gordon equation (as in, e.g.,~\cite{Agrawal:2019lmo} and~\cite{Smith:2019ihp}), avoiding the effective fluid description used in some past works (e.g.,~\cite{Poulin:2018cxd}). We implement adiabatic initial conditions for the scalar field fluctuations as described in~\cite{Smith:2019ihp}; see \cite{Smith:2019ihp} for a discussion of isocurvature initial conditions. The EDE potential is implemented as
\begin{equation}
V(\phi) = m^2 f^2\left( 1 - \cos{( \phi/f)} \right)^n + V_\Lambda,
\label{eq.EDEdef}
\end{equation}
where $V_\Lambda$ is a constant, which plays the role of the cosmological constant. Absorbing the cosmological constant into $V(\phi)$ allows efficient closure of the energy budget equation, $\sum \Omega_i=1$ in a flat universe, for arbitrary model parameters simply by tuning $V_\Lambda$ via the built-in shooting functionality of \texttt{CLASS}. 

\texttt{CLASS\_EDE} allows one to specify the EDE model parameters in terms of the particle physics parameters $f$ and $m$ or the effective EDE parameters $f_{\rm EDE}$ and $z_c$. If the latter set is specified, $\texttt{CLASS\_EDE}$ will find the corresponding $f$ and $m$ via a shooting algorithm, analogous to the shooting of $H_0$ from a user-specified $100\, \theta_s$ in \texttt{CLASS}. For both implementations the user must also specify the initial axion misalignment angle $\theta_i\equiv \phi_i/f$ and a value for $n$. To handle dynamics for small values of $f_{\rm EDE}$, we have increased the default time-step precision in \texttt{CLASS} to $7\times 10^{-4}$. The final update to the functionality of \texttt{CLASS} is a simple extraction of $f \sigma_8(z)$, where here $f \equiv d \log D/d \log a$ is the linear growth rate (not the axion decay constant), which is needed to implement the RSD likelihoods in our analysis below.  In all likelihoods requiring calculations of the non-linear matter power spectrum, we compute this quantity using the ``Halofit'' prescription implemented in \texttt{CLASS} \cite{Smith:2002dz,Takahashi2012}.

We perform Markov chain Monte Carlo (MCMC) analyses using the publicly available code \texttt{Cobaya} \cite{torrado_lewis_2019}. We sample from the posterior distributions using the Metropolis-Hastings algorithm implemented in \texttt{Cobaya}~\cite{LewisBridle2002,Lewis2013,Neal2005}, with a Gelman-Rubin \cite{Gelman:1992zz} convergence criterion $R-1 < 0.05$.  To determine best-fit parameter values, we use the ``BOBYQA'' likelihood maximization method implemented in \texttt{Cobaya}~\cite{Powell2009,Cartis2018a,Cartis2018b}.

\section{Imprint on Large-Scale Structure}
\label{sec:LSS}

The introduction of the EDE field into the cosmological model affects observables beyond the CMB temperature and polarization power spectra.  In particular, it affects the dynamics of all perturbation modes that are within the horizon during the epoch in which the EDE is relevant.  This change in the transfer function leaves signatures in the late-time matter power spectrum $P(k)$.  Moreover (and more significantly), the ``standard'' cosmological parameters must shift in the EDE scenario in order to maintain the fit to the primary CMB data while accommodating a higher $H_0$ value than possible in $\Lambda$CDM.  These shifts, particularly in $\omega_{\rm cdm}$, significantly affect $P(k)$.  As precise measurements of this observable are available from current surveys (e.g.,~\cite{Alam:2016hwk,Abbott:2017wau,Hildebrandt:2016iqg,Hikage:2018qbn}) and will dramatically grow in precision with near-future surveys (e.g.,~\cite{Levi:2019ggs,Ivezic:2008fe,Amendola:2016saw,2019arXiv190205569A}), it is an opportune time to examine their role in constraining the EDE scenario.

Following general conventions, we will often quantify LSS constraints by the $\sigma_8$ parameter, i.e., the RMS linear-theory mass fluctuation in a sphere of radius $8 \, {\rm Mpc}/h$ at $z = 0$. This is evaluated in Fourier space as an integral over the matter power spectrum with a spherical top-hat filter $W(kR)$ of radius $R = 8 \, {\rm Mpc}/h$, i.e.,
\be
\label{sigma8}
(\sigma_8 )^2 = \frac{1}{2 \pi^2}\int {\rm d}\log k \, W^2(kR) \, k^3 P(k) \, .
\ee
The value of $\sigma_8$ is predominantly determined by contributions to the integral from $0.1 \, h/{\rm Mpc} \lesssim k \lesssim 1 \, h/{\rm Mpc}$, due to a balancing of high-$k$ suppression of the filter and the small-$k$ suppression from the $k^3$ factor.

In recent years, CMB observations have consistently yielded best-fit values of $\sigma_8$, or similarly $S_8 \equiv \sigma_8 (\Omega_m/0.3)^{0.5}$, that are slightly greater than those found by LSS observations. In the fit to $\Lambda$CDM, the {\it Planck} 2018 analysis finds $S_8=0.830 \pm 0.013$ \cite{Aghanim:2018eyx}, while the DES-Y1 3x2pt analysis finds $S_8=0.773 ^{+0.026}
_{-0.020}$ \cite{Abbott:2017wau}, KiDS finds $S_8=0.745 \pm0.039$ \cite{Hildebrandt:2016iqg}, and HSC finds $S_8=0.780 ^{+0.030}
_{-0.033}$ \cite{Hikage:2018qbn}. Taken in conjunction as three independent measurements and combined with inverse-variance weights, these LSS experiments yield
$S_8=0.770 ^{+0.018}
_{-0.016}$, in $2.7\sigma$ tension with the {\it Planck} 2018 CMB result.

\begin{figure}[ht]
    \centering
    \includegraphics[width=\linewidth]{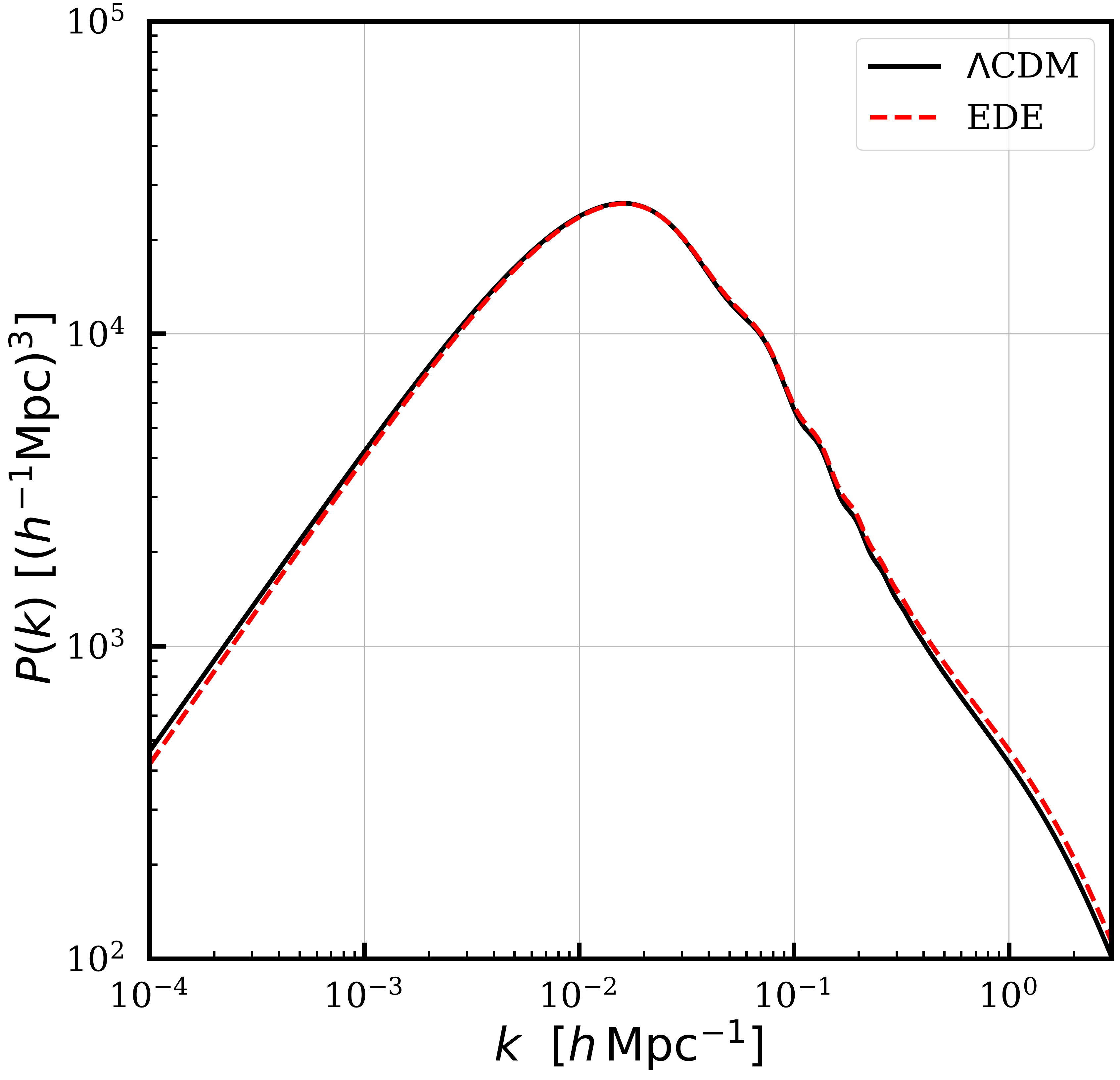}
    \caption{Non-linear matter power spectrum $P(k)$ at $z=0$ for $\Lambda$CDM and EDE models that fit the primary CMB, distances, and SH0ES data.  The change in $\sigma_8$ in the EDE scenario can be seen as the relative increase in $P(k)$ in the range $0.1 \, h/{\rm Mpc} \lesssim k \lesssim 1 \, h/{\rm Mpc}$ (although $\sigma_8$ is computed from the linear rather than non-linear power spectrum).  This increase is due primarily to shifts in the ``standard'' cosmological parameters in the EDE model, rather than the effects of the EDE itself.  The model parameters are the same as in Fig.~\ref{fig:CMB_TT} (see Eqs.~\eqref{smithparams} and~\eqref{smithparamsLCDM}).}
    \label{fig:Pk}
\end{figure}

This tension is worsened in the EDE scenario. For example, the \cite{Smith:2019ihp} results for the best-fit integer-$n$ EDE model give $S_8 = 0.842 \pm 0.014$, increasing the tension with the LSS result quoted above to $3.2\sigma$.  Moreover, $S_8$ is only a single number, while LSS data constrain the shape of $P(k)$ over a decade in $k$-space.  The value of $S_8$ depends on multiple $\Lambda$CDM parameters, which are shifted in the EDE scenario in order to maintain the fit to the CMB acoustic peaks and the damping tail of the power spectrum. The upward shift of $S_8$ is predominantly driven by the increase in the physical CDM density, which slightly shifts the peak of the matter power spectrum and increases the growth rate of perturbations at late times.  To illustrate this effect, in Fig.~\ref{fig:Pk} we plot the non-linear matter power spectrum, computed with the ``Halofit'' prescription implemented in \texttt{CLASS}~\cite{Smith:2002dz,Takahashi2012}, which is used in our analysis of LSS data in Sec.~\ref{sec:constraints}. The increase in power at $0.1 \, h/{\rm Mpc} \lesssim k \lesssim 1 \, h/{\rm Mpc}$ leads to an increase in $\sigma_8$ and $S_8$ (although these quantities are of course computed from the linear rather than non-linear power spectrum), and thus a worsening of the tension between these CMB-derived predictions and actual LSS data.  We emphasize that this primarily arises from the change to the $\Lambda$CDM parameters in the EDE scenario, as is required to produce CMB power spectra that match the \emph{Planck} fit to $\Lambda$CDM (see Fig.~\ref{fig:CMB_TT}).

On large scales ($k \lesssim 10^{-2}\, h/{\rm Mpc}$), outside the reach of current LSS experiments, the EDE $P(k)$ is suppressed relative to that in $\Lambda$CDM.  This difference is driven by the slight change in the primordial spectral index, with $n_s=0.9889$ and $n_s=0.9686$ in EDE and $\Lambda$CDM, respectively, while the amplitude at the pivot scale $k_{\rm piv} \equiv 0.05 \, {\rm Mpc}^{-1}$ remains essentially unchanged ($A_s=2.215\times10^{-9}$ and $A_s=2.216\times10^{-9}$ in EDE and $\Lambda$CDM, respectively).  

\begin{figure}[ht]
    \centering
    \includegraphics[width=\linewidth]{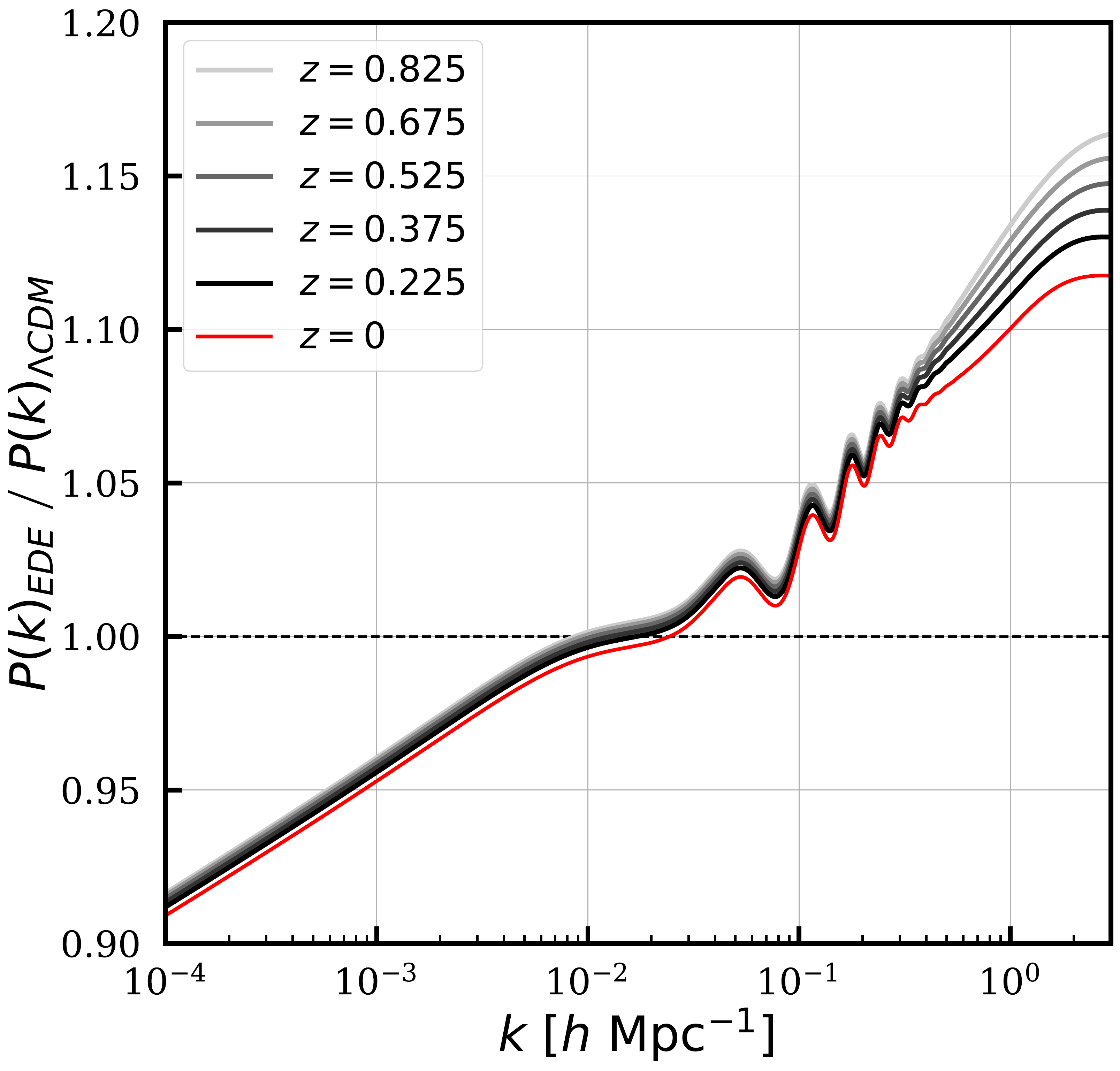}
    \caption{Ratio of the EDE and $\Lambda$CDM non-linear matter power spectra at $z$ values chosen to be the midpoints of the redshift bins used in the DES-Y1 analysis (and at $z=0$ in red).  The model parameters are the same as in Fig.~\ref{fig:CMB_TT}.}
    \label{fig:Pkratio}
\end{figure}

These differences persist across redshift: in Fig.~\ref{fig:Pkratio} we show the ratio of $P(k)$ in the two scenarios at $z=0$ (i.e., the ratio of the curves in Fig.~\ref{fig:Pk}) and at the midpoints of the DES-Y1 redshift bins. From this, one can see that the change in the matter power spectrum is substantial, up to $\approx 10\%$ for a wide range of wavenumbers that are well-measured in current data. The figure also shows that the quasi-linear and small-scale changes are more significant at higher redshift.

The redshift dependence of the deviations from $\Lambda$CDM are also encoded in the growth factor, $f(z)$, as well as $f\sigma_8(z)$. We include plots of these quantities in Appendix~\ref{app:EDEvsLCDMfigs} in Figs.~\ref{fig:fsigma8} and~\ref{fplot}.  The enhancement in EDE of $f\sigma_8$ at $z=1$ is twice that at $z=0$, given by $3\%$ and $1.5\%$, respectively. In comparison, BOSS RSD data provide a 6\% measurement of $f\sigma_8$ at $z = 0.38, 0.51$, and $0.61$ \cite{Alam:2016hwk}.  Upcoming measurements from DESI will significantly improve upon these constraints~\cite{Levi:2019ggs}.

\begin{figure}[ht]
    \centering
    \includegraphics[width=1.\linewidth]{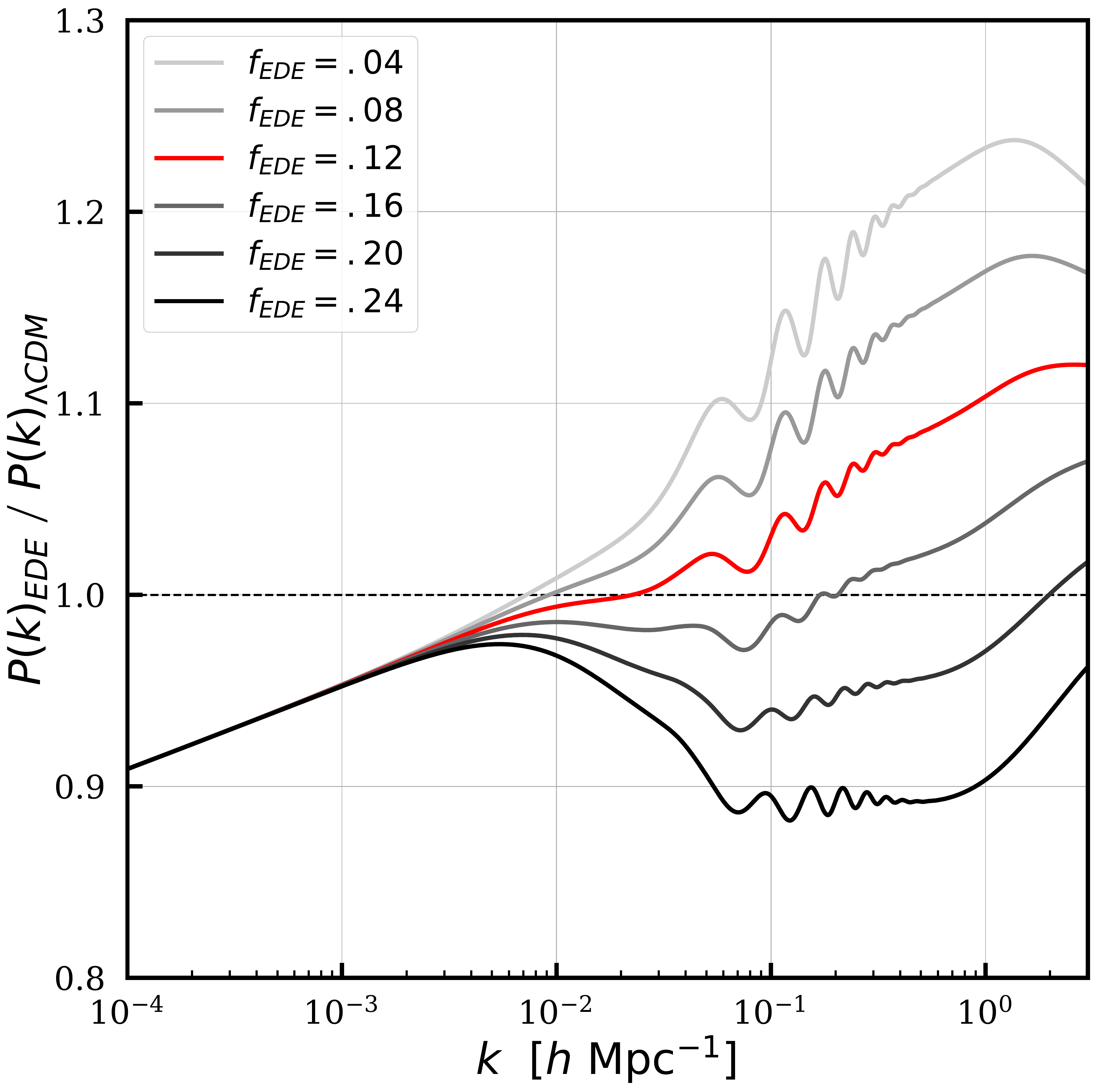}
    \caption{Ratio of $P(k)$ in EDE to that in $\Lambda$CDM as a function of the EDE fraction $f_{\rm EDE}$, at fixed $\log_{\rm 10}z_c=3.526$ and $\theta_i = 2.83$.  The other model parameters are given in Eq.~\eqref{smithparams}.  As $f_{\rm EDE}$ increases, the growth of perturbations that are within the horizon during the EDE epoch is suppressed by a progressively greater amount.  The red curve here is identical to that shown in Fig.~\ref{fig:Pkratio}.}
    \label{fig:fEDEPk}
\end{figure}

\begin{figure}[ht]
    \centering
    \includegraphics[width=1.\linewidth]{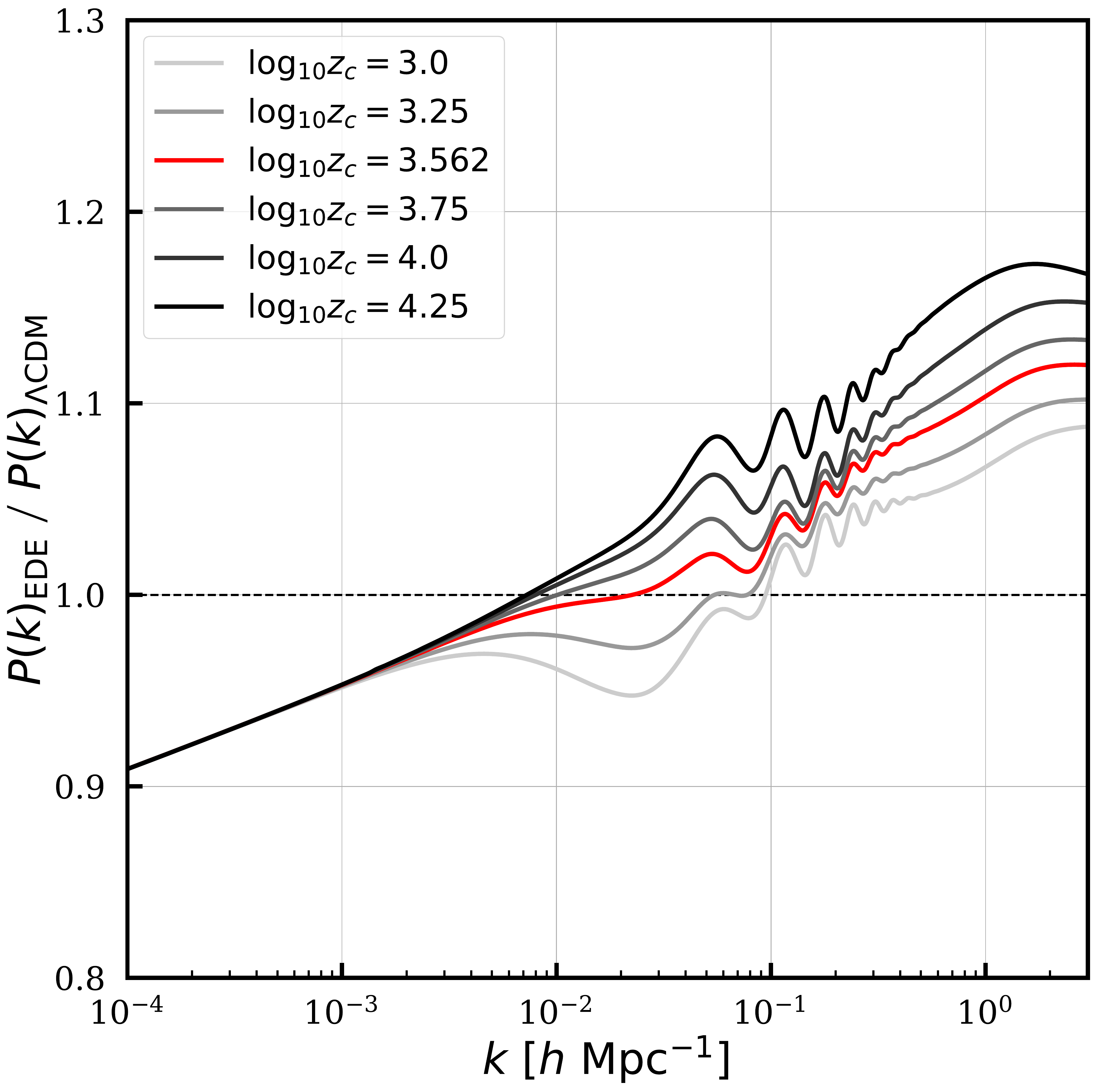}
    \caption{Ratio of $P(k)$ in EDE to that in $\Lambda$CDM as a function of the critical redshift $\log_{10}z_{c}$, at fixed $f_{\rm EDE}=0.12$ and $\theta_i=2.83$.  The other model parameters are given in Eq.~\eqref{smithparams}. As the critical redshift decreases, the growth suppression due to EDE is pushed to progressively later times, and thus affects modes on correspondingly larger scales (lower $k$).}
    \label{fig:zcPk}
\end{figure}

The $\sigma_8$ change does not fully capture the rich impact of EDE on the matter power spectrum.  The effects of the EDE field modify the dynamics of all modes within the horizon (or those that re-enter the horizon) during the epoch in which the EDE makes a non-negligible contribution to the cosmic energy budget, i.e., around $z_c$ (with a wider redshift window for larger $f_{\rm EDE}$).  These effects are sensitive to the amount and timing of EDE, as parameterized by $f_{\rm EDE}$ and $z_c$.  The imprint of $f_{\rm EDE}$ on $P(k)$ can be seen in Fig.~\ref{fig:fEDEPk}, while holding $z_c$ and $\theta_i$ fixed. Similarly, in Fig.~\ref{fig:zcPk} we show the matter power spectrum as a function of $\log_{10}(z_c)$, with $f_{\rm EDE}$ and $\theta_i$ held fixed. In both cases the $\Lambda$CDM parameters are also held fixed (to their values in Eq.~\eqref{smithparams}).

These figures show that $f_{\rm EDE}$ acts to suppress structure on small scales, with an effect that is compounded for small values of $z_c$, that is, models in which the EDE is relevant in the late universe.  Physically, this is due to the suppression of perturbation growth by the accelerated expansion, analogous to (but weaker than) that in late-time dark energy domination. Quantitatively, we confirm this intuition by computing the wavenumber $k_c$ corresponding to the size of the comoving horizon at $z_c$, when the EDE has maximal influence on the dynamics.  For the fiducial model considered in this section with $\log_{10}(z_c) = 3.562$, we find $k_c \approx 0.03 \, h$/Mpc.  Fig.~\ref{fig:fEDEPk} clearly shows increasing suppression for modes with $k>k_c$ as $f_{\rm EDE}$ increases, which makes sense as these modes are all within the horizon at that time.  There is also some suppression for modes with slightly lower $k$, as these modes re-enter the horizon while the EDE is still a non-negligible contribution to the cosmic energy budget.

\begin{figure*}[ht!]
\centering
\includegraphics[width=\linewidth]{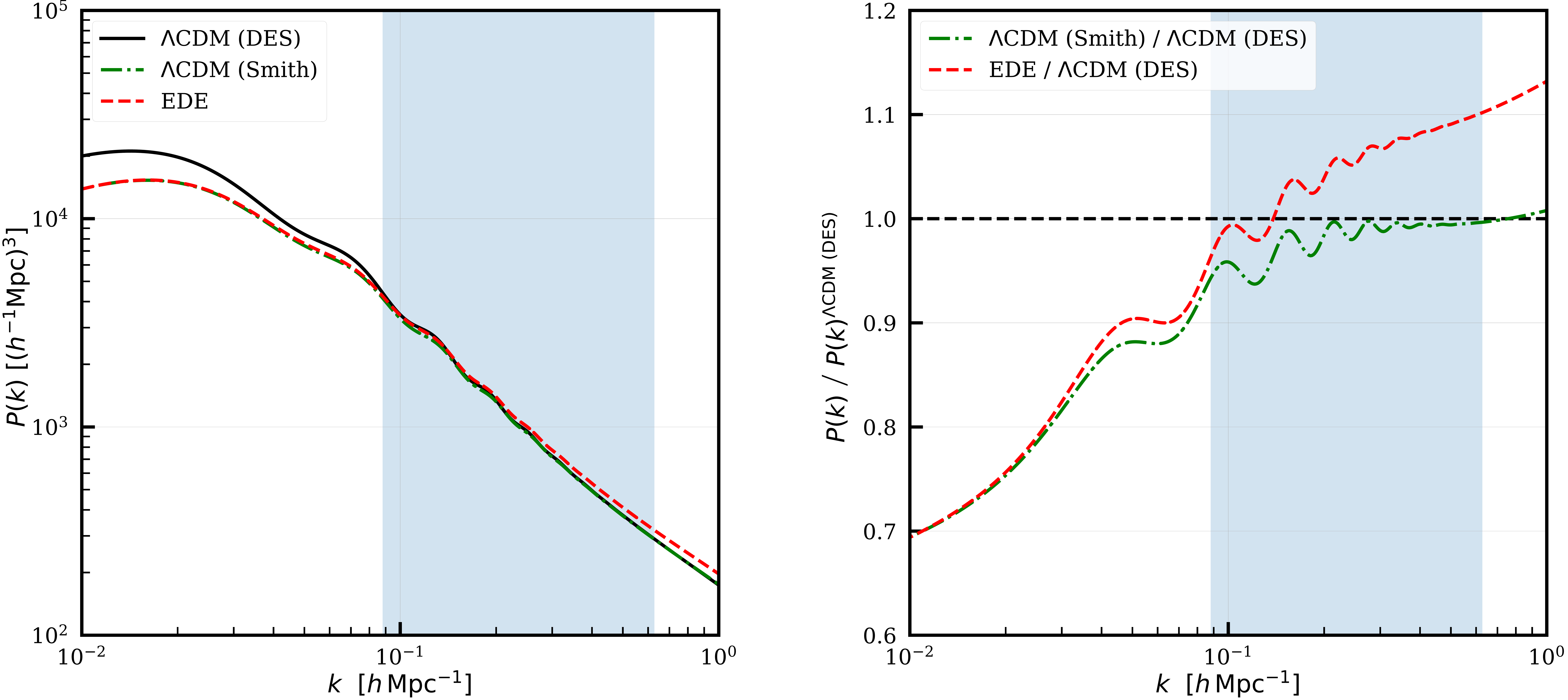}
\caption{Matter power spectra (left panel) and ratio of spectra (right panel) at the central redshift of DES observations, $z=0.525$, for $\Lambda$CDM with the DES-Y1 best-fit values $\Omega_m=0.267$ and $S_8=0.773$ (black, solid), the best-fit EDE model from~\cite{Smith:2019ihp} with parameters given in Eq.~\eqref{smithparams} (red, dashed), and the best-fit $\Lambda$CDM model from~\cite{Smith:2019ihp} with parameters given in Eq.~\eqref{smithparamsLCDM} (green, dash-dotted). The ratios in the right panel are computed with respect to DES-Y1 $\Lambda$CDM, thus giving an indication of how well the other two models' predictions match the DES-Y1 constraints. The blue shaded region is the approximate range of comoving wavenumbers probed by the angular scale cuts used in the DES-Y1 analysis.}
\label{fig:PkDES}
\end{figure*}

Finally, to contextualize the EDE impact on LSS, we consider a comparison between the matter power spectrum in EDE and a model consistent with DES-Y1 measurements of photometric galaxy clustering, galaxy-galaxy lensing, and cosmic shear two-point correlation functions~\cite{Abbott:2017wau}.  The latter yield constraints $S_8 = 0.773 ^{+0.026} _{-0.020}$ and $\Omega_m=0.267 ^{+0.030}
_{-0.017}$. The DES measurements are generally insensitive to the other $\Lambda$CDM parameters, and we adopt {\it Planck} 2018 \cite{Aghanim:2018eyx} TT+TE+EE best-fit values $n_s=0.9649$, $h=0.6727$, $\tau_{\rm reio}=0.0544$, and $\Omega_b h^2=0.02237$ to complete the model. The amplitude $A_s$ is set by CLASS to reproduce the DES measurement of $\sigma_8$, which gives $A_s=2.788 \times 10^{-9}$. We consider a redshift $z=0.525$, corresponding to the central redshift bin of DES.

A comparison of $P(k)$ in the best-fit EDE and $\Lambda$CDM models of \cite{Smith:2019ihp}, Eqs.~\eqref{smithparams} and~\eqref{smithparamsLCDM}, and the model consistent with DES-Y1 is shown in Fig.~\ref{fig:PkDES}. The blue shaded region corresponds to the approximate range of comoving wavenumbers probed the DES angular correlation functions, which span the range $2.5' < \theta < 250'$, with a lower scale cut imposed at comoving separations $R \approx 2$-$12 \, {\rm Mpc}/h$, depending on the observable. In particular, the right panel of Fig.~\ref{fig:PkDES} displays the ratio of $P(k)$ predicted by the EDE model to that inferred by DES in $\Lambda$CDM; this shows an even greater suppression of power on large scales than in Fig.~\ref{fig:Pk}, and an even greater enhancement on small scales. The enhancement on small scales can again be understood in terms of the physical CDM density $\Omega_c h^2$, which is $\Omega_c h^2=0.0984$ for DES (with $h$ and $\Omega_b h^2$ fixed by {\it Planck}), but $\Omega_c h^2 = 0.1306$ in the EDE model.

The suppression on large scales, which is beyond the observable range of DES or other current surveys, is driven by the significant shift in $A_s$, enhanced by the shift in $n_s$, and to a lesser extent by the significant shift in the matter density $\Omega_m$, which is $\Omega_m=0.267$ for DES and $\Omega_m=0.303$ for the EDE model parameters in Eq.~\eqref{smithparams}.

\section{Data Sets}
\label{sec:Data}

We consider the following data sets in our MCMC analyses:

\subsection{CMB}
\emph{Planck} 2018 \cite{Planck2018likelihood,Aghanim:2018eyx,2018arXiv180706210P} low-$\ell$ and high-$\ell$ [\texttt{Plik}] temperature, polarization, and CMB lensing power spectra. In comparison with the \emph{Planck} 2015 results \cite{Planck2015likelihood,Ade:2015xua}, the primary change in the fit to $\Lambda$CDM is a shift in the mean value and tightening in the error bar on the optical depth to reionization, from $\tau_{\rm reio}=0.066 \pm 0.016$ in 2015 to $\tau_{\rm reio}=0.054\pm 0.007$ in 2018, as well as a small shift downward of $n_s$ and a small shift upward in $\omega_{\rm cdm}$.

\subsection{LSS}
In addition to the {\it Planck} 2018 CMB lensing data set~\cite{2018arXiv180706210P}, which we consider to be an LSS data set as it probes the low-redshift universe, we include:\\

\indent {\it BAO:}
Measurements from the SDSS DR7 main galaxy sample~\cite{Ross:2014qpa} and the 6dF galaxy survey~\cite{2011MNRAS.416.3017B} at $z = 0.15$ and $z = 0.106$, respectively, as well as from the SDSS BOSS DR12~\cite{Alam:2016hwk} optimally combined LOWZ and CMASS galaxy samples at $z = 0.38$, $0.51$, and $0.61$.  \\

\indent{\it RSD:} SDSS BOSS DR12 \cite{Satpathy:2016tct,Alam:2016hwk} measurements of $f\sigma_8(z)$ from the imprint of peculiar velocities on the conversion between configuration- and redshift-space \cite{1987MNRAS.2271K}, at $z = 0.38$, $0.51$, and $0.61$. We include the full covariance of the joint BOSS DR12 BAO and RSD data.  In particular, we use the ``consensus'' final BAO+RSD results given in~\cite{Alam:2016hwk} in their Table 7 (final column) and the covariance given in their Table 8.\footnote{We note that, at the time our analysis was performed, there was a non-negligible bug in the implementation of the BOSS DR12 BAO+RSD likelihood in the \texttt{Monte Python} MCMC sampling code, which has afflicted previous EDE analyses in the literature (\url{https://github.com/brinckmann/montepython_public/issues/112}).  We have explicitly checked that this bug is not present in \texttt{Cobaya}, and that \texttt{Cobaya} reproduces the $\chi^2$ values for this likelihood found in the official \textit{Planck} 2018 MCMC chains.}\\ 

\indent{\it DES:} shear-shear, galaxy-galaxy, and galaxy-shear two-point correlation functions (often referred to as ``3x2pt''), measured from 26 million source galaxies in four redshift bins and 650,000 luminous red lens galaxies in five redshift bins, for the shear and galaxy correlation functions, respectively \cite{Abbott:2017wau}.  When analyzed in $\Lambda$CDM, the DES 3x2pt likelihood gives tight constraints on $S_8$ and $\Omega_m$, $S_8 = 0.773 ^{+0.026} _{-0.020}$  and $\Omega_m=0.267 ^{+0.030}_{-0.017}$, respectively. \\ 

\indent{\it Additional LSS data:} Weak gravitational lensing measurements from KiDS+VIKING-450 (hereafter KiDS or KV-450) \cite{Hildebrandt:2016iqg,2020A&A...633A..69H} and the Subaru Hyper Suprime-Cam (HSC) \cite{Hikage:2018qbn}, which provide complementary data sets to the Dark Energy Survey, and impose similar (though slightly weaker) constraints on $S_8$ and $\Omega_m$.  We do not include likelihoods for these datasets directly, but we approximately include their effect via priors on $S_8$.  For KV-450, we use the result from \cite{2020A&A...633A..69H}: $S_8 = 0.737^{+0.040}_{-0.036}$.  For HSC, we use the result from \cite{Hikage:2018qbn}: $S_8 = 0.780^{+0.030}_{-0.033}$.

\subsection{Supernovae}

The Pantheon dataset of 1048 SNe Ia in the redshift range $0.01 < z < 2.3$ \cite{Scolnic:2017caz}, which provide accurate relative luminosity distances.  We note that, as of writing, there is a discrepancy in the Pantheon likelihood implemented in \texttt{Cobaya} and that implemented in \texttt{Monte Python}, another popular cosmological MCMC sampling code.\footnote{\url{https://github.com/brinckmann/montepython_public/issues/131}}  The origin of this discrepancy is presently unknown, but the $\chi^2$ values computed for Pantheon using \texttt{Cobaya} match those found in the official \textit{Planck} 2018 MCMC chains.\footnote{\url{https://wiki.cosmos.esa.int/planck-legacy-archive/index.php/Cosmological_Parameters}}  We thus utilize the likelihood as implemented in \texttt{Cobaya}.

\subsection{Local Distance Ladder}
The most recent SH0ES measurement $H_0 = 74.03 \pm 1.42 \, {\rm km/s/Mpc}$ \cite{Riess:2019cxk}.  We correct a minor bug in the implementation of this likelihood in \texttt{Cobaya} that was present at the time of writing,\footnote{\url{https://github.com/CobayaSampler/cobaya/issues/105}} which produces a constant offset in the $\chi^2$ values.  This does not affect parameter constraints (as a constant offset in $\chi^2$ is irrelevant in MCMC analyses), but affects the best-fit $\chi^2$ values quoted.  We have checked that after fixing this bug, the $\chi^2$ values produced by the likelihood match those in the official \textit{Planck} 2018 MCMC chains.

\section{Constraints on the EDE Scenario}
\label{sec:constraints}

\begin{table*}[htb!]
Constraints from \emph{Planck} 2018 data only: TT+TE+EE \vspace{2pt} \\
  \centering
  \begin{tabular}{|l|c|c|}
    \hline\hline Parameter &$\Lambda$CDM~~&~~~EDE ($n=3$) ~~~\\ \hline \hline

    {\boldmath$\ln(10^{10} A_\mathrm{s})$} & $3.044 \, (3.055) \, \pm 0.016$ & $3.051 \, (3.056) \, \pm 0.017$ \\

    {\boldmath$n_\mathrm{s}$} & $0.9645 \, (0.9659) \, \pm 0.0043 $ & $0.9702 \, (0.9769)^{+0.0071}_{-0.0069}$ \\

    {\boldmath$100\theta_\mathrm{s}$} & $1.04185 \, (1.04200) \, \pm 0.00029 $ & $1.04164 \, (1.04168) \, \pm 0.00034$\\

    {\boldmath$\Omega_\mathrm{b} h^2$} & $0.02235 \, (0.02244) \, \pm 0.00015 $ &  $0.02250 \, (0.02250) \, \pm 0.00020$ \\

    {\boldmath$\Omega_\mathrm{c} h^2$} & $0.1202 \, (0.1201)\, \pm 0.0013 $ & $0.1234 \, (0.1268)^{+0.0031}_{-0.0030}$ \\

    {\boldmath$\tau_\mathrm{reio}$} & $0.0541 \, (0.0587) \, \pm 0.0076 $ & $0.0549 \, (0.0539) \, \pm 0.0078 $\\

    {\boldmath$\mathrm{log}_{10}(z_c)$} & $-$ & $3.66 \, (3.75)^{+0.28}_{-0.24}$ \\

    {\boldmath$f_\mathrm{EDE} $} & $-$ & $< 0.087 \, (0.068)$\\

    {\boldmath$\theta_i$} & $-$ & $> 0.36 \, (2.96) $\\

    \hline

    $H_0 \, [\mathrm{km/s/Mpc}]$ & $67.29 \, (67.44) \, \pm 0.59$ & $68.29 \, (69.13)^{+1.02}_{-1.00}$ \\

    $\Omega_\mathrm{m}$ & $0.3162 \, (0.3147) \, \pm 0.0083$ & $0.3145 \,(0.3138)\, \pm 0.0086$ \\

    $\sigma_8$ & $0.8114 \, (0.8156) \, \pm 0.0073$ & $0.8198 \, (0.8280)^{+0.0109}_{-0.0107}$ \\
    
    $S_8$ & $0.8331 \, (0.8355) \, \pm 0.0159$ & $0.8393 \, (0.8468) \, \pm 0.0173$ \\

    $\mathrm{log}_{10}(f/{\mathrm{eV}})$ & $-$ & $26.57 \,(26.36)^{+0.39}_{-0.36} $\\

    $\mathrm{log}_{10}(m/{\mathrm{eV}})$ & $-$ & $-26.94 \,(-26.90)^{+0.58}_{-0.53} $\\

    \hline
  \end{tabular} 
  \caption{The mean (best-fit) $\pm1\sigma$ constraints on the cosmological parameters in $\Lambda$CDM and in the EDE scenario with $n=3$, as inferred from \emph{Planck} 2018 primary CMB data only (TT+TE+EE).  Upper and lower limits are quoted at 95\% CL.  Although there is a small contribution to the constraining power in these data from acoustic-peak-smearing due to gravitational lensing, the constraints are dominated by information content from the recombination epoch.  The EDE component is not detected here; a two-tailed limit yields $f_\mathrm{EDE} = 0.033^{+0.027}_{-0.026}$ at 68\% CL, i.e., consistent with zero. }
  \label{table:params-P18-only}
\end{table*}

\enlargethispage{-10\baselineskip}

We focus exclusively on the $n=3$ EDE model (see Eq.~\eqref{eq.EDEdef}). This is the best-fit integer-valued $n$ reported in previous analyses \cite{Smith:2019ihp}. We do not consider non-integer values of $n$, for the reasons discussed below Eq.~\eqref{eq.PoulinEDE}.  We note that $n$ is not tightly constrained when allowed to vary \cite{Smith:2019ihp}.  We adopt uniform priors on the effective EDE parameters $f_{\rm EDE}=[0.001, 0.5]$ and $\log_{10}(z_c)=[3.1,4.3]$, and a uniform prior on the initial field displacement $\theta_i=[0.1,3.1]$. We examine the prior-dependence of these results in Sec.~\ref{sec:priors}, in particular by investigating results with uniform priors placed on the particle physics parameters $f$ and $\log_{10}(m)$.

As a benchmark comparison, we also fit the $\Lambda$CDM model to the above data sets.  We adopt broad uniform priors on the six standard $\Lambda$CDM parameters (in both the $\Lambda$CDM runs and the EDE runs, which of course include these parameters as well). Following the \emph{Planck} convention, we hold the sum of the neutrino masses fixed to 0.06 eV, assuming one massive eigenstate and two massless eigenstates, and fix the effective number of relativistic species $N_{\rm eff} = 3.046$.  We also sample and marginalize over the nuisance parameters for all likelihoods using standard methodology.  We analyze the MCMC chains using \texttt{GetDist} \cite{GetDist}.\footnote{\url{https://github.com/cmbant/getdist}}  All of our chains are publicly available for download.\footnote{\url{https://users.flatironinstitute.org/~chill/H20_data/}}  We encourage interested readers to analyze these data, as the full chains are more informative than any individual summary statistic.

A final comment is in order regarding our parameter constraints.  There are two common approaches to obtain marginalized parameter confidence intervals in the case of two-tailed limits. Considering $68\%$ intervals as an example case, the first approach is to compute a limit such that $32\%$ of the samples are outside the limit range (symmetrically in the tails), i.e., such that either tail of the marginalized posterior distribution contains $16\%$ of the samples beyond the quoted limit values. The second approach is to compute an interval between two points with highest equal marginalized probability density (the ``credible interval'') such that the enclosed region contains $68$\% of the samples.  The two approaches yield identical confidence intervals for Gaussian posterior distributions, but can non-negligibly differ if the distribution is skewed, which is indeed true for the EDE model parameter posterior distributions. Given this choice, we adopt the first approach in the main body of this paper and present estimates obtained with the second method in Appendix~\ref{app:margestats}. For either approach, the limits will be quoted as 
\[
\text{mean}^{+\text{(upper $68\%$ limit - mean)}}_{-\text{(mean - lower $68\%$ limit)}}\,,
\]
where ``mean'' refers to the mean of the marginalized posterior distribution for that parameter.

\subsection{Constraints on EDE from the Primary CMB Alone}
\label{sec:CMBonly}

We first consider the \emph{Planck} 2018 primary CMB TT, TE, and EE power spectrum data.  While there is a small contribution to the constraining power from acoustic-peak-smearing due to gravitational lensing, the overall constraints are dominated by information from the recombination epoch.  This analysis thus examines potential evidence for EDE from early-universe data considered on their own.

\begin{figure*}
\centering
 \includegraphics[width=\textwidth]{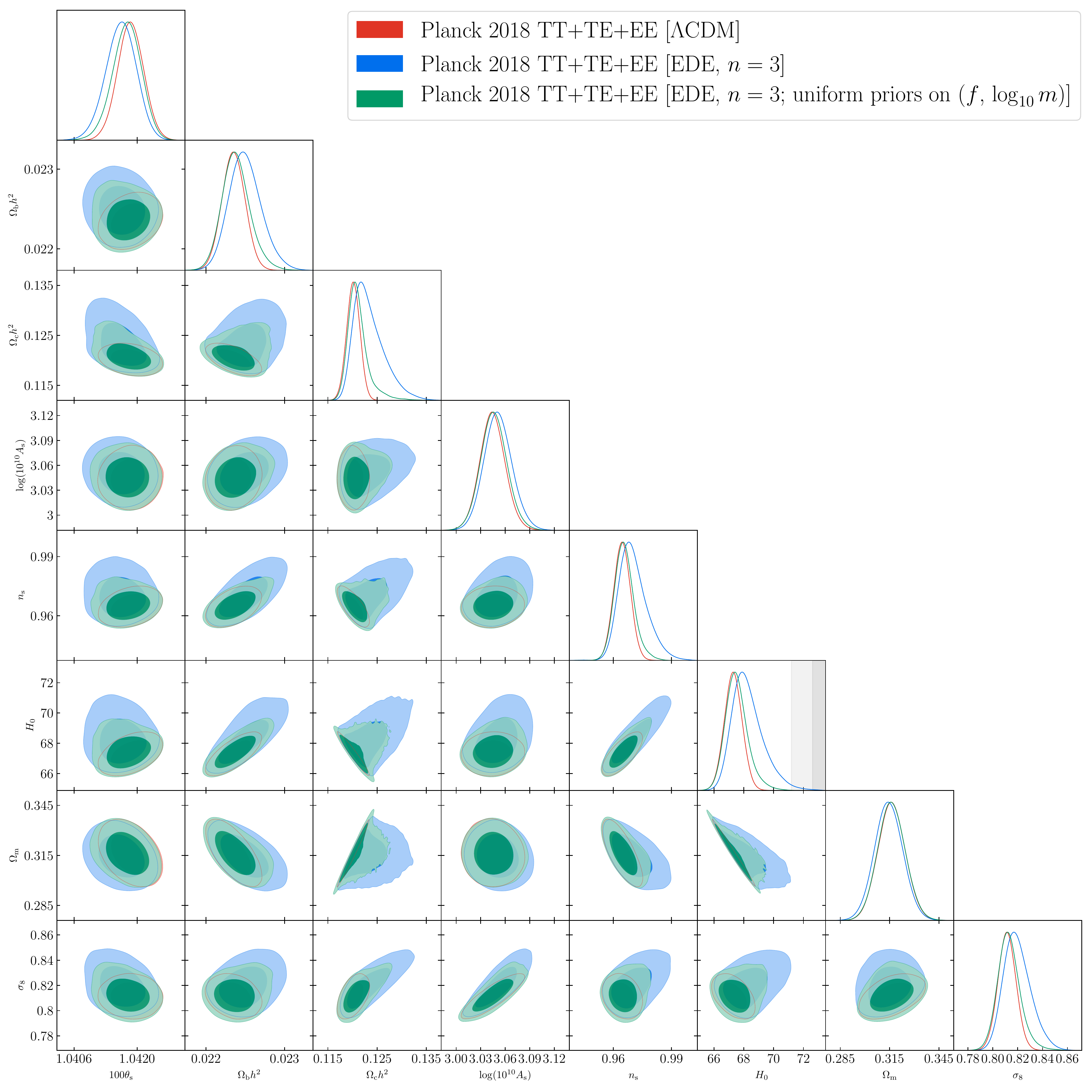}
    \caption{Cosmological parameter constraints from the \emph{Planck} 2018 primary CMB data alone (TT+TE+EE).  The red (blue) contours show $1\sigma$ and $2\sigma$ posteriors in the $\Lambda$CDM (EDE, $n=3$) models.  We do not plot $\tau$, as it is unchanged in the EDE fit.  The most significant changes in the EDE fit (compared to $\Lambda$CDM) are increases in $\Omega_c h^2$, $n_s$, $H_0$, and $\sigma_8$, as well as broadening of the error bars on these parameters.  However, the change in $H_0$ is not large enough to reconcile the tension with the SH0ES constraint (shown in the gray bands in the $H_0$ panel).  The green contours show posteriors for the EDE model, but with uniform priors placed on the (physical) particle physics parameters $f$ and $\log_{10}(m)$, rather than on the effective EDE parameters $f_{\rm EDE}$ and $\log_{10}(z_c)$.  Comparison of the blue and green contours indicates that the physical priors strongly downweight EDE models than can resolve the $H_0$ tension; see Sec.~\ref{sec:priors} for further discussion.}
    \label{fig:P18onlyconstraints}
\end{figure*}

We find no evidence for EDE in the \emph{Planck} 2018 primary CMB data alone.  Indeed, the data are powerful enough to set meaningful constraints on the possible existence of an EDE component.  The results are tabulated in Tables~\ref{table:params-P18-only}, \ref{table:chi2_CMB_alone}, and~\ref{table:params-P18-only-margestats} and in Figs.~\ref{fig:mainEDEconstraints} and~\ref{fig:P18onlyconstraints}.  We find an upper bound $f_{\rm EDE}<0.087$ at $95\%$ CL; a two-tailed limit yields $f_\mathrm{EDE} = 0.033^{+0.027}_{-0.026}$ at 68\% CL, i.e., consistent with zero. The initial EDE field displacement $\theta_i$ is poorly constrained, and we find a lower bound $\theta_i >0.36$ at $95\%$ CL, a reflection of the fact that at small $f_{\rm EDE}$ the model is indistinguishable from $\Lambda$CDM. The timing of EDE is constrained to $\mathrm{log}_{10}z_c= 3.66 \, ^{+0.28}_{-0.24}$, the only indication of a slight CMB preference for EDE. However, the posterior distribution shows significant support on the boundaries of the prior, indicating this result should not be considered to be physically meaningful.

A comparison of the posterior distributions in EDE and $\Lambda$CDM can be seen in Fig.~\ref{fig:P18onlyconstraints}. We find the Hubble constant in EDE to be $H_0 =68.29 ^{+1.02}_{-1.00}\, {\rm km/s/Mpc}$,  shifted slightly upwards relative to $\Lambda$CDM fit to the same data set, $H_0=67.29 \pm 0.59\, {\rm km/s/Mpc}$, and with a considerably larger error bar. This behaviour (slight upward shift,  posterior broadened and skewed towards larger values) is mirrored in $\Omega_bh^2$, $\Omega_ch^2$, and $n_s$. We find $S_8=0.8393\pm 0.0173$ in the EDE scenario, larger than the $\Lambda$CDM value $S_8=0.8331 \pm 0.0159$, both larger than the DES, HSC, and KV-450 constraints, but more so in the EDE case. We also note a considerable degeneracy between $f_{\rm EDE}$ and $H_0$, as well as between $f_{\rm EDE}$ and $\sigma_8$ (see Fig.~\ref{fig:mainEDEconstraints}).

The goodness-of-fit to the primary CMB anisotropies, as quantified by the $\chi^2$-statistic, is only marginally improved in the EDE three-parameter extension of $\Lambda$CDM. The $\chi^2$ statistics for each primary CMB likelihood are given in Table~\ref{table:chi2_CMB_alone}. We find an improvement $\Delta \chi^2=-4.1$, with nearly equal contributions from the low-$\ell$ TT, low-$\ell$ EE, and high-$\ell$ TT+TE+EE likelihoods.  With three additional parameters, this is not a significant improvement over $\Lambda$CDM.

\begin{figure*} 
\centering
 \includegraphics[width=\textwidth]{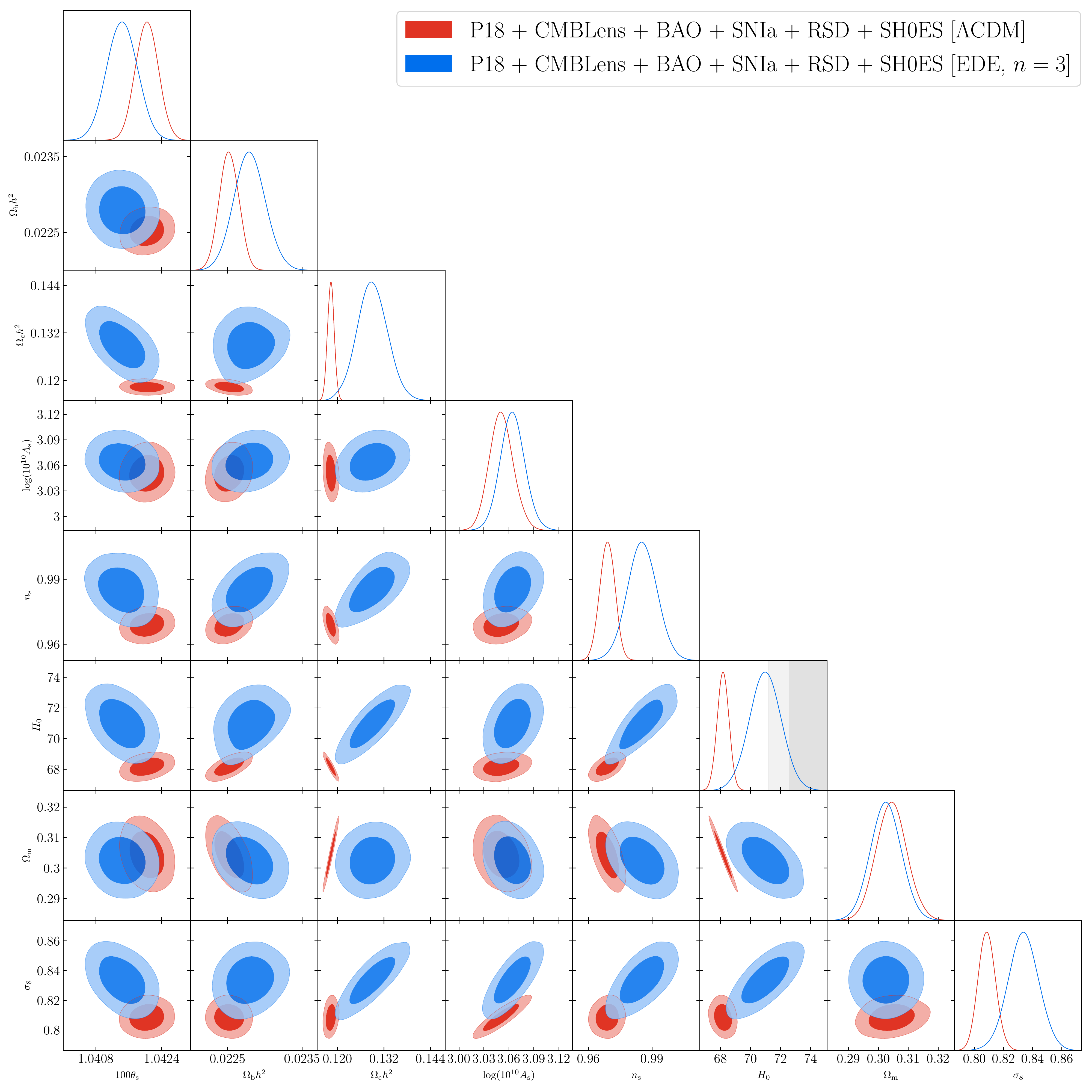}
    \caption{Cosmological parameter constraints from the combination of \emph{Planck} 2018 primary CMB data (TT+TE+EE); \emph{Planck} 2018 CMB lensing data; BAO data from 6dF, SDSS DR7, and SDSS DR12; Pantheon SNIa data; the latest SH0ES $H_0$ constraint; and SDSS DR12 RSD data.  This data set combination matches that used in~\cite{Smith:2019ihp}, with the exception of \emph{Planck} 2018 data used here in place of 2015 data.  The red (blue) contours show $1\sigma$ and $2\sigma$ posteriors in the $\Lambda$CDM (EDE, $n=3$) models.  We do not plot $\tau$, as it is essentially unchanged in the EDE fit.  Many parameters shift by a non-negligible amount in the EDE fit (compared to $\Lambda$CDM), including increases in $\Omega_b h^2$, $\Omega_c h^2$, $n_s$, $H_0$, and $\sigma_8$ and a decrease in $100\theta_s$, as well as broadening of the error bars on these parameters.  The increase in $H_0$ is large enough to mostly reconcile the tension with the SH0ES-only constraint (shown in the gray bands).  The EDE component is detected at $3.1\sigma$ significance using this combination of data sets, consistent with earlier work~\cite{Poulin:2018cxd,Smith:2019ihp}.}
    \label{fig:Smithcomb-constraints}
\end{figure*}

\begin{table}[t!]
\centering
$\chi^2$ statistics from {\it Planck} 2018 data only: TT+TE+EE
  \begin{tabular}{|l|c|c|}
    \hline\hline
    Datasets &~~$\Lambda$CDM~~&~~~EDE ~~~\\ \hline \hline
         \textit{Planck} 2018 low-$\ell$ TT & 23.4 & 22.1\\
        \textit{Planck} 2018 low-$\ell$ EE &397.2 & 396.0\\
    \textit{Planck} 2018 high-$\ell$ TT+TE+EE &  2344.9   &  2343.3 \\

    \hline
    Total $\chi^2 $   & 2765.5 & 2761.4\\
     $\Delta \chi^2 $ &  & -4.1 \\ 
    \hline
  \end{tabular}
  \caption{$\chi^2$ values for the best-fit $\Lambda$CDM and EDE models to the primary CMB alone. The reduction in $\chi^2$ is 4.1 for the three-parameter EDE extension of $\Lambda$CDM.}
  \label{table:chi2_CMB_alone}
\end{table}

\subsection{Constraints from Primary CMB, CMB Lensing, BAO, RSD, SNIa, and SH0ES Data} 
\label{sec:constraints2}

We now supplement the {\it Planck} 2018 primary CMB anisotropy data with {\it Planck} 2018 CMB lensing, BAO, RSD, supernova, and local distance ladder data. This combination of data sets, with the exception of our use of {\it Planck} 2018 data in place of 2015 data, was the basis for the conclusions in~\cite{Poulin:2018cxd} and~\cite{Smith:2019ihp}.  The posterior distributions are shown in Figs.~\ref{fig:mainEDEconstraints} and~\ref{fig:Smithcomb-constraints}, the best-fit parameters and constraints are tabulated in Tables~\ref{table:params-Smithcomb} and~\ref{table:params-Smithcomb-margestats}, and the $\chi^2$ values are given in Table~\ref{table:chi2_CMB_BAO_RSD_SNIA}.

\begin{table*}
Constraints from \emph{Planck} 2018 TT+TE+EE + CMB Lensing, BAO, SNIa, SH0ES, and RSD \vspace{2pt} \\
  \centering
  \begin{tabular}{|l|c|c|}
    \hline\hline Parameter &$\Lambda$CDM~~&~~~EDE ($n=3$) ~~~\\ \hline \hline

    {\boldmath$\ln(10^{10} A_\mathrm{s})$} & $3.051 \, (3.047) \, \pm 0.014$ & $3.064 \, (3.058) \, \pm 0.015$ \\

    {\boldmath$n_\mathrm{s}$} & $0.9689 \, (0.9686) \, \pm 0.0036 $ & $0.9854 \, (0.9847)^{+0.70}_{-0.69}$ \\

    {\boldmath$100\theta_\mathrm{s}$} & $1.04204 \, (1.04209) \, \pm 0.00028 $ & $1.04144 \, (1.04119) \, \pm 0.00037$\\

    {\boldmath$\Omega_\mathrm{b} h^2$} & $0.02252 \, (0.02249) \, \pm 0.00013 $ &  $0.02280 \, (0.02286) \pm 0.00021$ \\

    {\boldmath$\Omega_\mathrm{c} h^2$} & $0.11830 \, (0.11855)\, \pm 0.00085 $ & $0.12899 \, (0.12999) \pm 0.00390$ \\

    {\boldmath$\tau_\mathrm{reio}$} & $0.0590 \, (0.0566) \, \pm 0.0072 $ & $0.0573 \, (0.0511) \, \pm 0.0071 $\\

    {\boldmath$\mathrm{log}_{10}(z_c)$} & $-$ & $3.63 \, (3.59)^{+0.17}_{-0.10}$ \\

    {\boldmath$f_\mathrm{EDE} $} & $-$ & $ 0.098 \, (0.105) \pm 0.032 $\\

    {\boldmath$\theta_i$} & $-$ & $ 2.58 \, (2.71)^{+0.29}_{-0.09} $\\

    \hline

    $H_0 \, [\mathrm{km/s/Mpc}]$ & $68.17 \, (68.07) \, \pm 0.39$ & $70.98 \, (71.15) \pm 1.05$ \\

    $\Omega_\mathrm{m}$ & $0.3044 \, (0.3058) \, \pm 0.0051$ & $0.3025 \,(0.3032)\, \pm 0.0051$ \\

    $\sigma_8$ & $0.8086 \, (0.8081) \, \pm 0.0060$ & $0.8337 \, (0.8322) \pm 0.0105$ \\

    $S_8$ & $0.8145 \, (0.8158) \, \pm 0.0099$ & $0.8372 \, (0.8366) \pm 0.0127$ \\

    $\mathrm{log}_{10}(f/{\mathrm{eV}})$ & $-$ & $26.64 \,(26.63)^{+0.08}_{-0.15} $\\

    $\mathrm{log}_{10}(m/{\mathrm{eV}})$ & $-$ & $-27.15 \,(-27.27)^{+0.34}_{-0.22} $\\

    \hline
  \end{tabular}
  \caption{The mean (best-fit) $\pm1\sigma$ constraints on the cosmological parameters in $\Lambda$CDM and in the EDE scenario with $n=3$, as inferred from the combination of \emph{Planck} 2018 primary CMB data (TT+TE+EE); \emph{Planck} 2018 CMB lensing data; BAO data from 6dF, SDSS DR7, and SDSS DR12; Pantheon SNIa data; the latest SH0ES $H_0$ constraint; and SDSS DR12 RSD data.  The EDE component is detected at $3.1 \sigma$ significance.}
  \label{table:params-Smithcomb}
\end{table*}

\begin{table}
\centering
$\chi^2$ statistics from the fit to \emph{Planck} 2018 TT+TE+EE + CMB Lensing, BAO, SNIa, SH0ES, and RSD
  \begin{tabular}{|l|c|c|}
    \hline\hline
    Datasets &~~$\Lambda$CDM~~&~~~EDE ~~~\\ \hline \hline
    CMB TT, EE, TE: & &\\
      \;\;\;\; \textit{Planck} 2018 low-$\ell$ TT & 22.8 & 21.4\\
       \;\;\;\; \textit{Planck} 2018 low-$\ell$ EE & 396.4 & 395.8\\
  \hspace{.16cm} \; \begin{tabular}[t]{@{}c@{}}\textit{Planck} 2018 high-$\ell$ \\ TT+TE+EE\end{tabular}   &  2346.8   &  2346.9 \\

    LSS:& &\\
    \;\;\;\;\,\textit{Planck} CMB lensing& 8.9 & 9.5\\
   \;\;\;\; BAO (6dF)& 0.0015 & 0.000002\\
   \;\;\;\; BAO (DR7 MGS)& 1.6 & 1.7 \\
      \;\;\;\; BAO+RSD (DR12 BOSS)& 5.9 & 6.5 \\
    Supernovae:& &\\
   \;\;\;\; Pantheon & 1034.8 & 1034.8 \\
   SH0ES & 17.6 & 4.1 \\
    \hline
    Total $\chi^2$   & 3834.8 & 3820.7 \\
    $\Delta \chi^2$ &  & -14.1 \\ 
    \hline
  \end{tabular}
 \caption{$\chi^2$ values for the best-fit $\Lambda$CDM and EDE models to CMB + CMB Lensing + BAO + SNIa + SH0ES + RSD data. The reduction in $\chi^2$ is $14.1$ for the three-parameter EDE extension of $\Lambda$CDM, driven almost entirely by the improved fit to SH0ES.  However, it is also notable that the fit to the CMB is not degraded, even as the fit to SH0ES improves.}
  \label{table:chi2_CMB_BAO_RSD_SNIA}
\end{table}

\begin{table*}[ht!]
  \hspace{.4cm}  Constraints from \emph{Planck} 2018 TT+TE+EE + CMB Lensing, BAO, SNIa, SH0ES, RSD, and DES-Y1 \vspace{2pt} \\
  \centering
  \begin{tabular}{|l|c|c|}
    \hline\hline Parameter &$\Lambda$CDM~~&~~~EDE ($n=3$) ~~~\\ \hline \hline

    {\boldmath$\ln(10^{10} A_\mathrm{s})$} & $3.049 \, (3.049) \, \pm 0.014$ & $3.058 \, (3.064) \, \pm 0.015$ \\

    {\boldmath$n_\mathrm{s}$} & $0.9704 \, (0.9698) \, \pm 0.0035 $ & $0.9838 \, (0.9909)^{+0.0074}_{-0.0075}$ \\

    {\boldmath$100\theta_\mathrm{s}$} & $1.04208 \, (1.04183) \, \pm 0.00028 $ & $1.04162 \, (1.04172) \, \pm 0.00036$\\

    {\boldmath$\Omega_\mathrm{b} h^2$} & $0.02258 \, (0.02260) \, \pm 0.00013 $ &  $0.02285 \, (0.02304) \pm 0.00021$ \\

    {\boldmath$\Omega_\mathrm{c} h^2$} & $0.11752 \, (0.11810)\, \pm 0.00078 $ & $0.1251 \, (0.1254)^{+0.0035}_{-0.0037}$ \\

    {\boldmath$\tau_\mathrm{reio}$} & $0.0590 \, (0.0584) \, \pm 0.0072 $ & $0.0581 \, (0.0626) \, \pm 0.0072 $\\

    {\boldmath$\mathrm{log}_{10}(z_c)$} & $-$ & $3.69 \, (3.84)^{+0.18}_{-0.15}$ \\

    {\boldmath$f_\mathrm{EDE} $} & $-$ & $ 0.077 \, (0.088)^{+0.032}_{-0.034} $\\

    {\boldmath$\theta_i$} & $-$ & $ 2.58 \, (2.89)^{+0.32}_{-0.15} $\\

    \hline

    $H_0 \, [\mathrm{km/s/Mpc}]$ & $68.52 \, (68.24) \, \pm 0.36$ & $70.75 \, (71.05)^{+1.05}_{-1.09}$ \\

    $\Omega_\mathrm{m}$ & $0.2998 \, (0.3035) \, \pm 0.0046$ & $0.2970 \,(0.2954)\, \pm 0.0047$ \\

    $\sigma_8$ & $0.8054 \, (0.8067) \, \pm 0.0057$ & $0.8228 \, (0.8263)^{+0.0099}_{-0.0101}$ \\

    $S_8$ & $0.8051 \, (0.8115) \, \pm 0.0087$ & $0.8186 \, (0.8199) \pm 0.0109$ \\

    $\mathrm{log}_{10}(f/{\mathrm{eV}})$ & $-$ & $26.57 \,(26.47)^{+0.11}_{-0.16} $\\

    $\mathrm{log}_{10}(m/{\mathrm{eV}})$ & $-$ & $-27.03 \,(-26.76)^{+0.33}_{-0.32} $\\

    \hline
  \end{tabular} 
  \caption{The mean (best-fit) $\pm1\sigma$ constraints on the cosmological parameters in $\Lambda$CDM and in the EDE scenario with $n=3$, as inferred from the combination of \emph{Planck} 2018 primary CMB data (TT+TE+EE); \emph{Planck} 2018 CMB lensing data; BAO data from 6dF, SDSS DR7, and SDSS DR12; Pantheon SNIa data; the latest SH0ES $H_0$ constraint; SDSS DR12 RSD data; and DES-Y1 3x2pt data.  With the inclusion of the DES data, the evidence for the EDE component decreases to $2.3\sigma$.  A one-sided upper limit on the EDE fraction yields $f_{\rm EDE} < 0.127$ at 95\% CL.  The non-negligible discrepancy between the best-fit and mean posterior values is a reflection of the underlying tension between these data sets (i.e., SH0ES and all other data sets) in the context of this model.}
  \label{table:params-uberlikelihood}
\end{table*}

We find $f_\mathrm{\rm EDE}=0.098 \pm 0.032$, i.e., a $3.1 \sigma$ detection of EDE, when using this combination of data sets. This value is larger than the 95\% CL upper limit from the CMB alone, as can be appreciated from Fig.~\ref{fig:mainEDEconstraints}, indicating minor tension between the data sets in the context of EDE. The shift in $f_{\rm EDE}$ is driven by the $H_0$ tension combined with the $f_{\rm EDE}-H_0$ degeneracy of the EDE fit to the primary CMB, which allows larger $H_0$ values without substantially degrading the fit to the latter.  These results are consistent with those presented in previous work \cite{Poulin:2018cxd,Smith:2019ihp}.  While the RSD and CMB lensing likelihoods provide some LSS information in this data set combination, their error bars are large enough so as to not strongly disfavor the region of parameter space that can resolve the Hubble tension.  The somewhat higher value of $\sigma_8$ found in the \emph{Planck} CMB lensing analysis \cite{2018arXiv180706210P} in comparison to the DES, HSC, and KV-450 galaxy weak lensing analyses also plays an important role here.

We find $H_0 = 70.98 \, \pm \, 1.05 \,{\rm km/s/Mpc}$, reducing the tension with SH0ES to $\approx 1.7\sigma$, in comparison with the $\Lambda$CDM value $H_0 = 68.17 \, \pm 0.39\, {\rm km/s/Mpc}$, the latter in $4.0\sigma$ tension with SH0ES. The EDE reduction in the tension is consistent with the conclusions of past works. However we note the relative decrease in tension is in part driven by a tripling of the error bar in the broadened parameter space. 

As anticipated, the tension with low-redshift LSS $S_8$ constraints is worsened in the EDE fit, as compared to $\Lambda$CDM. We find $\sigma_8=0.8337 \pm 0.0105$ for EDE, $\sigma_8=0.8086 \pm 0.0060$ for $\Lambda$CDM, and $\Omega_\mathrm{m}=0.3025 \pm 0.0051$ and $\Omega_\mathrm{m}=0.3044 \pm 0.0051$ for EDE and $\Lambda$CDM, respectively. This drives an enhanced $S_8$ tension in EDE; $S_8=0.8372 \pm 0.0127$, in moderate $2.2\sigma$ tension with DES-Y1, for example. In contrast, for $\Lambda$CDM fit to this combination of data sets, we find $S_8=0.8145 \pm 0.0099$, which differs from DES at $1.5\sigma$, i.e., the two are statistically consistent.

The impact of EDE on the fit to the other $\Lambda$CDM parameters is similar to that observed in the fit to the primary CMB alone. Relative to the $\Lambda$CDM fit to the same data sets, we find a shift upwards of $\Omega_c h^2$, $\Omega_b h^2$, $n_s$, and a significant broadening of the posteriors. The shift is most noticeable in $\Omega_c h^2$, which is the driver of the changes to the matter power spectrum $P(k)$ observed in Sec.~\ref{sec:LSS}. We find $\Omega_\mathrm{c}h^2=0.12899 \pm 0.00390$ in EDE and $\Omega_\mathrm{c} h^2=0.11830 \pm 0.00085$ in $\Lambda$CDM.

The EDE parameters $z_c$ and $\theta_i$ are well constrained in comparison to the fit to the primary CMB alone (see Fig.~\ref{fig:mainEDEconstraints}). The EDE critical redshift $z_c$ is found to be $\log_{10}(z_c)=3.63 \, ^{+0.17}_{-0.10}$. The posterior exhibits a weakly bimodal distribution, mirroring the results of \cite{Smith:2019ihp}, with a tail towards large $\log_{10}(z_c)$. As discussed in \cite{Smith:2019ihp}, the subdominant peak is driven by high-$\ell$ polarization data; it does not have an obvious physical interpretation and could simply be due to a noise fluctuation. On the other hand, the tail correlates with large $f_{\rm EDE}$ and small $\Omega_c h^2$; this is simply the region of parameter space where the sound horizon and subsequent cosmology is left unaffected by the EDE, even for fairly large values of $f_{\rm EDE}$. We find a strong preference for a large initial field displacement, $\theta_i = 2.58^{+0.29}_{-0.09}$, consistent with the finding of \cite{Smith:2019ihp} that the best-fit models lie outside the regime wherein the scalar field potential can be expanded as a monomial.

The $\chi^2$ statistics for each likelihood in this fit are given in Table~\ref{table:chi2_CMB_BAO_RSD_SNIA}. The $3.1 \sigma$ detection of EDE is accompanied by a marked increase in the goodness-of-fit as compared to $\Lambda$CDM. The EDE improvement in the total $\chi^2$-statistic is $\Delta \chi^2=-14.1$, driven almost entirely by the improved fit to SH0ES, $\Delta \chi^2 _{\rm SH0ES}=-13.5$, counteracted by a slightly worsened fit to the LSS data, $\Delta \chi^2 _{\rm LSS}=+1.3$ in total. The latter hints at the potential for additional LSS likelihoods to substantially constrain EDE, particularly via the sensitivity to $P(k)$ as motivated by Fig.~\ref{fig:PkDES}.

\subsection{Including the (Early) Dark Energy Survey}
\label{sec:constraintsDES}

We now expand our analysis to include likelihoods from the DES-Y1 data set, in particular the ``3x2pt'' likelihood comprised of photometric galaxy clustering, galaxy-galaxy lensing, and cosmic shear two-point correlation functions \cite{Abbott:2017wau}.  We jointly analyze all likelihoods described in the previous two subsections and the DES-Y1 3x2pt likelihood.

As mentioned earlier, we use the ``Halofit'' prescription \cite{Smith:2002dz,Takahashi2012} to compute the non-linear matter power spectrum, following the DES methodology \cite{Abbott:2017wau}. Thus we assume that the ``Halofit'' fitting function calibration remains valid in the EDE models under consideration.  To justify this assumption, we note that in the models capable of addressing the $H_0$ tension and fitting the CMB data, the deviation from a $\Lambda$CDM $P(k)$ is not particularly large, since $f_{\rm EDE} \lesssim 0.1$.  The test that we perform in Appendix~\ref{app:S8-validation} (see Figs.~\ref{fig:S8-validation-EDE-params},~\ref{fig:S8-validation}, and~\ref{fig:S8-validation-LCDM}, which are described near the beginning of the next subsection) provides a further justification for the validity of using Halofit.  There, we compare results obtained when using the full DES-Y1 3x2pt likelihood, which relies on the Halofit non-linear $P(k)$, to results obtained when imposing a Gaussian prior on $S_8$ corresponding to the DES-Y1 result, which only requires linear theory to compute.  If the Halofit prediction of the non-linear $P(k)$ were highly inaccurate in the EDE models under consideration, then the posteriors obtained for the EDE parameters would a priori be very different in the two approaches.  The test thus not only verifies that the information content in the DES-Y1 data is almost entirely contained in the $S_8$ result, as discussed further in the next subsection, but also verifies that the non-linear modeling used in the 3x2pt likelihood is sufficiently accurate, even for the EDE models.

\begin{figure*}
\centering
 \includegraphics[width=\textwidth]{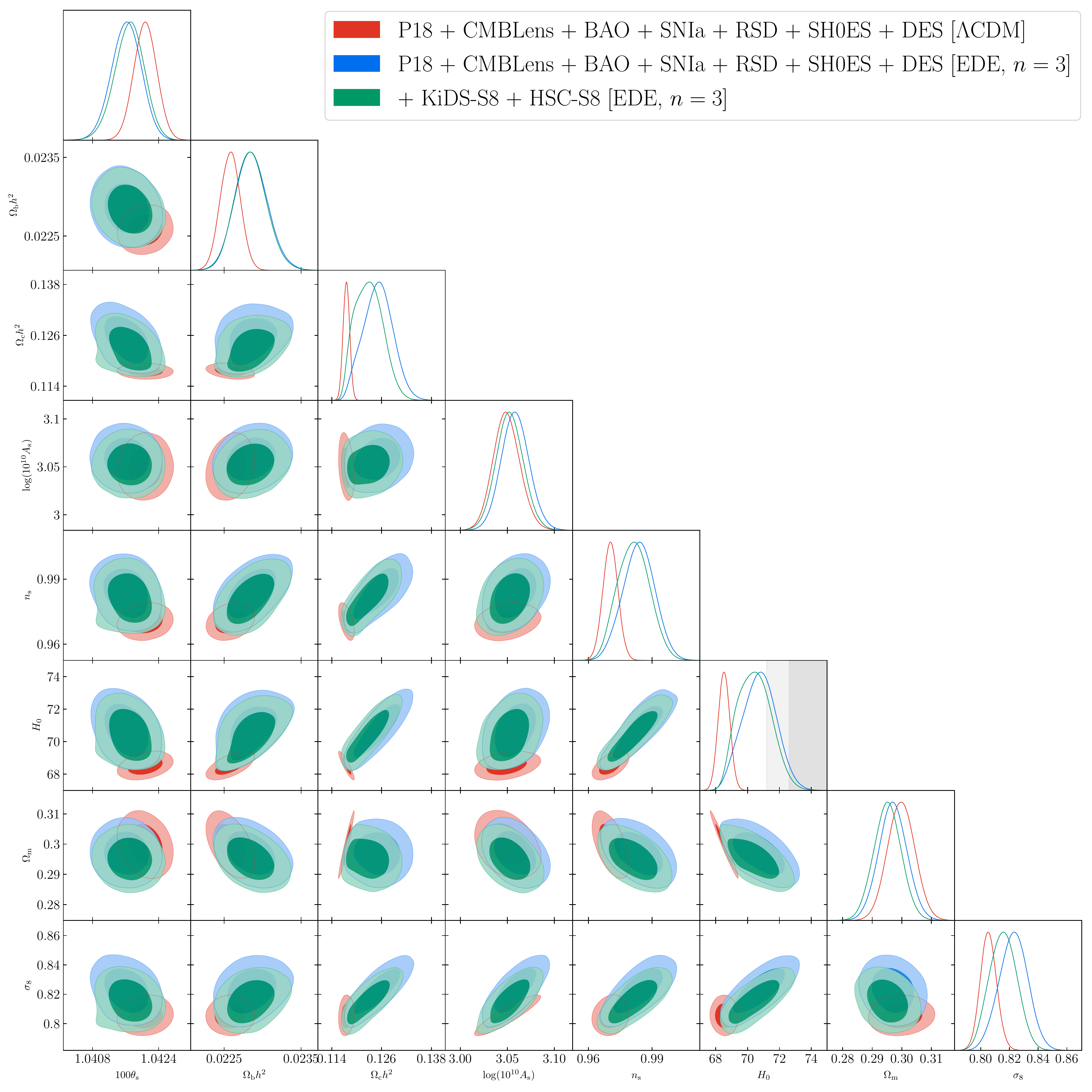}
    \caption{Cosmological parameter constraints from the combination of \emph{Planck} 2018 primary CMB data (TT+TE+EE); \emph{Planck} 2018 CMB lensing data; BAO data from 6dF, SDSS DR7, and SDSS DR12; Pantheon SNIa data; the latest SH0ES $H_0$ constraint; SDSS DR12 RSD data; and the DES-Y1 3x2pt data.  The red (blue) contours show $1\sigma$ and $2\sigma$ posteriors in the $\Lambda$CDM (EDE, $n=3$) models.  We do not plot $\tau$, as it is essentially unchanged in the EDE fit.  The green contours show posteriors in the EDE model when further including priors on $S_8$ derived from KiDS and HSC data (as an approximation to the use of full likelihoods from these data sets).  Inclusion of the DES data weakens the preference for EDE seen in Fig.~\ref{fig:Smithcomb-constraints}; this shift is due to the discrepancy between the DES matter power spectrum constraints and the predictions of the EDE model fit to the data sets in Fig.~\ref{fig:Smithcomb-constraints} (see Fig.~\ref{fig:PkDES}).  Inclusion of the KiDS and HSC $S_8$ priors further weakens the evidence for EDE, bringing it well under $2\sigma$.  The upshot is that the increase in $H_0$ seen in the EDE fit in Fig.~\ref{fig:Smithcomb-constraints} is now reduced, and the increased tension with SH0ES alone (shown in the gray bands) is apparent. }
    \label{fig:uber-constraints}
\end{figure*}

The posterior distributions for our analysis including the full DES-Y1 3x2pt likelihood are shown in Figs.~\ref{fig:mainEDEconstraints} and~\ref{fig:uber-constraints}, the parameter constraints are tabulated in Tables~\ref{table:params-uberlikelihood} and~\ref{table:params-uberlikelihood-margestats}, and the $\chi^2$ values are given in Table~\ref{table:chi2_CMB_BAO_RSD_SNIA_DES}.  With the inclusion of the DES-Y1 3x2pt likelihood, the evidence for EDE decreases to $2.3\sigma$.  We find $f_{\rm EDE}=0.077 ^{+0.032}_{-0.034}$, shifted downwards from the result without DES (Sec.~\ref{sec:constraints2}) to come into statistical agreement with the upper bound from the primary CMB anisotropies (Sec.~\ref{sec:CMBonly}).  A one-sided upper limit yields $f_{\rm EDE} < 0.127$ at 95\% CL. The initial field displacement and critical redshift are constrained to be $\theta_i=2.58^{+0.32}_{-0.15}$, and $\log_{10}(z_c)=3.69^{+0.18}_{-0.15}$.  Broadly speaking, the EDE parameter posteriors move towards the CMB-only results, as can be appreciated from Fig.~\ref{fig:mainEDEconstraints}.  It is notable that the best-fit parameter values in Table~\ref{table:params-uberlikelihood} differ by a non-negligible amount from the mean of the posteriors.  This reflects the fact that the data sets are pulling the model parameters in opposite directions: the SH0ES data can only be fit by increasing $f_{\rm EDE}$ (and thus moving other parameters along their degeneracies with it), but the CMB and LSS data do not prefer large $f_{\rm EDE}$.  In a proper Bayesian sense, it is likely that the data sets are in tension and should not be combined in the first place~(e.g.,~\cite{Raveri:2018wln,Raveri:2019gdp,Lemos:2019txn}).

The downward shift in $f_{\rm EDE}$ when DES-Y1 3x2pt data are added to the combined data set of Sec.~\ref{sec:constraints2} can be understood in terms of the interplay between $\sigma_8$, $\Omega_m$, $H_0$, and $f_{\rm EDE}$. As discussed in \cite{Abbott:2017wau} in the context of $\Lambda$CDM, the precise DES measurement of $\Omega_m$ breaks the $\Omega_m-H_0$ degeneracy in the $\Lambda$CDM fit to the CMB, shifting $H_0$ to larger values. In the EDE scenario, the impact of DES measurements on $H_0$ is the reverse, caused by a marked correlation between $\sigma_8$ and $H_0$, which can be observed in both Fig.~\ref{fig:P18onlyconstraints} and Fig.~\ref{fig:Smithcomb-constraints}. This is manifested in the discrepancy between the DES matter power spectrum constraints and the predictions of the EDE model fit to the data sets in Fig.~\ref{fig:Smithcomb-constraints} (see Fig.~\ref{fig:PkDES}). As a result, the DES likelihood drives $H_0$ to \emph{lower} values. The tight correlation between $H_0$ and $f_{\rm EDE}$ then leads to a smaller value for $f_{\rm EDE}$.

This is borne out in the $H_0$ constraints. We find $H_0=70.75^{+1.05}_{-1.09}\,{\rm km/s/Mpc}$, in mild $1.9\sigma$ tension with SH0ES. This is shifted slightly downwards from the value in the fit without DES, $H_0=70.98 \, \pm \, 1.05 \,{\rm km/s/Mpc}$.  In contrast, the $\Lambda$CDM constraint is raised to $H_0= 68.52 \pm 0.36 \,{\rm km/s/Mpc}$, decreasing the $\Lambda$CDM tension with SH0ES to $3.8\sigma$. This contrary motion in $H_0$ values is reflected in the SH0ES contribution to the $\chi^2$-statistic, $\Delta\chi^2_{\rm SH0ES}=-12.2$, as seen in Table~\ref{table:chi2_CMB_BAO_RSD_SNIA_DES}, which is slightly lower than the improvement when DES was not included, $\Delta\chi^2_{\rm SH0ES}=-13.5$, as seen in Table \ref{table:chi2_CMB_BAO_RSD_SNIA}.

\begin{table}[h!]
\centering
$\chi^2$ statistics from the fit to \emph{Planck} 2018 TT+TE+EE + CMB Lensing, BAO, SNIa, SH0ES, RSD, and DES-Y1 
  \begin{tabular}{|l|c|c|}
    \hline\hline
    Datasets &~~$\Lambda$CDM~~&~~~EDE ~~~\\ \hline \hline
    CMB TT, EE, TE: & &\\
      \;\;\;\;  \textit{Planck} 2018 low-$\ell$ TT & 22.7 &  20.5\\
       \;\;\;\; \textit{Planck} 2018 low-$\ell$ EE & 396.8 & 397.7\\
 \hspace{.16cm} \; \begin{tabular}[t]{@{}c@{}}\textit{Planck} 2018 high-$\ell$ \\ TT+TE+EE\end{tabular}&  2347.6   &  2350.2 \\
    LSS:& &\\
    \;\;\;\;\,\textit{Planck} CMB lensing& 9.0 & 9.6 \\
   \;\;\;\; BAO (6dF)& 0.0001 & 0.04\\
   \;\;\;\; BAO (DR7 MGS)& 1.8 & 2.4 \\
      \;\;\;\; BAO+RSD (DR12 BOSS)& 5.9 & 6.9 \\
         \;\;\;\; DES-Y1 &507.8 & 508.7 \\ 
    Supernovae:& &\\
   \;\;\;\; Pantheon & 1034.8 & 1034.8 \\
   SH0ES & 16.6 & 4.4 \\
    \hline
    Total $\chi^2 $   & 4343.0 & 4335.2 \\
    $\Delta \chi^2 $ &  & -7.8 \\ 
    \hline
  \end{tabular}
  \caption{The $\chi^2$ statistics in the fit to CMB + CMB Lensing + BAO + SNIa + SH0ES + RSD + DES data for the best-fit $\Lambda$CDM and EDE models. The reduction in $\chi^2$ is $\Delta \chi^2=-7.8$ for the three-parameter EDE extension of $\Lambda$CDM. As in Table~\ref{table:chi2_CMB_BAO_RSD_SNIA}, this decrease is driven almost entirely by the improved fit to SH0ES ($\Delta \chi^2_{\rm SH0ES} = -12.2$).  The fit to the primary CMB actually worsens slightly ($\Delta \chi^2_{\rm CMB} = 1.3$), as does the fit to LSS data ($\Delta \chi^2_{\rm LSS} = 3.1$).}
  \label{table:chi2_CMB_BAO_RSD_SNIA_DES}
\end{table}

It is also notable that the $\sigma_8$ posterior matches closely that of the fit to the primary CMB alone, as seen in Fig. \ref{fig:mainEDEconstraints}, erasing the shift observed in the fit without DES (Sec.~\ref{sec:constraints2}). Generated by the degeneracy between $f_{\rm EDE}$ and $\sigma_8$, this is a further indication that both LSS and CMB observations are statistically consistent with $f_{\rm EDE}=0$. This is matched by shifts in $S_8$ and $\Omega_m$; we find $S_8=0.8186 \pm 0.0109$ and $\Omega_m=0.2970\, \pm 0.0047$, both in statistical agreement with DES-only measurements. 

The $\chi^2$-statistic for each likelihood (in the joint-best-fit model) is given in Table~\ref{table:chi2_CMB_BAO_RSD_SNIA_DES}. The EDE improvement in the total $\chi^2$-statistic with respect to $\Lambda$CDM is $\Delta \chi^2=-7.8$, noticeably worse than the improvement in fit with DES excluded (Table~\ref{table:chi2_CMB_BAO_RSD_SNIA}).  The $\chi^2$ improvement here is again driven almost fully by the improved fit to SH0ES data ($\Delta \chi^2_{\rm SH0ES} = -12.2$); the fit to the primary CMB and LSS data actually worsen slightly ($\Delta \chi^2_{\rm CMB} = 1.3$ and $\Delta \chi^2_{\rm LSS} = 3.1$).  The EDE model does not appear to provide a region of parameter space that is in concordance with all cosmological data sets.

\subsection{Additional LSS Data: KiDS+VIKING-450 and Hyper Suprime-Cam} 
\label{sec:constraintsKV450HSC}

The KV-450 \cite{Hildebrandt:2016iqg,2020A&A...633A..69H} and HSC \cite{Hikage:2018qbn} surveys provide complementary data sets to DES. However, rather than perform the computationally expensive MCMC analysis of directly sampling from these likelihoods in addition to DES-Y1, we opt instead to approximate the KV-450 and HSC data sets by priors on $S_8$, corresponding to the constraints $S_8 = 0.737^{+0.040}_{-0.036}$ and $S_8 = 0.780^{+0.030}_{-0.033}$, respectively.\footnote{In addition, as of this writing, we are not aware of a publicly available likelihood for the HSC data.}

To validate this procedure, we first test it with the DES-Y1 3x2pt data, for which we have the full likelihood, as well as the $S_8$ constraint given in~\cite{Abbott:2017wau}. In Figs.~\ref{fig:S8-validation-EDE-params} and \ref{fig:S8-validation} in Appendix~\ref{app:S8-validation}, we compare the posterior distributions for cosmological parameters in the EDE scenario fit to the combined data set with DES 3x2pt data included (i.e., the results from Sec.~\ref{sec:constraintsDES}) to those with the DES 3x2pt data replaced by a Gaussian prior on $S_8$ given by the DES result $S_8 = 0.773 ^{+0.026} _{-0.020}$, i.e., an $S_8$ prior imposed on the results of Sec.~\ref{sec:constraints2}.  The posterior distributions are in near perfect agreement, for both the EDE parameters (Fig.~\ref{fig:S8-validation-EDE-params}) and standard $\Lambda$CDM parameters (Fig.~\ref{fig:S8-validation}).  Quantitatively, for example, the marginalized constraints on the EDE fraction and critical redshift are $f_{\rm EDE} = 0.076^{+0.032}_{-0.033}$ and $\log_{10}(z_c) = 3.66^{+0.18}_{-0.14}$, which are essentially identical to those found in the full analysis in Table~\ref{table:params-uberlikelihood}, $f_{\rm EDE} = 0.077^{+0.032}_{-0.034}$ and $\log_{10}(z_c) = 3.69^{+0.18}_{-0.15}$.  Quantitative results for the other parameters are of similar fidelity.

As a further test, in Fig.~\ref{fig:S8-validation-LCDM} we also perform the same comparison in the $\Lambda$CDM model fit to these data sets.  We again find excellent agreement in the posteriors, albeit with a small shift visible in $\Omega_m$ due to the fact that the DES 3x2pt likelihood does constrain this parameter, but the $S_8$ prior approach neglects this information.  The correlation of $\Omega_m$ with $\Omega_c h^2$ and $H_0$ then produces small ($\ll 1\sigma$) shifts in these parameters.  Overall, we conclude that the agreement between the $S_8$ prior approach and the full DES 3x2pt likelihood is excellent in both the EDE and $\Lambda$CDM models.  Fundamentally, this arises from the fact that DES only weakly constrains the non-$S_8$ parameters in comparison to \emph{Planck}.

Guided by this excellent agreement for DES-Y1, we include HSC and KV-450 constraints by imposing appropriate Gaussian priors on $S_8$ using the values given above.  We apply this methodology to both the $\Lambda$CDM and EDE models.  The results of this analysis are shown in Figs.~\ref{fig:mainEDEconstraints} and~\ref{fig:uber-constraints}, and tabulated in Tables~\ref{table:params-uberlikelihoodKiDSHSC} and~\ref{table:params-uberlikelihoodKiDSHSC-margestats}.  We treat the three surveys as independent, which is accurate since the sky overlap between the DES-Y1, KV-450, and HSC-Y1 survey regions is small.  There is roughly $20$ deg${}^2$ overlap between DES-Y1 and HSC-Y1; no overlap between DES-Y1 and KV-450; and roughly $70$ deg${}^2$ overlap between HSC-Y1 and KV-450.  Any covariance due to the latter overlap is further weakened by the significantly greater depth of the HSC survey compared to KV-450 (roughly double the effective number of source galaxies), i.e., the redshift window functions of the modes probed by the surveys are quite different.  These sky overlap numbers are approximate estimates based on the footprints given in~\cite{2018ApJS..235...33D},~\cite{2018PASJ...70S..25M}, and~\cite{2019A&A...632A..34W} for DES-Y1, HSC-Y1, and KV-450, respectively.  Regarding possible common systematics, the surveys also differ in their photo-$z$ calibration: KV-450 uses a combined data set comprised of a large number of spectroscopic surveys \cite{2020A&A...633A..69H}, while DES-Y1 and HSC-Y1 both use the COSMOS photo-$z$ catalogue \cite{2016ApJS..224...24L}.

The evidence for EDE is weakened to below 2$\sigma$ by the inclusion of HSC and KV-450 information.  We find $f_{\rm EDE}=0.062 ^{+0.032} _{-0.033}$, statistically consistent with $f_{\rm EDE}=0$. The $f_{\rm EDE}$ posterior exhibits substantial support at the boundary $f_{\rm EDE}=0$, and we find a one-sided upper limit $f_{\rm EDE}<0.112$ at $95\%$ CL. The Hubble constant shifts downward to $H_0 = 70.45 ^{+1.05} _{-1.08}$ km/s/Mpc, in mild $2\sigma$ tension with the SH0ES-only constraint.

The EDE posterior distributions for the standard $\Lambda$CDM parameters are shown in Fig.~\ref{fig:uber-constraints} (green).  The HSC and KV-450 $S_8$ priors lead to a substantial shift to smaller values of $\Omega_c h^2$ and $\sigma_8$, while $\Omega_b h^2$ is unaffected. This is partnered with the shift in $H_0$ to reduce $S_8$ to $S_8=0.8090\pm0.0100$, in statistical agreement with the $S_8$ measurements of DES, HSC, and KV-450. The EDE parameter posteriors (Fig.~\ref{fig:mainEDEconstraints}) move towards those in the fit of EDE to the primary CMB alone, both consistent with $f_{\rm EDE}=0$.

\begin{table}[ht!]
  \hspace{-0.26cm} Constraints from \emph{Planck} 2018 TT+TE+EE + CMB Lensing, BAO, SNIa, SH0ES, RSD, DES-Y1, KiDS-$S_8$, \\ 
\mbox{} \hfill  and  HSC-$S_8$ \hfill \mbox{}
\vspace{2pt} \\
  \centering
  \begin{tabular}{|l|c|c|c|}
    \hline\hline Parameter &$\Lambda$CDM~~&~~~EDE ($n=3$) ~~~\\ \hline \hline

    {\boldmath$\ln(10^{10} A_\mathrm{s})$} & $3.046 \, \pm 0.014$ & $3.053 \, \pm 0.014$ \\

    {\boldmath$n_\mathrm{s}$} & $0.9710 \, \pm 0.0035 $ & $0.9814 \, ^{+0.0075}_{-0.0077}$ \\

    {\boldmath$100\theta_\mathrm{s}$} & $1.04209 \, \pm 0.00028 $ & $1.04169 \, \pm 0.00036$\\

    {\boldmath$\Omega_\mathrm{b} h^2$} & $0.02260 \, \pm 0.00013 $ &  $0.02285 \, \pm 0.00020$ \\

    {\boldmath$\Omega_\mathrm{c} h^2$} & $0.11718 \, \pm 0.00075 $ & $0.1230 \, ^{+0.0034}_{-0.0035}$ \\

    {\boldmath$\tau_\mathrm{reio}$} & $0.0581 \, \pm 0.0070 $ & $0.0574 \, \pm 0.0071 $\\

    {\boldmath$\mathrm{log}_{10}(z_c)$} & $-$ & $3.73 \, ^{+0.20}_{-0.19}$ \\

    {\boldmath$f_\mathrm{EDE} $} & $-$ & $ 0.062 \, ^{+0.032}_{-0.033} $\\

    {\boldmath$\theta_i$} & $-$ & $ 2.49 \, ^{+0.40}_{-0.38} $\\

    \hline

    $H_0 \, [\mathrm{km/s/Mpc}]$ & $68.67 \, \pm 0.35$ & $70.45 \,^{+1.05}_{-1.08}$ \\

    $\Omega_\mathrm{m}$ & $0.2978 \, \pm 0.0044$ & $0.2952 \, \pm 0.0046$ \\

    $\sigma_8$ & $0.8032 \, \pm 0.0055$ & $0.8157 \, \pm 0.0096$ \\

    $S_8$ & $0.8002 \, \pm 0.0082$ & $0.8090 \, \pm 0.0100$ \\

    $\mathrm{log}_{10}(f/{\mathrm{eV}})$ & $-$ & $26.55 \,^{+0.13}_{-0.18} $\\

    $\mathrm{log}_{10}(m/{\mathrm{eV}})$ & $-$ & $-26.94 \,^{+0.39}_{-0.40} $\\

    \hline
  \end{tabular} 
  \caption{The mean $\pm1\sigma$ constraints on the cosmological parameters in $\Lambda$CDM and in the EDE scenario with $n=3$, as inferred from the combination of \emph{Planck} 2018 primary CMB data (TT+TE+EE); \emph{Planck} 2018 CMB lensing data; BAO data from 6dF, SDSS DR7, and SDSS DR12; Pantheon SNIa data; the latest SH0ES $H_0$ constraint; SDSS DR12 RSD data; DES-Y1 3x2pt data; and priors on $S_8$ derived from KiDS and HSC data.  The KiDS and HSC $S_8$ priors serve as approximations to the full likelihood functions for these data sets; we have validated the accuracy of this approximate approach using the DES data, for which we can compare the full likelihood and $S_8$-prior approaches, in Fig.~\ref{fig:S8-validation-EDE-params}.  However, we do not provide best-fit parameter values here due to the use of these approximations for the true likelihoods.  With the inclusion of the KiDS and HSC $S_8$ priors, evidence for the EDE component weakens further beyond that found in Table~\ref{table:params-uberlikelihood}, to below $2\sigma$.  This arises from the inability of the EDE model to accommodate the ``low’’ $\sigma_8$ and $\Omega_m$ values found by the weak lensing experiments.  A one-sided upper limit on the EDE fraction yields $f_{\rm EDE} < 0.112$ at 95\% CL.}
  \label{table:params-uberlikelihoodKiDSHSC}
\end{table}

\subsection{Walking Barefoot: EDE without SH0ES}
\label{sec:no-SH0ES}

\begin{table}[h!]
   \hspace{-0.24cm} 
   Constraints from \emph{Planck} 2018 TT+TE+EE + CMB Lensing, BAO, SNIa, RSD, DES-Y1, KiDS-$S_8$, and  HSC-$S_8$ \\
\mbox{} \hfill (No-SH0ES) \hfill \mbox{}
\vspace{2pt} \\
  \centering
  \begin{tabular}{|l|c|c|}
    \hline\hline Parameter &$\Lambda$CDM~~&~~~EDE ($n=3$) ~~~\\ \hline \hline

    {\boldmath$\ln(10^{10} A_\mathrm{s})$} & $3.041 \, \pm 0.014$ & $3.044 \, \pm 0.014$ \\

    {\boldmath$n_\mathrm{s}$} & $0.9692 \, \pm 0.0035 $ & $0.9718 \, \pm 0.0049$ \\

    {\boldmath$100\theta_\mathrm{s}$} & $1.04200 \, \pm 0.00028 $ & $1.04177 \, ^{+0.00035}_{-0.00033}$\\

    {\boldmath$\Omega_\mathrm{b} h^2$} & $0.02253 \, \pm 0.00013 $ &  $0.02264 \, \pm 0.00017$ \\

    {\boldmath$\Omega_\mathrm{c} h^2$} & $0.11785 \, \pm 0.00076 $ & $0.1196 \, \pm 0.0016$ \\

    {\boldmath$\tau_\mathrm{reio}$} & $0.0552 \, \pm 0.0069 $ & $0.0558 \, \pm 0.0069 $\\

    {\boldmath$\mathrm{log}_{10}(z_c)$} & $-$ & $ > 3.28$ \\

    {\boldmath$f_\mathrm{EDE} $} & $-$ & $ < 0.060 $\\

    {\boldmath$\theta_i$} & $-$ & $ > 0.35 $\\

    \hline

    $H_0 \, [\mathrm{km/s/Mpc}]$ & $68.33 \, \pm 0.36$ & $68.92 \, ^{+0.57}_{-0.59}$ \\

    $\Omega_\mathrm{m}$ & $0.3021 \, \pm 0.0045$ & $0.3008 \, \pm 0.0047$ \\

    $\sigma_8$ & $0.8032 \, \pm 0.0053$ & $0.8064 \, \pm 0.0065$ \\

    $S_8$ & $0.8060 \, \pm 0.0082$ & $0.8074 \, \pm 0.0089$ \\

    $\mathrm{log}_{10}(f/{\mathrm{eV}})$ & $-$ & $26.52 \,^{+0.38}_{-0.36} $\\

    $\mathrm{log}_{10}(m/{\mathrm{eV}})$ & $-$ & $-26.67 \,^{+0.65}_{-0.69} $\\

    \hline
  \end{tabular} 
  \caption{The mean $\pm1\sigma$ constraints on the cosmological parameters in $\Lambda$CDM and in the EDE scenario with $n=3$, as inferred from the combination of \emph{Planck} 2018 primary CMB data (TT+TE+EE); \emph{Planck} 2018 CMB lensing data; BAO data from 6dF, SDSS DR7, and SDSS DR12; Pantheon SNIa data; SDSS DR12 RSD data; DES-Y1 3x2pt data; and priors on $S_8$ derived from KiDS and HSC data, i.e., all data sets used in Table~\ref{table:params-uberlikelihoodKiDSHSC} but with SH0ES excluded.  As in that table, we do not provide best-fit parameter values here due to the use of the approximate likelihoods for KiDS and HSC.  As in the \emph{Planck} primary CMB-only analysis (Table~\ref{table:params-P18-only}), no evidence for the EDE component is seen here.  A two-tailed analysis yields $f_{\rm EDE} = 0.023 \pm 0.017$.  The upper limit on $f_{\rm EDE}$ here is in significant tension with the values preferred when including SH0ES in the analysis.}
  \label{table:params-uberlikelihoodKiDSHSCNoSH0ES}
\end{table}

To complete our analysis of constraints on EDE from these data sets and data set combinations, we consider the cosmological constraints when the SH0ES measurement $H_0 = 74.03 \pm 1.42 \, {\rm km/s/Mpc}$ \cite{Riess:2019cxk} is excluded from the combined data set. We impose an additional inverse-Gaussian $H_0$ prior on the results of Sec.~\ref{sec:constraintsDES} to effectively remove the SH0ES likelihood, which itself is effectively a prior on $H_0$. The resulting posterior distributions correspond to the fit of the $\Lambda$CDM or EDE model to \emph{Planck} 2018 primary CMB, \emph{Planck} 2018 CMB lensing, BAO data from 6dF, SDSS DR7, and SDSS DR12, Pantheon SNIa, SDSS DR12 RSD, DES-Y1 3x2pt data, and the KiDS and HSC $S_8$ priors.  Note that the original MCMC sampling for these constraints was performed with the SH0ES likelihood included, which obviates the concerns related to parameter-space volume effects expressed in, e.g., \cite{Smith:2019ihp}. For a complementary discussion of this issue, see Appendix~B of \cite{Ivanov:2020ril}.

\begin{figure*}
\centering
 \includegraphics[width=\textwidth]{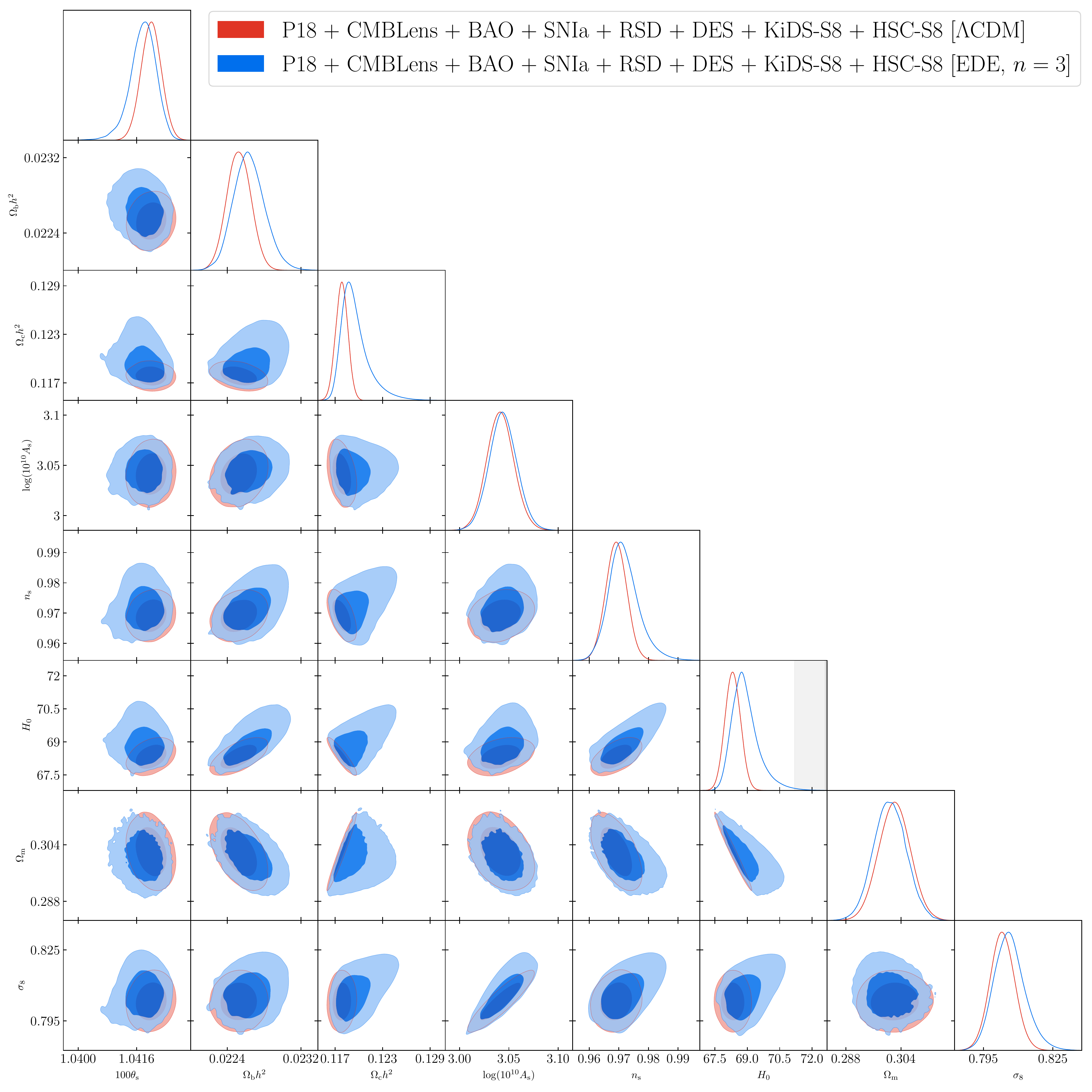}
    \caption{Cosmological parameter constraints with SH0ES excluded from the combined data set (see Sec. \ref{sec:no-SH0ES}), in $\Lambda$CDM and EDE. We include \emph{Planck} 2018 primary CMB data (TT+TE+EE); \emph{Planck} 2018 CMB lensing data; BAO data from 6dF, SDSS DR7, and SDSS DR12; Pantheon SNIa data; SDSS DR12 RSD data; the DES-Y1 3x2pt data; and KiDS and HSC data approximated by an $S_8$ prior. There is little difference in $H_0$ between $\Lambda$CDM and EDE, and the EDE $H_0$ posterior has negligible support in the 2$\sigma$ region of the SH0ES measurement (indicated by the gray band), indicating that the EDE scenario appears unlikely to resolve the Hubble tension. }
    \label{fig:no-SH0ES}
\end{figure*}


We find no evidence for the EDE component. The results are shown in Fig. \ref{fig:mainEDEconstraints} and Fig.~\ref{fig:no-SH0ES} for the EDE and standard $\Lambda$CDM posteriors respectively, and parameter constraints are given in Tables~\ref{table:params-uberlikelihoodKiDSHSCNoSH0ES} and~\ref{table:params-uberlikelihoodKiDSHSCNoSH0ES-margestats}.  We find an upper bound $f_\mathrm{EDE} < 0.060$ at 95\% CL; a two-tailed analysis gives $f_{\rm EDE} = 0.023 \pm 0.017$. This constraint is substantially stronger than found with the primary CMB alone (Table~\ref{table:params-P18-only}, $f_\mathrm{EDE} < 0.087$ at 95\% CL), and indicates clear tension with the values preferred when including SH0ES in the analysis.

We find $H_0=68.92^{+0.57}_{-0.59}$ km/s/Mpc, in substantial ($3.3\sigma$) tension with SH0ES. The EDE value for $H_0$ is extremely close to the value found in $\Lambda$CDM fit to the same combination of data sets, $H_0=68.33 \, \pm 0.36$ km/s/Mpc, consistent with the non-detection of the EDE component in the former.

The posterior distributions for the EDE parameters, Fig.~\ref{fig:mainEDEconstraints}, are in broad agreement with those found in the fit to the primary CMB alone. The initial field displacement $\theta_i$ is bounded by $\theta_i > 0.35$ at 95\% CL, comparable to the result from the fit to the primary CMB alone, $\theta_i > 0.36$. Similarly, we find only a lower bound on $z_c$: $\log_{10}(z_c) > 3.28$ at at 95\% CL. The $f_{\rm EDE}$ posterior is even more weighted towards $f_{\rm EDE}=0$ than the fit to the CMB alone, as is reflected in the tighter upper bound, $f_\mathrm{EDE} < 0.060$ at 95\% CL.  This upper limit is well below the mean and best-fit values found when including SH0ES and excluding DES, HSC, and KV-450 (see Table~\ref{table:params-Smithcomb} or~\cite{Smith:2019ihp}), indicating clear discordance.

\vspace{-.4cm}

\section{Prior Dependence}
\label{sec:priors}

All analyses up until this point have followed past work (e.g.,~\cite{Smith:2019ihp}) and assumed uniform prior probability distributions for the effective EDE parameters $f_{\rm EDE}$ and $\log_{10}(z_c)$, as well as for the initial misalignment angle $\theta_i$. No consideration has been given to the particle physics parameters, namely the axion mass $m$ and decay constant $f$.

\begin{figure}[h!]
    \centering
    \includegraphics[width=\linewidth]{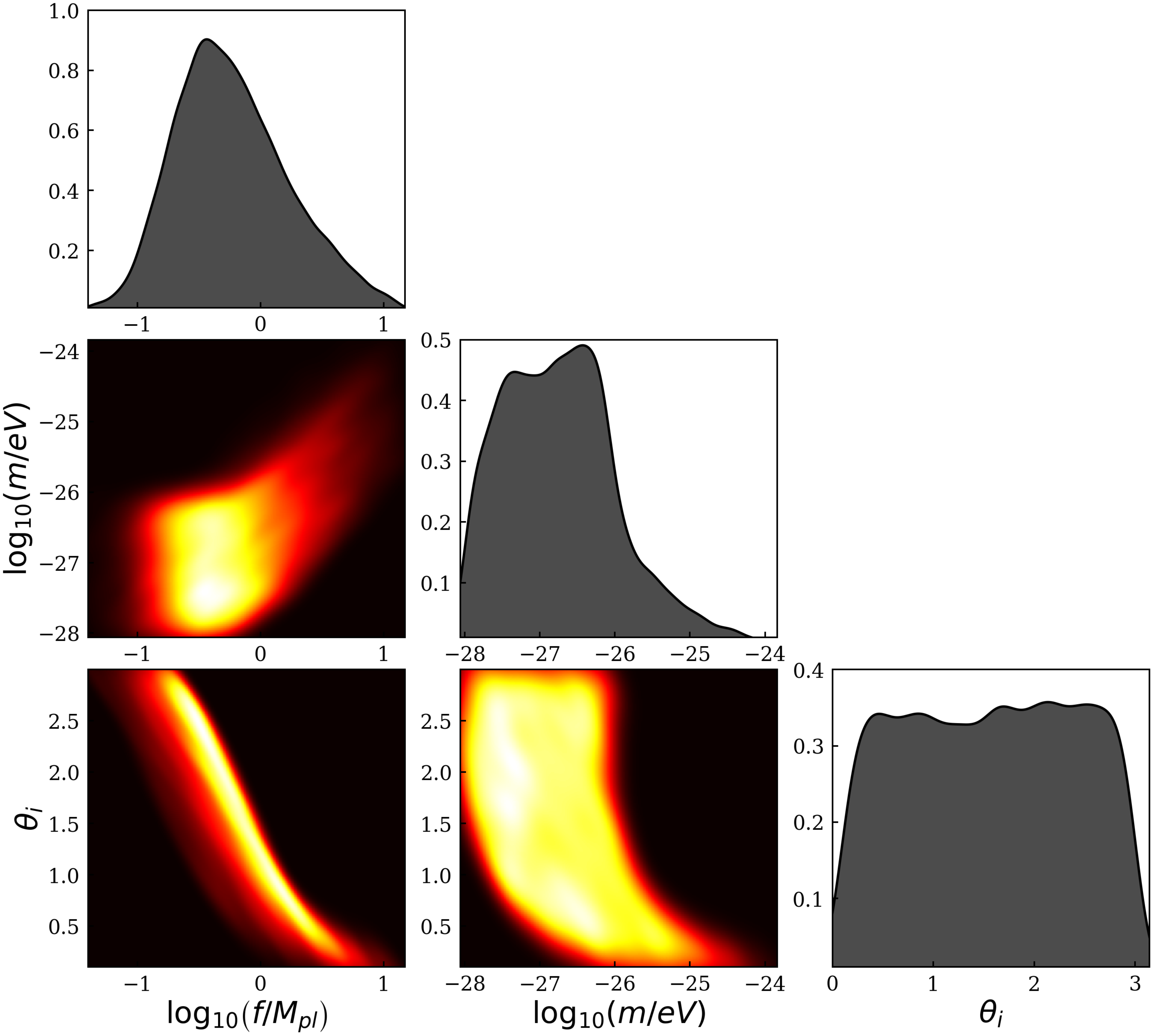}
    \caption{Effective priors on $f$ and $m$ in EDE with flat priors on $f_{\rm EDE}$, $\log_{10}(z_c)$, and $\theta_i$. The distribution of $f$ is peaked at $f=0.59 M_{pl}$, where $M_{pl}=2.435 \times 10^{27} {\rm eV}$ is the reduced Planck mass.}
    \label{fig:smithpriors}
\end{figure}

However, implicit assumptions have been made about the particle physics parameters: uniform priors on $\{ f_{\rm EDE},\log_{10} z_c,\theta_i \}$, translated into implicit priors on $\{f,m,\theta_i \}$, correspond to strongly \emph{non-uniform} priors on $f$ and $m$. To illustrate this issue, in Fig.~\ref{fig:smithpriors} we plot the implicit effective priors imposed on $f$ and $m$ by \cite{Smith:2019ihp}, computed by sampling a uniform probability distribution on $f_{\rm EDE}$, $\log_{10}(z_c)$, and $\theta_i$, in the range $[0.01,0.25]$, $[3.1,4.2]$, and $[0.1,3.0]$, respectively.

\begin{figure}[h!]
    \centering
    \includegraphics[width=\linewidth]{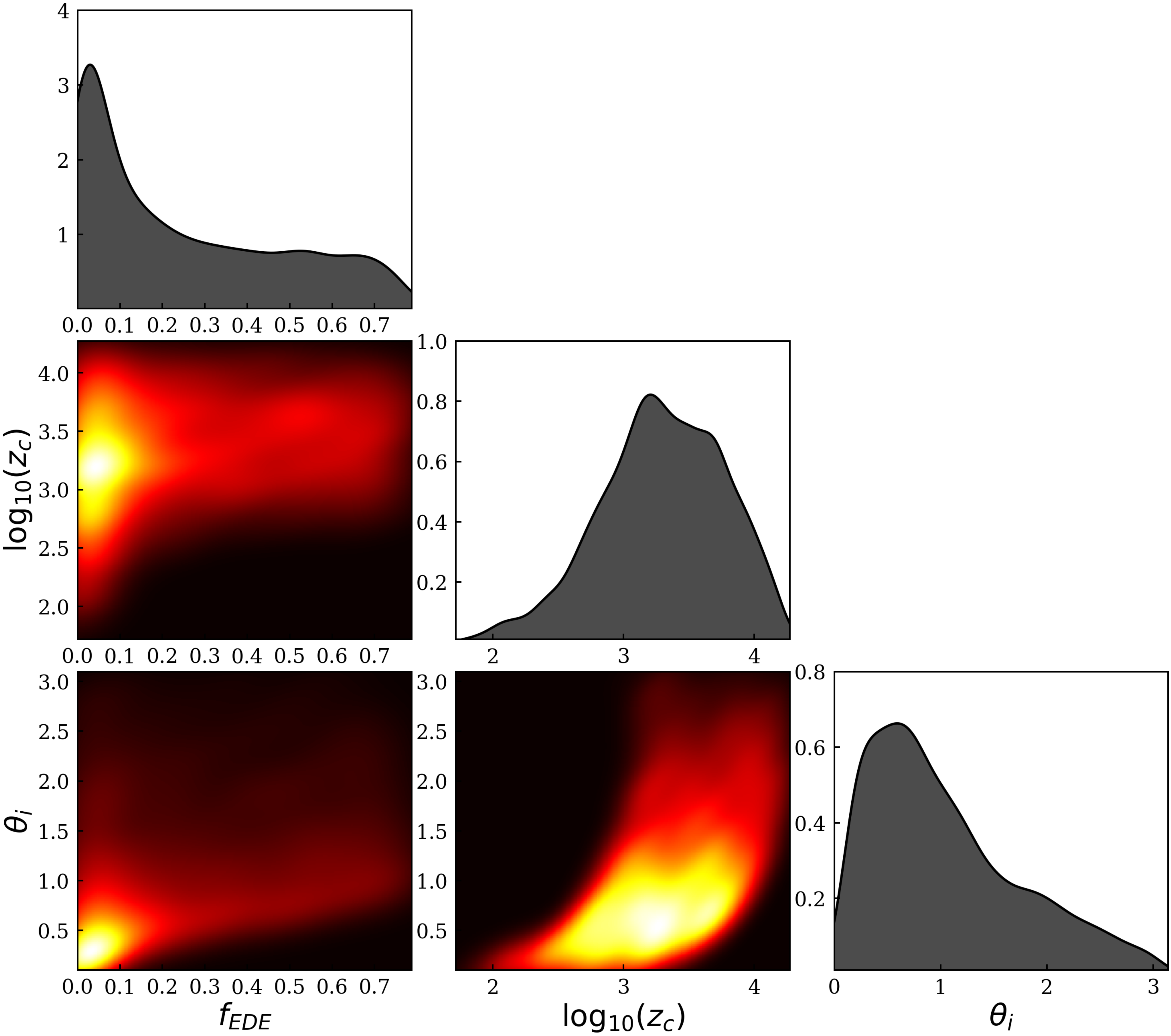}
    \caption{Effective priors on $f_{\rm EDE}$ and
$\log_{10}(z_c)$ in EDE with flat priors on $f$, $\log_{10}(m)$, and $\theta_i$. The $f_{\rm EDE}$ distribution is peaked at $f_{\rm EDE}=0.029$, corresponding to a maximal fraction of $ \approx 3\%$ of the cosmic energy budget in the EDE field. The deviation from a flat distribution of $\theta_i$ results from the rejection of samples with $f_{\rm EDE}>0.8$.}
    \label{fig:uniformpriors}
\end{figure}

One can see in Fig.~\ref{fig:smithpriors} that the distribution of axion decay constants is highly peaked at $f\sim M_{pl}$, in stark contrast with theory expectations, both from statistical arguments \cite{Halverson:2019cmy}, and the Weak Gravity Conjecture \cite{ArkaniHamed:2006dz}; see, e.g.,~\cite{Rudelius:2014wla}.  There is a tight correlation between $f$ and $\theta_i$, with small $\theta_i$ ($\theta_i \lesssim 1.0$) correlating with super-Planckian decay constants $f>M_{pl}$.

An obvious concern is the dependence of the EDE posterior distributions on the MCMC sampler exploring super-Planckian axion decay constants ($f>M_{pl}$), where the theory may no longer be under control.  To quantify this effect, we impose an additional prior on the axion decay constant, $f \leq M_{pl}$, on the results of Sec.~\ref{sec:constraints2}, i.e., the combined data set that does not include DES, HSC, or KV-450. The posterior distributions are shown in Appendix \ref{app:fmplcut}, Fig.~\ref{fig:fmplcut}. We find the restriction to $f \leq M_{pl}$ has a modest effect, with a small overall impact, including on $f_{\rm EDE}$.

\begin{table}[h!]
Constraints from \emph{Planck} 2018 data only (TT+TE+EE) with uniform priors on $f$ and $\log_{10}(m)$ \vspace{2pt} \\
  \centering
  \begin{tabular}{|l|c|}
    \hline\hline Parameter & ~~~EDE ($n=3$) ~~~\\ \hline \hline

    {\boldmath$\ln(10^{10} A_\mathrm{s})$} & $3.046 \, (3.042) \, \pm 0.016$ \\

    {\boldmath$n_\mathrm{s}$} & $0.9657 \, (0.9626)^{+0.0048}_{-0.0049}$ \\

    {\boldmath$100\theta_\mathrm{s}$} & $1.04177 \, (1.04184) \, \pm 0.00032$\\

    {\boldmath$\Omega_\mathrm{b} h^2$} &  $0.02238 \, (0.02222) \, \pm 0.00017$ \\

    {\boldmath$\Omega_\mathrm{c} h^2$} & $0.1212 \, (0.1218)^{+0.0017}_{-0.0019}$ \\

    {\boldmath$\tau_\mathrm{reio}$} & $0.0541 \, (0.0532) \, \pm 0.0075 $\\

    {\boldmath$f/(10^{27} \, {\mathrm{eV}})$} & $2.25 \,(0.80)^{+2.49}_{-1.96} $\\

    {\boldmath$\mathrm{log}_{10}(m/{\mathrm{eV}})$} & $-26.98 \,(-26.11)^{+0.66}_{-0.64} $\\

    {\boldmath$\theta_i$} & $< 2.31 \, (0.15) $\\

    \hline

    $H_0 \, [\mathrm{km/s/Mpc}]$ & $67.53 \, (66.63)^{+0.71}_{-0.73}$ \\

    $\Omega_\mathrm{m}$ & $0.3162 \,(0.3258)\, \pm 0.0083$ \\

    $\sigma_8$ & $0.8132 \, (0.8153)^{+0.0086}_{-0.0089}$ \\
    
    $S_8$ & $0.8349 \, (0.8497) \, \pm 0.0166$ \\

    $\mathrm{log}_{10}(z_c)$ & $< 3.89 \, (3.16)$ \\

    $f_\mathrm{EDE} $ & $< 0.041 \, (0.0003)$\\

    $\mathrm{log}_{10}(f/{\mathrm{eV}})$ & $27.07 \,(26.92) \, \pm 0.60 $\\

    \hline
  \end{tabular} 
  \caption{The mean (best-fit) $\pm1\sigma$ constraints on the cosmological parameters in the EDE model with $n=3$, as inferred from \emph{Planck} 2018 primary CMB data only (TT+TE+EE), with uniform priors placed on the fundamental physics parameter $f$ and $\log_{10}(m)$ rather than the effective EDE parameters $f_{\rm EDE}$ and $\log_{10}(z_c)$ (as used in all other analyses in this paper, and previously in the literature).  Upper limits are quoted at 95\% CL.  The upper bound on the EDE component is even tighter than seen in Table~\ref{table:params-P18-only}, due to the stronger prior weight placed on low values of $f_{\rm EDE}$ by the physical priors (see Fig.~\ref{fig:uniformpriors}).}
  \label{table:params-P18-only-uniform-f-logm}
\end{table}

However, the drastic departure from a flat distribution of decay constants in Fig.~\ref{fig:smithpriors} raises a more general concern regarding the dependence of the EDE posteriors on the choice to impose flat priors on the EDE parameters $f_{\rm EDE}$ and $\log_{10}(z_c)$ instead of the particle physics parameters $f$ and $m$, the latter arguably being the physically reasonable choice of priors. To further elucidate the prior dependence, we perform an MCMC analysis with uniform priors imposed directly on the particle physics parameters $f$ and $\log_{10}(m)$. We consider $f/{\rm eV}=[10^{26},10^{28}]$, $\log_{10}(m/{\rm eV})=[-26,-28]$, $\theta_i=[0.1,3.1]$, and impose an upper bound $f_{\rm EDE} < 0.8$ to ensure a physically reasonable cosmology. We choose to impose a flat prior on $f$ rather than $\log_{10}(f)$ in order to allow a non-negligible probability for large $f_{\rm EDE}$ values; if we had imposed a flat prior on $\log_{10}(f)$ instead (as one could argue is more physically reasonable), then the results below would even more strongly favor small $f_{\rm EDE}$ values. To illustrate the difference in priors, in Fig.~\ref{fig:uniformpriors} we plot the implied prior probability distributions for $f_{\rm EDE}$ and $z_c$. From this one can appreciate that uniform priors on $f$ and $\log_{10}(m)$ impose a strong prior preference for \emph{small} $f_{\rm EDE}$ values, $f_{\rm EDE}<0.1$, and the distribution is peaked at $f_{\rm EDE}=0.029$. The $\theta_i$ distribution deviates significantly from a flat distribution, due solely to the restriction to samples with $f_{\rm EDE}<0.8$.

\begin{figure*}
\centering
 \includegraphics[width=\textwidth]{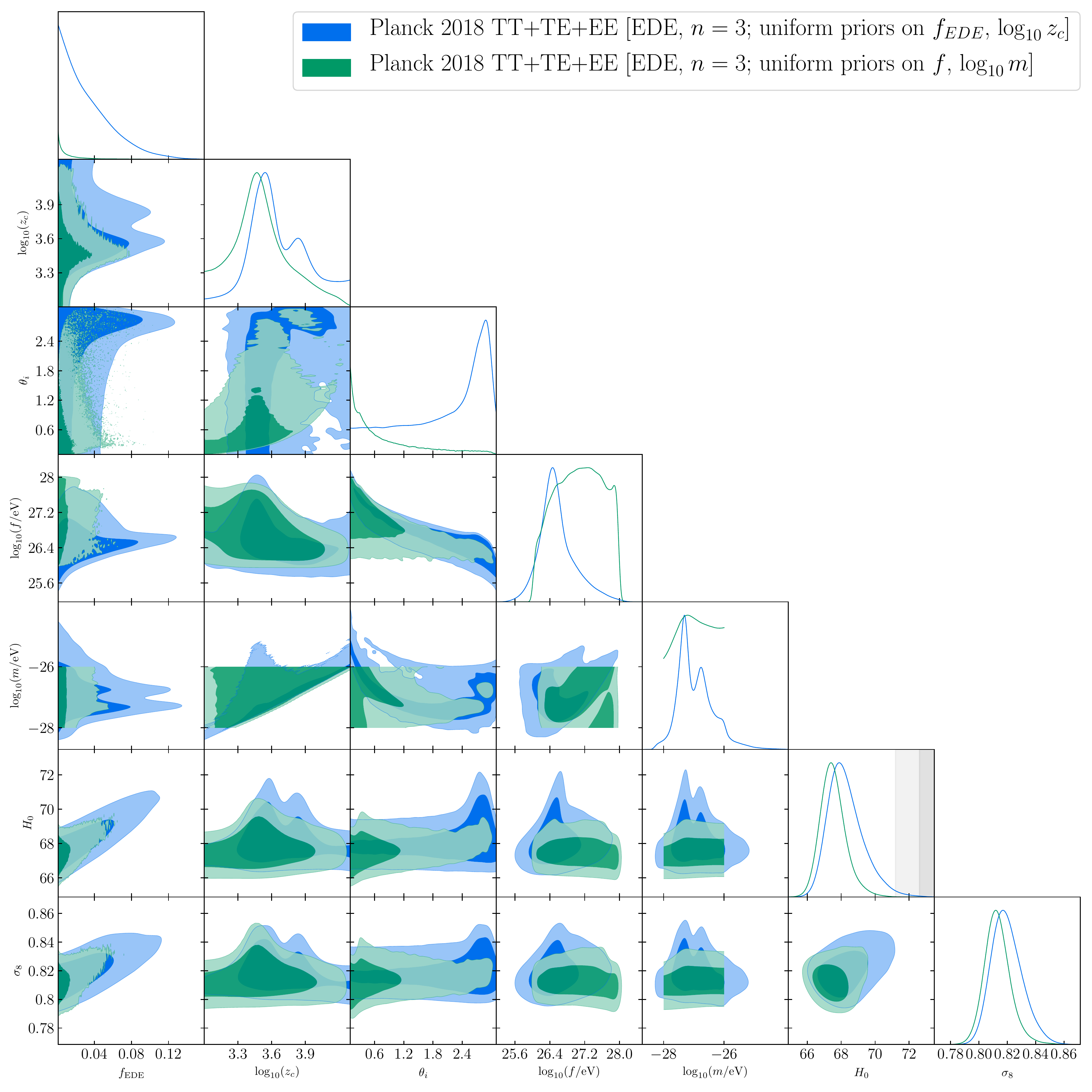}
    \caption{Constraints on the EDE scenario from \emph{Planck} 2018 primary CMB data alone, with varying choices of priors on the EDE model parameters. The blue contours show results with uniform priors placed on the effective EDE parameters $f_{\rm EDE}$ and $\log_{10}(z_c)$, while the green contours show results with uniform priors placed on the fundamental physics EDE parameters $f$ and $\log_{10}(m)$.  The physical priors strongly downweight large $f_{\rm EDE}$ values, and thus significantly disfavor large $H_0$ values.  The possibility of resolving the tension with SH0ES (gray bands) is thus significantly weakened when adopting these physical priors, rather than uniform priors on the effective EDE parameters.}
    \label{fig:EDEconstraints_priors}
\end{figure*}

As a first analysis of the effect of these physical priors, we consider the fit to primary CMB data alone. We recompute the EDE parameter constraints of Sec.~\ref{sec:CMBonly} with the above uniform priors imposed on $f$ and $\log_{10}(m)$.  The posterior distributions are shown in Fig.~\ref{fig:EDEconstraints_priors} for the EDE parameters and Fig.~\ref{fig:P18onlyconstraints} for the standard $\Lambda$CDM parameters. The parameter constraints are tabulated in Table~\ref{table:params-P18-only-uniform-f-logm}.

The upper bound on the EDE fraction is even stronger than found in Sec.~\ref{sec:CMBonly}, which further exacerbates the discrepancy of EDE fit to the primary CMB with the fit of EDE to primary CMB, CMB lensing, BAO, RSD, SNIa, and SH0ES (Sec.~\ref{sec:constraints2}). We find $f_{\rm EDE} < 0.041$ at 95\% CL; the best-fit model lies nearly exactly at $f_{\rm EDE} = 0$.  The posterior distribution for $\theta_i$ in Fig.~\ref{fig:EDEconstraints_priors} is heavily skewed towards $\theta_i=0$, which is strongly correlated with small $f_{\rm EDE}$ (see Fig.~\ref{fig:uniformpriors}). At small values of $f_{\rm EDE}$ the model is poorly constrained, and we observe this in the posteriors of $f$ and $m$, which are visibly prior-dominated. This is despite the fact that the priors imposed on $\log_{10}(m)$ and $f$ encompass the peak values found with uniform priors on $f_{\rm EDE}$ and $\log_{10}(z_c)$ in Sec.~\ref{sec:CMBonly}. In turn, the best-fit parameter values given in  Table~\ref{table:params-P18-only-uniform-f-logm} reflect the posterior preference for small $f_{\rm EDE}$: while the true maximum of the likelihood should be near the best-fit parameter values quoted in Table~\ref{table:params-P18-only}, any reasonable numerical search for a maximum will converge on $f_{\rm EDE} \simeq 0$ here, and not on $f_{\rm EDE}=0.068$ as quoted in Table~\ref{table:params-P18-only}, due to the posteriors' strong weight toward small $f_{\rm EDE}$ values in this analysis.

The posterior distributions for the standard $\Lambda$CDM parameters in the EDE scenario (green contours in Fig.~\ref{fig:P18onlyconstraints}), are nearly identical to the $\Lambda$CDM contours (red). The only hint of EDE is in a slight broadening of the $\Lambda$CDM parameter posteriors. The slight shift in $H_0$ observed in Sec.~\ref{sec:CMBonly} is absent, and we find $H_0$ is nearly identical to that in $\Lambda$CDM.

\section{Discussion and Conclusions}
\label{sec:Discussion}

The Hubble tension (e.g.,~\cite{Verde:2019ivm}) poses a challenge for precision cosmology.  While systematic errors in the calibration of the cosmic distance ladder and/or other data sets may be the ultimate explanation, the growing severity of the tension (modulo the recent TRGB result in~\cite{Freedman:2020dne}) provides impetus to examine theoretical explanations and explore alternative cosmologies that may restore cosmological concordance.

A logical possibility is the presence of new physics in the early universe. One well-motivated scenario is the introduction of a new source of energy density that increases $H(z)$ just prior to recombination, decreasing the sound horizon, and thereby raising the value of $H_0$ inferred from early-universe probes. A popular proposal of this type is the EDE scenario \cite{Poulin:2018cxd}, and its variations \cite{Agrawal:2019lmo,Alexander:2019rsc,Lin:2019qug,Sakstein:2019fmf,Niedermann:2019olb,Kaloper:2019lpl,Berghaus:2019cls,Smith:2019ihp}. 

In this work we reanalyze the EDE scenario accounting for \emph{Planck} 2018 primary CMB data (TT+TE+EE) and LSS data from \emph{Planck} 2018 CMB lensing, BAO (6dF, SDSS DR7, and SDSS DR12), SDSS DR12 RSD, DES-Y1 3x2pt, HSC, and KV-450, as well as supernova distance data from the Pantheon compilation and the SH0ES distance-ladder $H_0$ measurement.  We obtain strong constraints on the existence of an EDE component in the early universe, as seen in Fig.~\ref{fig:mainEDEconstraints} and Table~\ref{table:EDE-params-full}.

In the region of parameter space capable of addressing the Hubble tension, the impact of EDE on LSS observables is substantial. To quantify and contextualize this, in Fig.~\ref{fig:PkDES} we consider the imprint of EDE models on the matter power spectrum in the range of wavenumbers probed by DES, in comparison to the matter power spectrum inferred from DES-Y1 measurements \cite{Abbott:2017wau}. In Fig.~\ref{fig:fEDEPk} and Fig.~\ref{fig:zcPk} we show the fractional change in the matter power spectrum as a function of the effective EDE parameters $f_{\rm EDE}$ and $z_c$, corresponding to the amount and timing of EDE. For $f_{\rm EDE} \approx 12\%$, as proposed in past works, the increase in $P(k)$ is $\mathcal{O}(10\%)$ on small scales, in particular those that are probed by the DES, HSC, and KV-450 data sets.  This change is primarily driven by the significant shifts in the standard $\Lambda$CDM parameters (especially $\omega_{\rm cdm}$ and $n_s$) that are seen when fitting the CMB and SH0ES data to the EDE model, although the scale-dependent suppression of growth induced by the EDE itself also affects the shape of $P(k)$.

Our main results are given in Fig.~\ref{fig:mainEDEconstraints} and Table~\ref{table:EDE-params-full}.  We find no evidence for EDE in the primary CMB anisotropies alone: the fit to {\it Planck} 2018 TT+TE+EE gives an upper bound $f_{\rm EDE}<0.087$ at $95\%$ CL, with $H_0 = 68.29 ^{+1.02}_{-1.00} \, {\rm km/s/Mpc}$, shifted only slightly upwards relative to the $\Lambda$CDM value $H_0=67.29 \pm 0.59 \, {\rm km/s/Mpc}$, and with a considerably larger error bar.  Both $H_0$ values are in significant tension with the SH0ES measurement $H_0 = 74.03 \pm 1.42 \, {\rm km/s/Mpc}$ \cite{Riess:2019cxk}. When the primary CMB data set is supplemented with data from {\it Planck} 2018 CMB lensing, BAO distance measurements, RSD data, the Pantheon SNIa distance measurements, and the SH0ES $H_0$ measurement, we instead find $3.1\sigma$ evidence for EDE, $f_\mathrm{\rm EDE}=0.098 \pm 0.032$. We find $H_0 = 70.98 \, \pm \, 1.05 \,{\rm km/s/Mpc}$, reducing the tension with SH0ES to $\approx 1.7\sigma$. This is consistent with past results in the literature for this combination of data sets \cite{Poulin:2018cxd,Smith:2019ihp}.

However, this is not the end of the story: these shifts in $f_{\rm EDE}$ and $H_0$ are reversed upon the inclusion of additional LSS data, in particular measurements of weak gravitational lensing and galaxy clustering. These data strongly constrain $S_8$, which in the EDE scenario is highly correlated with $f_{\rm EDE}$ and $H_0$. The tight correlation between $\sigma_8$, $f_{\rm EDE}$, and $H_0$ is clearly visible in the posterior distributions in Fig.~\ref{fig:mainEDEconstraints}, and is manifested in the matter power spectrum. We find that additional LSS data substantially weaken the evidence for EDE, as a result of the tension between the larger values of $S_8$ needed to fit the CMB and SH0ES in these models and the lower values of this parameter measured in LSS surveys. Including the full DES-Y1 3x2pt likelihoods, we find the evidence for EDE is reduced to $2.3\sigma$, with $f_{\rm EDE}= 0.077 ^{+0.032} _{-0.034}$. The inclusion of HSC and KV-450 data, approximated as priors on $S_8$, shrinks this further to $f_{\rm EDE}=0.062 \, ^{+0.032}_{-0.033}$, corresponding to a one-sided upper bound $f_{\rm EDE} < 0.112$ at 95\% CL.  We find $H_0=70.45 \,^{+1.05}_{-1.08}$ in the fit with DES, HSC, and KV-450 included, in moderate 2$\sigma$ tension with SH0ES alone (note that SH0ES is included in all of the aforementioned fits).  There is no sign of concordance amongst these data sets: the LSS data pull the parameters in the \emph{opposite} direction to that required to simultaneously fit the CMB and SH0ES data.

To understand the apparent conflict between LSS data and SH0ES-tension-resolving EDE cosmologies, we fit the EDE model to the combined data set with SH0ES excluded in Sec.~\ref{sec:no-SH0ES} (note that Pantheon SNIa relative distances are still included). We find a constraint on EDE even tighter than found with the primary CMB alone, $f_{\rm EDE}<0.060$ at 95\% CL, with no hint of a preference for EDE. The corresponding Hubble constraint is $H_0 = 68.92 ^{+0.57} _{-0.59}$ km/s/Mpc, in significant tension with SH0ES.  The tight constraint found here indicates that CMB and LSS data do not show any hint of moving toward a cosmology that can accommodate the SH0ES $H_0$ value, even in the broadened EDE parameter space.  Physically, this arises from the fact that a higher $H_0$ value in the CMB fit requires a higher $f_{\rm EDE}$ value, which in turn requires higher $\omega_{\rm cdm}$ and $n_s$ values, thereby increasing $\sigma_8$ and conflicting with LSS constraints.  There does not appear to be a viable swath of parameter space to satisfy all existing constraints.

Finally, we examine the choice of priors and the role of the axion decay constant $f$. Uniform priors imposed on $f_{\rm EDE}$, $\log_{10} z_c$, and $\theta_i$ effectively impose a non-uniform prior on $f$ and $\log_{10}(m)$, as seen in Fig.~\ref{fig:smithpriors}. Notably, the effective prior for the axion decay constant $f$ is peaked near the Planck scale, $f\approx M_{pl}$, in conflict with theoretical expectations from particle physics and quantum gravity. When the fit to primary CMB data is repeated with uniform priors imposed directly on $f$ and $\log_{10}(m)$, the upper bound on $f_{\rm EDE}$ becomes $f_\mathrm{\rm EDE}<0.041$ at $95\%$ CL. This is significantly lower than the corresponding result for uniform priors on the effective parameters $f_{\rm EDE}$ and $z_c$, suggestive of a prior dependence for EDE results more generally.  The use of such physical priors in the other analyses presented in this paper would only further tighten the upper limits on EDE.

Taken in conjunction, these results paint a bleak picture for the viability of the EDE scenario as a candidate to restore cosmological concordance. More generally, it is likely that {\it any model} that attempts to decrease the sound horizon by increasing $H(z)$ through a new dark-energy-like component that is active at early times will encounter the problems identified here.  All such models, insofar as they can accommodate a close fit to both the CMB and SH0ES measurement, will do so at the cost of a shift in $\Lambda$CDM parameters that is not compatible with LSS data. Furthermore, we have not utilized all possible data sets that constrain LSS in this paper; in particular, galaxy cluster number counts~(e.g.,~\cite{2013JCAP...07..008H,2015MNRAS.446.2205M,2016A&A...594A..24P,2019ApJ...878...55B}) and thermal Sunyaev-Zel'dovich power spectrum measurements~(e.g.,~\cite{2016A&A...594A..22P,2018MNRAS.477.4957B}), amongst other probes, tightly constrain $\sigma_8$ and $\Omega_m$.  Cluster number counts tend to favor ``low'' values of $\sigma_8$, consistent with weak lensing and other LSS probes. Thus, their inclusion would likely strengthen the conclusions drawn here.  However, it may be important to explicitly verify the accuracy of current fitting functions~(e.g.,~\cite{2008ApJ...688..709T}) or emulators that are used for the halo mass function in the context of EDE cosmologies, prior to applying this methodology to constrain the EDE scenario.

Broadening the model space, one possible solution to the tensions identified in this work may be to introduce a larger neutrino mass, which would suppress small-scale power in $P(k)$ and thereby allow larger $f_{\rm EDE}$ (and hence $H_0$) than found here.  However, whether a large enough neutrino mass is consistent with the CMB and distance probes is unclear. The coming decade will see the launch of several powerful LSS experiments (e.g., WFIRST \cite{2019arXiv190205569A}, DESI \cite{Levi:2019ggs}, VRO \cite{Ivezic:2008fe} (formerly LSST), and Euclid \cite{Amendola:2016saw}), and with these, an abundance of data from a range of redshifts.  In the absence of significant shifts with respect to current LSS data, it seems unlikely that these next-generation data sets will reverse the negative trajectory we have seen here in the evidence for EDE as LSS data are included in the analysis.  However, the additional statistical power will allow for tight constraints on EDE, even when additional degrees of freedom are allowed to vary (e.g., neutrino masses, $N_{\rm eff}$, the primordial power spectrum, etc.).

Regardless of implications for the Hubble tension, ultralight axions are of cosmological interest in their own right (see, e.g.,~\cite{Grin:2019mub,Hlozek:2014lca,Hlozek:2016lzm}), and the EDE variant of this idea may leave interesting cosmological signatures even in the region of parameter space where the impact on the inferred $H_0$ value is minimal. For example, interesting effects arise in birefringence of CMB polarization \cite{Capparelli:2019rtn}, in principle yielding a signal in CMB circular polarization \cite{Alexander:2019sqb,Padilla:2019dhz}.  Other interesting signals may arise due to the parametric resonance effects described in~\cite{Smith:2019ihp}.  Orthogonal to these considerations, it may be interesting to perform an appraisal of the discordance of the EDE model along the lines proposed in, e.g., \cite{Raveri:2018wln,Raveri:2019gdp,Lemos:2019txn}. Looking towards alternative approaches to the $H_0$ tension, the results presented here potentially motivate further study of new physics in the cosmic distance ladder itself \cite{Sakstein:2019qgn,Desmond:2019ygn,Desmond:2020wep}. We leave these interesting directions to future work.

\acknowledgments
The authors thank Eric Baxter, Jo Dunkley, Daniel Foreman-Mackey, Gil Holder, Mikhail Ivanov, Antony Lewis, Julien Lesgourgues, Blake Sherwin, Marko Simonovic, David Spergel, and Matias Zaldarriaga for helpful discussions, as well as the Scientific Computing Core staff at the Flatiron Institute for computational support. EM and MT thank the Simons Foundation and the Flatiron Institute Center for Computational Astrophysics for hospitality while a portion of this work was completed.  JCH thanks the Simons Foundation for support.

\bibliographystyle{JHEP}
\bibliography{EDE-LSS-refs}

\onecolumngrid

\pagebreak
\appendix

\section{Constraints on EDE scalar field decay constant and mass}
\label{app:mfconstraints}

\begin{figure}[h!]
\centering
 \includegraphics[scale=0.7]{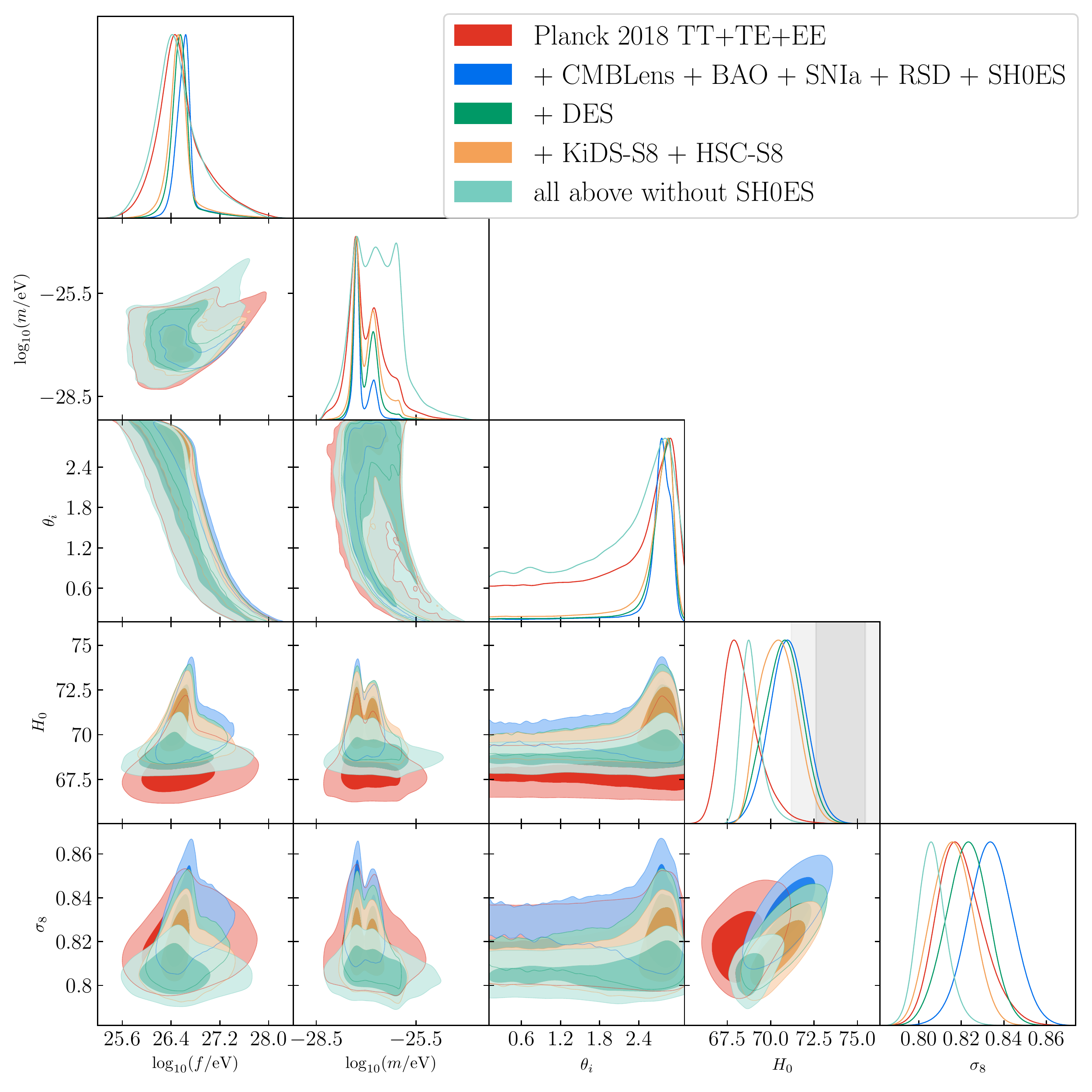}
    \caption{Constraints on the EDE scenario from a variety of cosmological data sets. The analyses shown are identical to those in Fig.~\ref{fig:mainEDEconstraints}, but with the fundamental physics parameters $\log_{10}(f)$ and $\log_{10}(m)$ displayed instead of the effective EDE parameters $f_{\rm EDE}$ and $\log_{10}(z_c)$.}
    \label{fig:no-SH0ES-logf-logm}
\end{figure}


\newpage

\section{Validation of $S_8$ prior approach}
\label{app:S8-validation}

In this appendix, we test the approximation of replacing the full DES 3x2pt likelihood with a Gaussian prior on $S_8$ corresponding to the DES measurement. While the DES data most tightly constrain $S_8$, they also yield a reasonably strong constraint on $\Omega_m$ \cite{Abbott:2017smn}. It is thus not {\it a priori} clear that a simple $S_8$ prior is an adequate substitute for the full likelihood.  However, as we will see, since DES only weakly constrains any non-$S_8$ parameter in comparison to \emph{Planck} (including $\Omega_m$), this approximation turns out to be excellent when performing a joint analysis of these data sets.

\begin{figure}[h!]
\centering
 \includegraphics[scale=0.5]{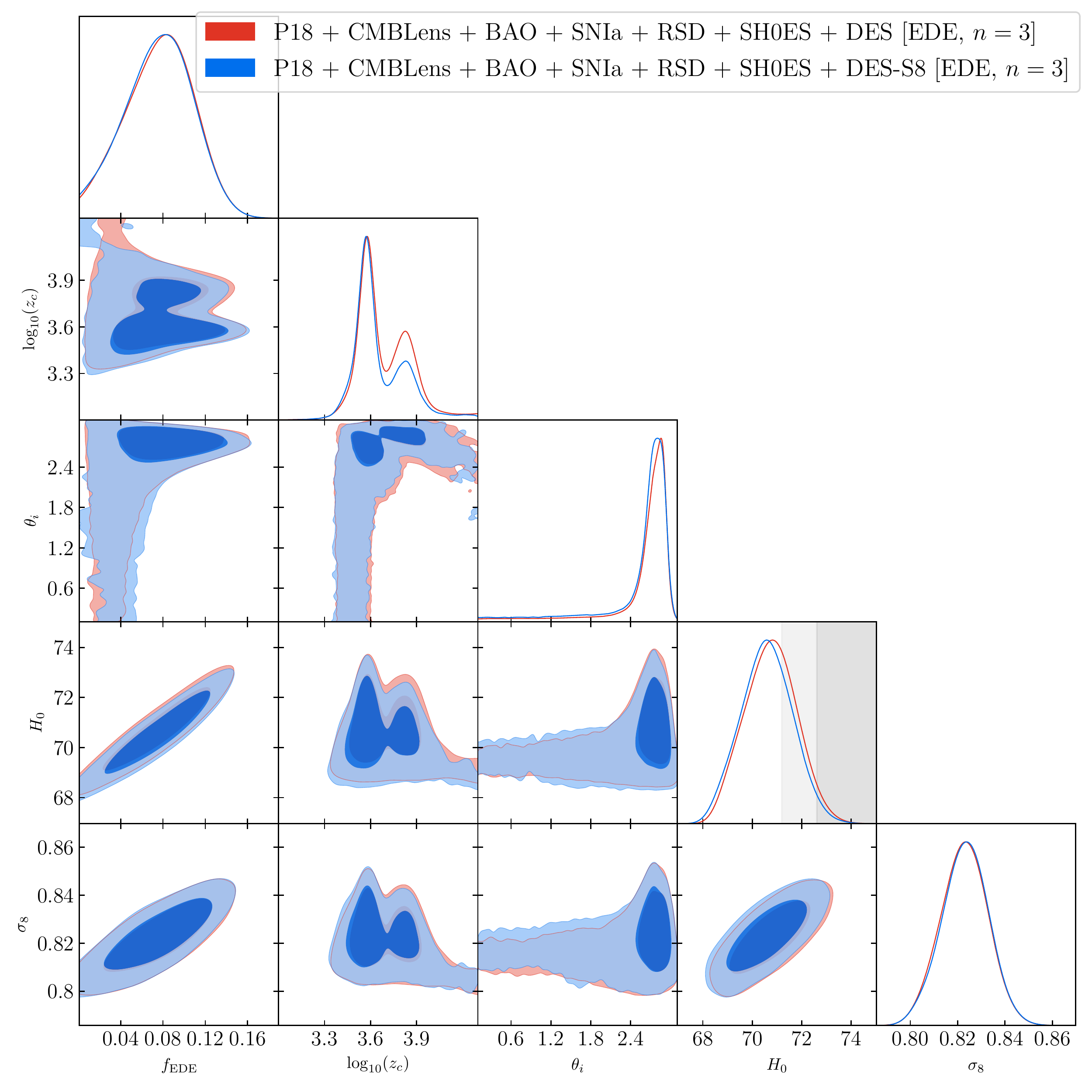}
    \caption{Validation of an $S_8$ prior as a good approximation to the inclusion of the full DES-Y1 3x2pt likelihood in the EDE model. We compare the posterior distributions in the fit of EDE to \emph{Planck} 2018 primary CMB data (TT+TE+EE); \emph{Planck} 2018 CMB lensing data; BAO data from 6dF, SDSS DR7, and SDSS DR12; Pantheon SNIa data; the latest SH0ES $H_0$ constraint; SDSS DR12 RSD data; and the DES-Y1 3x2pt data (red), to the posterior distributions in the fit to same data set but with DES-Y1 3x2pt data replaced by a prior on $S_8$ (blue), $S_8 = 0.773 ^{+0.026} _{-0.020}$. The close agreement of the posteriors indicates a Gaussian prior on $S_8$ is an excellent approximation to the inclusion of the full DES-Y1 3x2pt likelihood in this data combination in the context of EDE. }
    \label{fig:S8-validation-EDE-params}
\end{figure}

\begin{figure}[h!]
\centering
 \includegraphics[width=\textwidth]{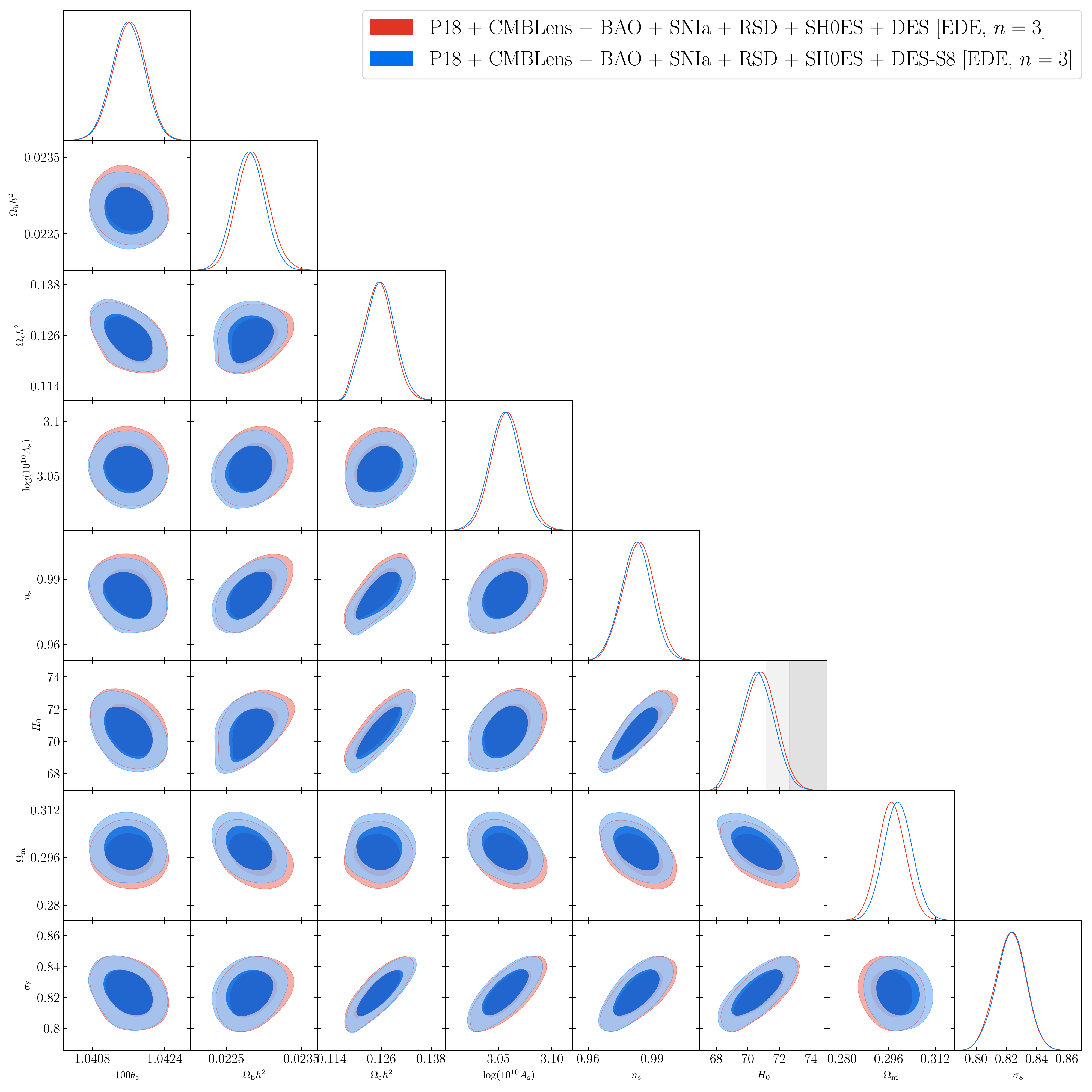}
    \caption{Validation of an $S_8$ prior as a good approximation to the inclusion of the full DES likelihood in the EDE model, analogous to Fig.~\ref{fig:S8-validation-EDE-params} but here showing the standard cosmological parameters. We compare the posterior distributions in the fit of EDE to \emph{Planck} 2018 primary CMB data (TT+TE+EE); \emph{Planck} 2018 CMB lensing data; BAO data from 6dF, SDSS DR7, and SDSS DR12; Pantheon SNIa data; the latest SH0ES $H_0$ constraint; SDSS DR12 RSD data; and the DES-Y1 3x2pt data (red), to the posterior distributions in the fit to same data set but with DES 3x2pt data replaced by a prior on $S_8$ (blue), $S_8 = 0.773 ^{+0.026} _{-0.020}$. The close agreement of the posteriors indicates a Gaussian prior on $S_8$ is an excellent approximation to the inclusion of the DES 3x2pt likelihood in this data combination in the context of EDE. }
    \label{fig:S8-validation}
\end{figure}

\begin{figure}[h!]
\centering
 \includegraphics[width=\textwidth]{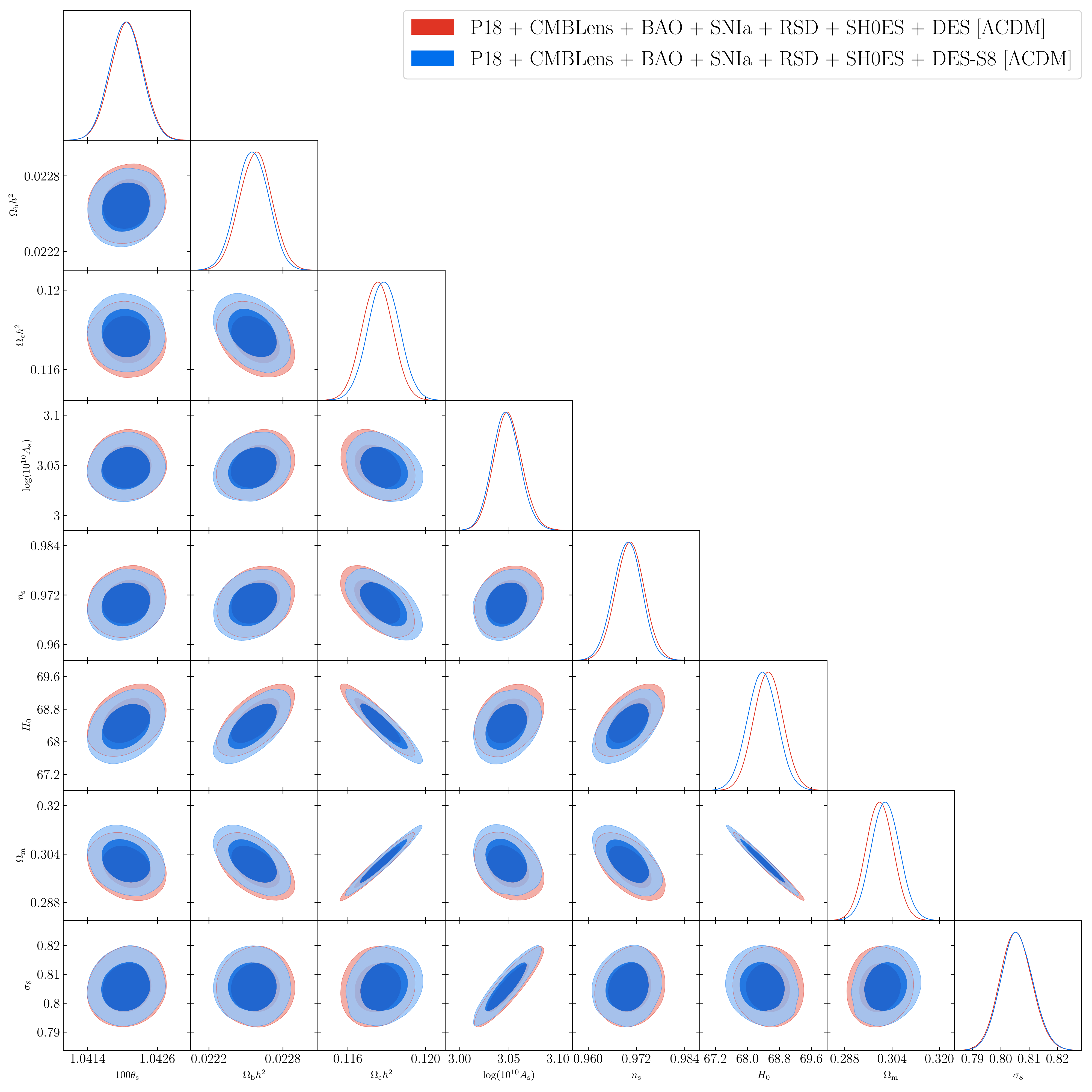}
    \caption{Validation of an $S_8$ prior as a good approximation to the inclusion of the full DES likelihood in $\Lambda$CDM. We compare the posterior distributions in the fit of $\Lambda$CDM to \emph{Planck} 2018 primary CMB data (TT+TE+EE); \emph{Planck} 2018 CMB lensing data; BAO data from 6dF, SDSS DR7, and SDSS DR12; Pantheon SNIa data; the latest SH0ES $H_0$ constraint; SDSS DR12 RSD data; and the DES-Y1 3x2pt data (red), to the posterior distributions in the fit to same data set but with DES 3x2pt data replaced by a prior on $S_8$ (blue), $S_8 = 0.773 ^{+0.026} _{-0.020}$. The close agreement of the posteriors indicates a Gaussian prior on $S_8$ is an excellent approximation to the inclusion of the DES 3x2pt likelihood in this data set combination in the context of $\Lambda$CDM. The slight ($\ll 1 \sigma$) shift in $H_0$ arises due to the constraining power of DES on $\Omega_m$, which is not captured in the simple $S_8$ prior approach, and the tight correlation between $\Omega_m$ and $H_0$ in $\Lambda$CDM. }
    \label{fig:S8-validation-LCDM}
\end{figure}

We are particularly interested in the validity of this approach in the context of the EDE model, and in the context of the data set combination utilized here. Thus, we consider the posterior distributions obtained for the full DES likelihood and for a DES prior on $S_8$. The results for the EDE model are shown in Figs.~\ref{fig:S8-validation-EDE-params} and \ref{fig:S8-validation}. The figures display a near-perfect match of the posteriors, thus demonstrating that the $S_8$ prior is an excellent approximation to the full DES likelihood in the EDE model, within the context of this data set combination. For completeness, we perform the same comparison for $\Lambda$CDM, shown in Fig.~\ref{fig:S8-validation-LCDM}. The match is again excellent, although we note a slight ($\ll 1\sigma$) shift in the peak of the $H_0$ posterior. This is caused by the $\Omega_m$-$H_0$ degeneracy in $\Lambda$CDM, and the fact that the $S_8$ prior approach does not include the DES constraining power on $\Omega_m$.

From these comparisons we conclude that the $S_8$ prior is an excellent approximation to the full DES 3x2pt likelihood in the EDE and $\Lambda$CDM models. However, as evidenced by the small yet visible shift of the $H_0$ posterior in $\Lambda$CDM, this claim should be evaluated on a model-by-model basis.

\vfill

\pagebreak

\clearpage

\vspace{5cm}
\section{Restriction to Sub-Planckian Axion Decay Constants}
\label{app:fmplcut}

\begin{figure}[h!]
\centering
 \includegraphics[scale=0.5]{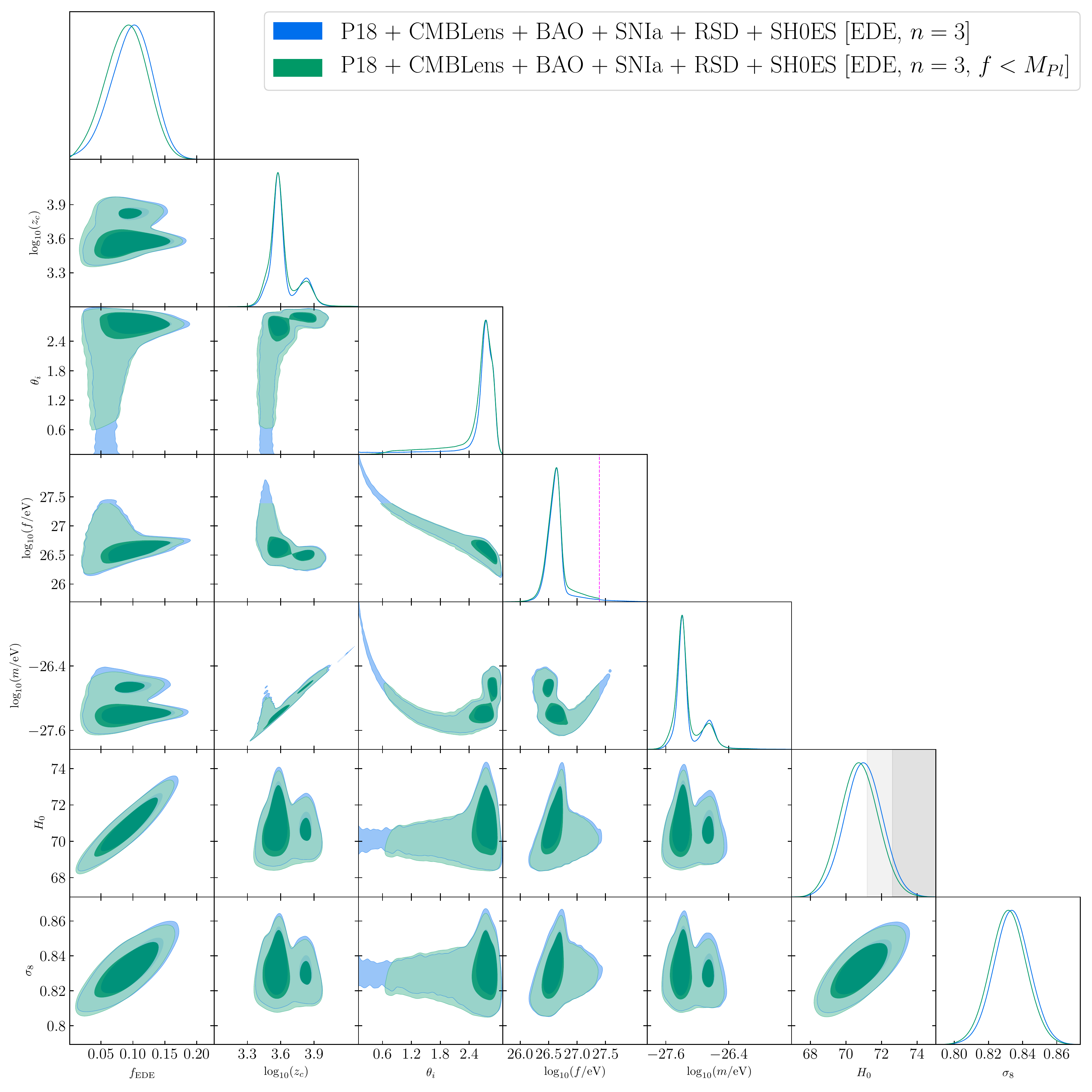}
    \caption{Cosmological parameter constraints with an upper bound imposed on the axion decay constant, $f\leq M_{pl}$. See Sec.~\ref{sec:priors}.}
    \label{fig:fmplcut}
\end{figure}

\pagebreak

\clearpage

\section{Additional Figures Comparing EDE and $\Lambda$CDM Predictions}
\label{app:EDEvsLCDMfigs}

\begin{figure}[h!]
    \centering
    \includegraphics[width=\textwidth]{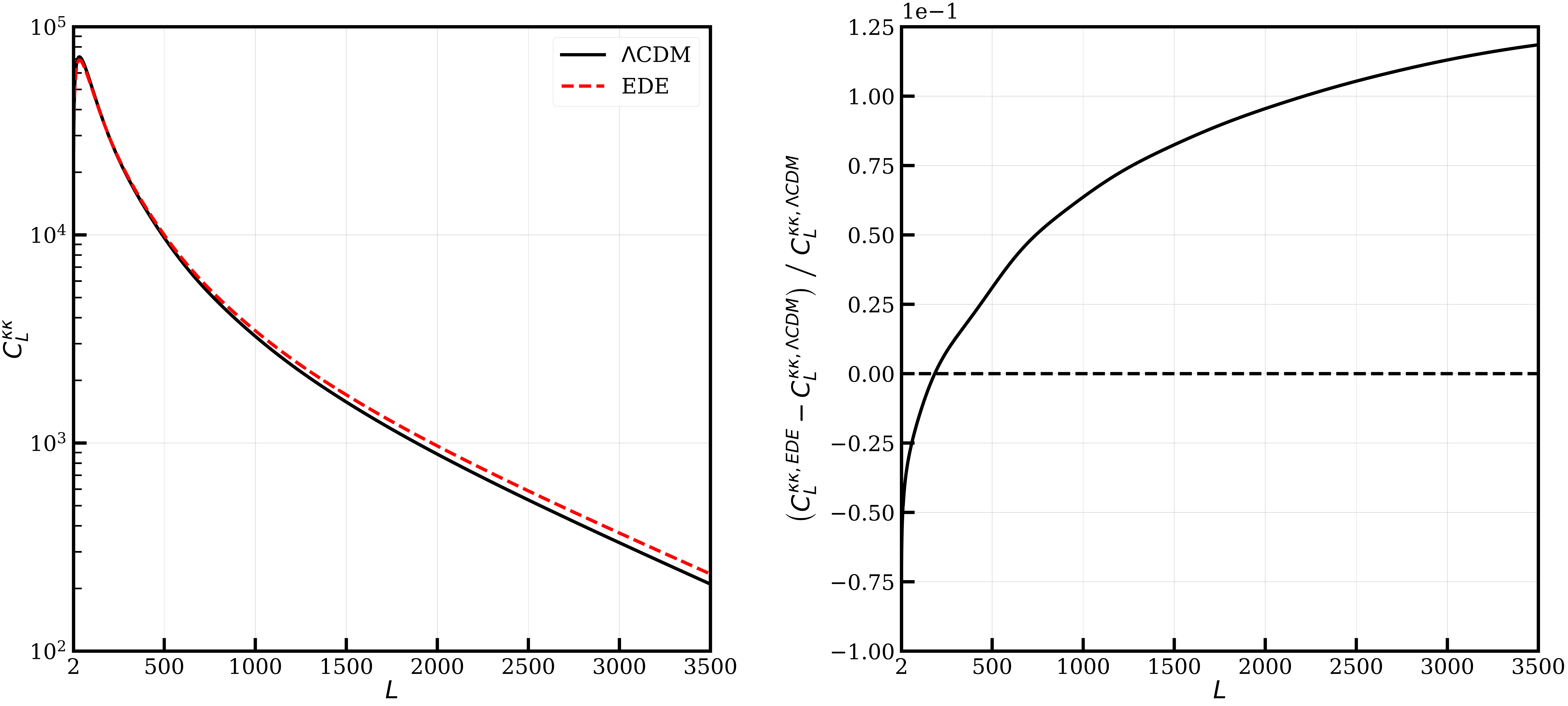} 
  \caption{CMB lensing convergence auto-power spectrum in EDE with parameters given in Eq.~\eqref{smithparams} and $\Lambda$CDM with parameters given in Eq.~\eqref{smithparamsLCDM}, and the fractional difference between them. The standard $\Lambda$CDM parameters differ non-negligibly between the models, and similar to the matter power spectrum observed in Sec.~\ref{sec:LSS}, this generates a substantial change in the CMB lensing power spectrum. The change is $\mathcal{O}(10\%)$ at high-$L$, driven primarily by the shift in $\Lambda$CDM parameters, and not effects of the EDE itself.}
  \label{CMBTEgg}
\end{figure}

\begin{figure}[!h]
    \centering
\includegraphics[width=\textwidth]{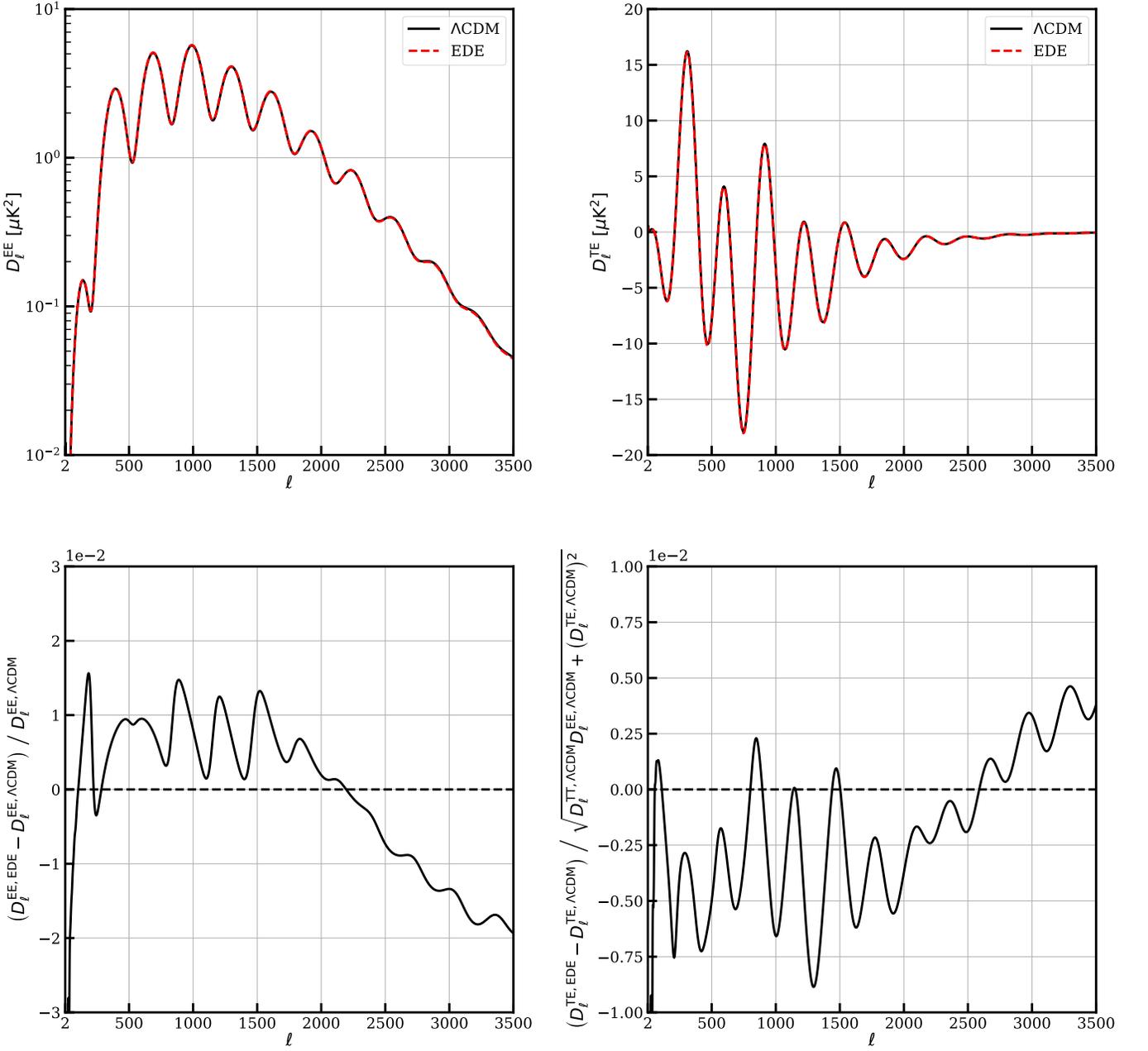}
  \caption{CMB EE (left) and TE (right) power spectra in EDE with parameters given in Eq.~\eqref{smithparams} and $\Lambda$CDM with parameters given in Eq.~\eqref{smithparamsLCDM}, and fractional difference between EDE and $\Lambda$CDM (bottom). The parameters for both models correspond to the best-fit values from Table 1 of \cite{Smith:2019ihp}. The standard $\Lambda$CDM parameters differ non-negligibly between the models, while producing remarkably similar CMB temperature and polarization power spectra.  Note that in the fractional difference plot for TE we have normalized by the variance, differing in our convention relative to other figures, because of the zero crossings of the TE cross-correlation.} 
  \label{CMBTTEE}
\end{figure}

\begin{figure}
    \centering
    \includegraphics[width=\textwidth]{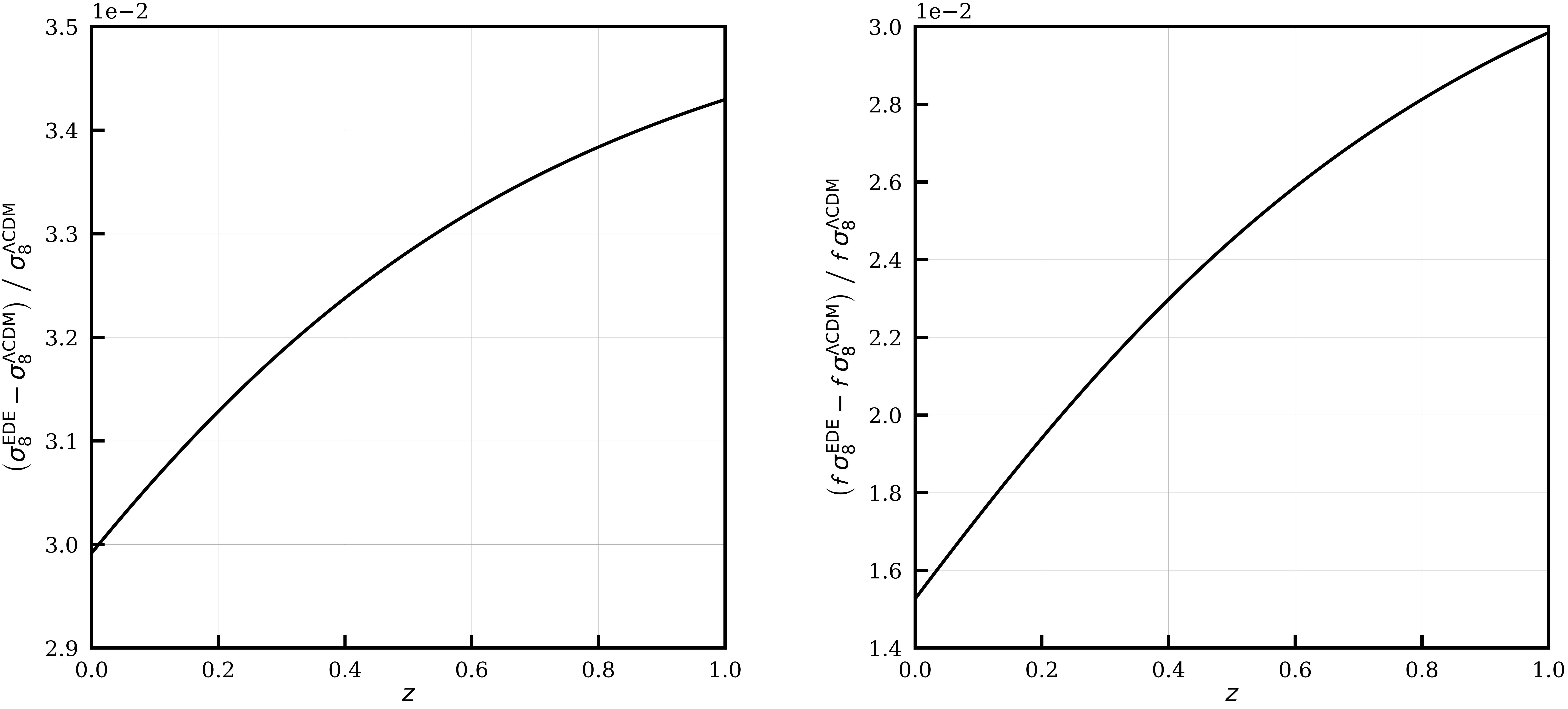}
 \caption{Fractional difference of $\sigma_8(z)$ and $f\sigma_8(z)$ in EDE with parameters given in Eq.~\eqref{smithparams} and $\Lambda$CDM with parameters given in Eq.~\eqref{smithparamsLCDM}. The parameters for both models correspond to the best-fit values from Table 1 of \cite{Smith:2019ihp}. The standard $\Lambda$CDM parameters differ non-negligibly between the models, leading to the changes seen here, while producing remarkably close CMB power spectra (see Figs.~\ref{fig:CMB_TT} and~\ref{CMBTTEE}).}
 \label{fig:fsigma8}
\end{figure}

\begin{figure}[h!]
\includegraphics[width=\textwidth]{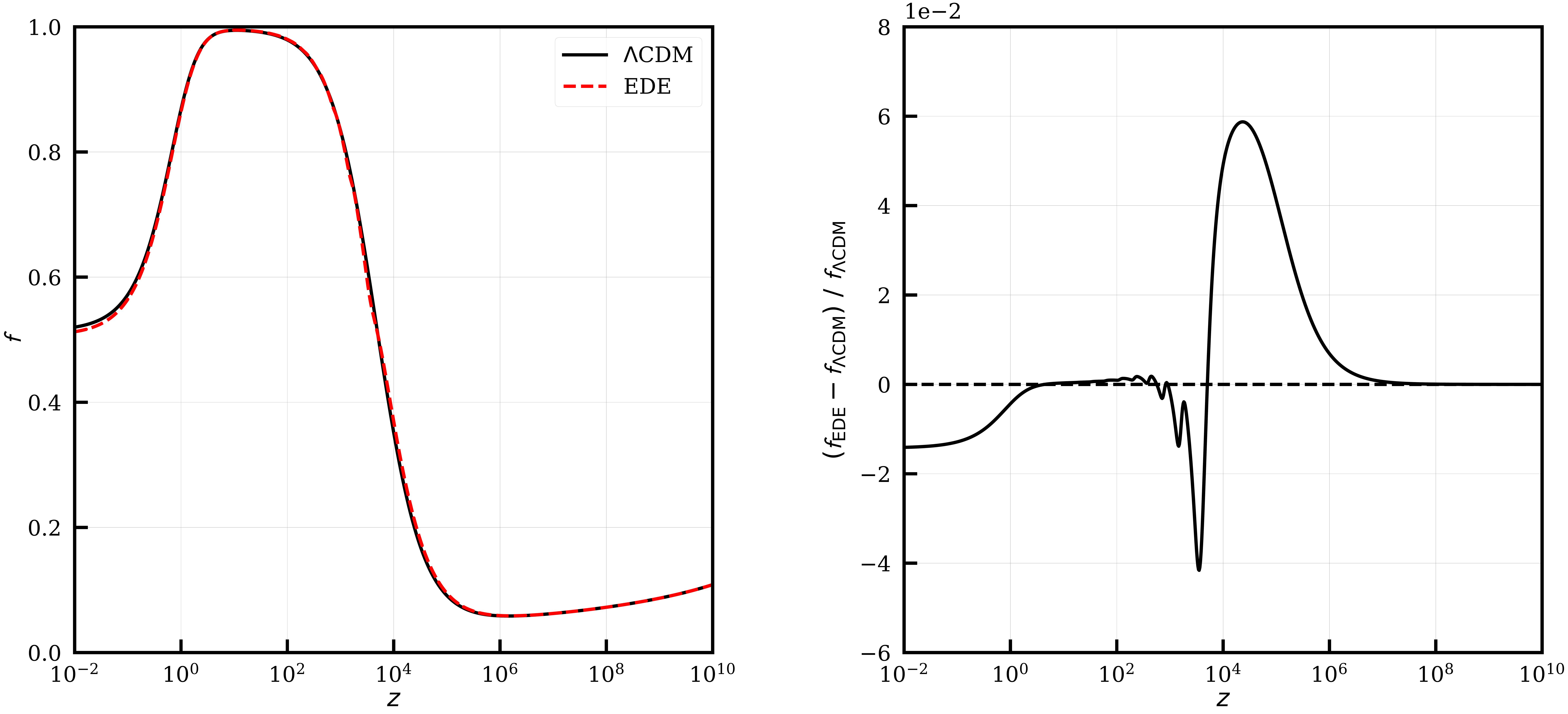}
\caption{Growth factor $f$ in EDE vs. $\Lambda$CDM (left) and fractional difference between the two (right).  The models and parameters are identical to those used in Fig.~\ref{fig:fsigma8} (and elsewhere in this appendix).  Note that this plot covers a very wide range in redshift, in order to show the impact of the EDE field on the growth of perturbations over all of cosmic history.}
\label{fplot}
\end{figure}

In what follows we include additional figures displaying the difference between the best-fit EDE and $\Lambda$CDM models to non-LSS data, i.e., CMB, SH0ES, and distance data (RSD data were also included in the fits, but play very little role due to their relatively large error bars).  The parameters used correspond to the best-fit values from Table 1 of \cite{Smith:2019ihp}. For EDE, these are given in Eq.~\eqref{smithparams},
 \begin{equation}
     H_0=72.19 \, {\rm km/s/Mpc} \;\; , \;\; 100\omega_b=2.253 \;\; , \;\; \omega_{\rm cdm} =0.1306 \;\; , \;\; 10^9 A_s = 2.215 \;\; , \;\;  n_s=0.9889 \;\; , \;\; 
     \tau_{\rm reio}=0.072   \nonumber
\end{equation}
\begin{equation}
         f_{\rm EDE}=0.122 \;\;,\;\;\log_{10}(z_c)=3.562 \;\;,\;\;\theta_i = 2.83.   \nonumber
\end{equation}
 while for $\Lambda$CDM, these are given in Eq.~\eqref{smithparamsLCDM},
\begin{equation}
     H_0=68.21 \, {\rm km/s/Mpc} \;\; , \;\;   100\omega_b=2.253 \;\; , \;\;  \omega_{\rm cdm}=0.1177  \;\; , \;\; 10^9 A_s = 2.216\;\; , \;\;  n_s=0.9686 \;\; , \;\;    \tau_{\rm reio}=0.085. \nonumber 
 \end{equation}
We show the CMB lensing convergence auto-power spectrum, $C_L^{\kappa \kappa}$, and the fractional difference between EDE and $\Lambda$CDM in Fig.~\ref{CMBTEgg}. The impact here is substantial at high $L$, giving rise to shifts from $\Lambda$CDM of $\mathcal{O}$(10\%), consistent with the changes induced in the matter power spectrum (see Sec.~\ref{sec:LSS}).  It should be noted that the changes in the CMB lensing and matter power spectra are driven by the sizeable shift in the physical CDM density $\omega_{\rm cdm}$, as well as the shift in the scalar spectral index $n_s$; these shifts are also what preserve the fit to the primary CMB power spectra.  Fig.~\ref{CMBTTEE} shows the primary CMB power spectra $D^{EE}_\ell$ (left) and $D^{TE}_\ell$ (right) and the fractional difference between the two models. These results further emphasize the remarkable agreement between the two models in the CMB, as was displayed in Fig.~\ref{fig:CMB_TT} for the temperature power spectrum. We additionally include the fractional difference for $f\sigma_8$ and $\sigma_8$ in Fig.~\ref{fig:fsigma8}.  We conclude with a comparison of the growth factors and their fractional differences for both cosmologies in Fig.~\ref{fplot}, which illustrate the effects of the EDE on the growth of perturbations over nearly all of cosmic history.

\clearpage

\section{Parameter Constraints from Credible-Interval Approach}
\label{app:margestats}

\begin{table*}[htb!]
Constraints from \emph{Planck} 2018 data only: TT+TE+EE \vspace{2pt} \\
  \centering
  \begin{tabular}{|l|c|c|c|c|}
    \hline\hline Parameter &$\Lambda$CDM Best-Fit~~&$\Lambda$CDM Marg.~~&~~~EDE ($n=3$) Best-Fit~~~&~~~EDE ($n=3$) Marg.\\ \hline \hline

{\boldmath$\ln(10^{10} A_\mathrm{s})$} & $3.055$ & $3.044\pm 0.016$ & $3.056$ & $3.051\pm 0.017$\\

{\boldmath$n_\mathrm{s}$} & $0.9659$ & $0.9645\pm 0.0043$ & $0.9769$ & $0.9702^{+0.0053}_{-0.0083}$\\

{\boldmath$100\theta_\mathrm{s}$} & $1.04200$ & $1.04185\pm 0.00029$ & $1.04168$ & $1.04164\pm 0.00034$\\

{\boldmath$\Omega_\mathrm{b} h^2$} & $0.02244$ & $0.02235\pm 0.00015$ & $0.02250$ & $0.02250^{+0.00018}_{-0.00022}$\\

{\boldmath$\Omega_\mathrm{c} h^2$} & $0.1201$ & $0.1202\pm 0.0013$ & $0.1268$ & $0.1234^{+0.0019}_{-0.0038}$\\

{\boldmath$\tau_\mathrm{reio}$} & $0.0587$ & $0.0541\pm 0.0076$ & $0.0539$ & $0.0549\pm 0.0078$\\

{\boldmath$\mathrm{log}_{10}(z_c)$} & $-$ & $-$ & $3.75$ & $3.66^{+0.24}_{-0.28}$\\

{\boldmath$f_\mathrm{EDE} $} & $-$ & $-$ & $0.068$ & $< 0.087$\\

{\boldmath$\theta_i$} & $-$ & $-$ & $2.96$ & $> 0.36$\\

    \hline

$H_0 \, [\mathrm{km/s/Mpc}] $ & $67.44$ & $67.29\pm 0.59$ & $69.13$ & $68.29^{+0.73}_{-1.2}$\\

$\Omega_\mathrm{m}         $ & $0.3147$ & $0.3162\pm 0.0083$ & $0.3138$ & $0.3145\pm 0.0086$\\

$\sigma_8                  $ & $0.8156$ & $0.8114\pm 0.0073$ & $0.8280$ & $0.8198^{+0.0090}_{-0.012}$\\

$S_8$                        & $0.8355$ & $0.8331\pm 0.0159$ & $0.8468$ & $0.8393\pm 0.0175$\\

$\mathrm{log}_{10}(f/{\mathrm{eV}})$ & $-$ & $-$ & $26.36$ & $26.57^{+0.26}_{-0.46}     $\\

$\mathrm{log}_{10}(m/{\mathrm{eV}})$ & $-$ & $-$ & $-26.90$ & $-26.94^{+0.39}_{-0.65}    $\\

    \hline
  \end{tabular} 
  \caption{The best-fit and mean $\pm1\sigma$ constraints on the cosmological parameters in $\Lambda$CDM and in the EDE scenario with $n=3$, as inferred from \emph{Planck} 2018 primary CMB data only (TT+TE+EE).  Upper and lower limits are quoted at 95\% CL.  The EDE component is consistent with zero.  These constraints are computed with the credible-interval approach discussed near the beginning of Sec.~\ref{sec:constraints}, and can be compared with the equal-tail limits presented in Table~\ref{table:params-P18-only} (the best-fit and mean values are, of course, identical in the two tables).}
  \label{table:params-P18-only-margestats}
\end{table*}

\begin{table*}[htb!]
Constraints from \emph{Planck} 2018 TT+TE+EE + CMB Lensing, BAO, SNIa, SH0ES, and RSD \vspace{2pt} \\
  \centering
  \begin{tabular}{|l|c|c|c|c|}
    \hline\hline Parameter &$\Lambda$CDM Best-Fit~~&$\Lambda$CDM Marg.~~&~~~EDE ($n=3$) Best-Fit~~~&~~~EDE ($n=3$) Marg.\\ \hline \hline

{\boldmath$\ln(10^{10} A_\mathrm{s})$} & $3.047                   $ & $3.051\pm 0.014            $ & $3.058                    $ & $3.064\pm 0.015            $\\

{\boldmath$n_\mathrm{s}$} & $0.9686                   $ & $0.9689\pm 0.0036          $ & $0.9847                    $ & $0.9854\pm 0.0070          $\\

{\boldmath$100\theta_\mathrm{s}$} & $1.04209                  $ & $1.04204\pm 0.00028        $ & $1.04119                  $ & $1.04144\pm 0.00037        $\\

{\boldmath$\Omega_\mathrm{b} h^2$} & $0.02249                  $ & $0.02252\pm 0.00013        $ & $0.02286                  $ & $0.02280\pm 0.00021        $\\

{\boldmath$\Omega_\mathrm{c} h^2$} & $0.11855                   $ & $0.11830\pm 0.00086        $ & $0.12999                   $ & $0.1290\pm 0.0039          $\\

{\boldmath$\tau_\mathrm{reio}$} & $0.0566                    $ & $0.0590^{+0.0067}_{-0.0076}$ & $0.0511                    $ & $0.0573\pm 0.0073          $\\

{\boldmath$\mathrm{log}_{10}(z_c)$} & $-$ & $-$ & $3.59                     $ & $3.63^{+0.22}_{-0.14}      $\\

{\boldmath$f_\mathrm{EDE} $} & $-$ & $-$ & $0.105                    $ & $0.098^{+0.034}_{-0.029}   $\\

{\boldmath$\theta_i$} & $-$ & $-$ & $2.71                     $ & $2.58^{+0.35}_{+0.05}     $\\

    \hline

$H_0 \, [\mathrm{km/s/Mpc}] $ & $68.07                    $ & $68.17\pm 0.39             $ & $71.15                     $ & $71.0\pm 1.1               $\\

$\Omega_\mathrm{m}         $ & $0.3058                    $ & $0.3044\pm 0.0051          $ & $0.3032                    $ & $0.3025\pm 0.0051          $\\

$\sigma_8                  $ & $0.8081                    $ & $0.8086\pm 0.0060          $ & $0.8322                    $ & $0.834\pm 0.011            $\\

$S_8$                        & $0.8158                    $ & $0.8145\pm 0.0099          $ & $0.8366                    $ & $0.837\pm 0.013            $\\

$\mathrm{log}_{10}(f/{\mathrm{eV}})$ & $-$ & $-$ & $26.63                    $ & $26.64^{+0.08}_{-0.15}   $\\

$\mathrm{log}_{10}(m/{\mathrm{eV}})$ & $-$ & $-$ & $-27.27                  $ & $-27.15^{+0.41}_{-0.29}    $\\

    \hline
  \end{tabular} 
    \caption{The best-fit and mean $\pm1\sigma$ constraints on the cosmological parameters in $\Lambda$CDM and in the EDE scenario with $n=3$, as inferred from the combination of \emph{Planck} 2018 primary CMB data (TT+TE+EE); \emph{Planck} 2018 CMB lensing data; BAO data from 6dF, SDSS DR7, and SDSS DR12; Pantheon SNIa data; the latest SH0ES $H_0$ constraint; and SDSS DR12 RSD data.  The EDE component is detected at $3.4 \sigma$ significance (using the credible-interval-derived error bar here). These constraints are computed with the credible-interval approach discussed near the beginning of Sec.~\ref{sec:constraints}, and can be compared with the equal-tail limits presented in Table~\ref{table:params-Smithcomb} (the best-fit and mean values are, of course, identical in the two tables).}
  \label{table:params-Smithcomb-margestats}
\end{table*}

\begin{table*}[htb!]
Constraints from \emph{Planck} 2018 TT+TE+EE + CMB Lensing, BAO, SNIa, SH0ES, RSD, and DES-Y1 \vspace{2pt} \\
  \centering
  \begin{tabular}{|l|c|c|c|c|}
    \hline\hline Parameter &$\Lambda$CDM Best-Fit~~&$\Lambda$CDM Marg.~~&~~~EDE ($n=3$) Best-Fit~~~&~~~EDE ($n=3$) Marg.\\ \hline \hline

{\boldmath$\ln(10^{10} A_\mathrm{s})$} & $3.049                    $ & $3.049\pm 0.014            $ & $3.064                    $ & $3.058\pm 0.015            $\\

{\boldmath$n_\mathrm{s}$} & $0.9698                   $ & $0.9704\pm 0.0035          $ & $0.9909                    $ & $0.9838\pm 0.0074          $\\

{\boldmath$100\theta_\mathrm{s}$} & $1.04183                  $ & $1.04208\pm 0.00028        $ & $1.04172                  $ & $1.04162\pm 0.00036        $\\

{\boldmath$\Omega_\mathrm{b} h^2$} & $0.02260                  $ & $0.02258\pm 0.00013        $ & $0.02304                  $ & $0.02285\pm 0.00021        $\\

{\boldmath$\Omega_\mathrm{c} h^2$} & $0.11810                   $ & $0.11752\pm 0.00078        $ & $0.1254                  $ & $0.1251\pm 0.0035          $\\

{\boldmath$\tau_\mathrm{reio}$} & $0.0584                    $ & $0.0590^{+0.0067}_{-0.0076}$ & $0.0626                    $ & $0.0581\pm 0.0074          $\\

{\boldmath$\mathrm{log}_{10}(z_c)$} & $-$ & $-$ & $3.84                     $ & $3.69\pm 0.18              $\\

{\boldmath$f_\mathrm{EDE} $} & $-$ & $-$ & $0.088                    $ & $0.077^{+0.035}_{-0.030}   $\\

{\boldmath$\theta_i$} & $-$ & $-$ & $2.89                     $ & $2.58^{+0.38}_{+0.035}     $\\

    \hline

$H_0 \, [\mathrm{km/s/Mpc}] $ & $68.24                    $ & $68.52\pm 0.36             $ & $71.05                     $ & $70.7\pm 1.0               $\\

$\Omega_\mathrm{m}         $ & $0.3035                   $ & $0.2998\pm 0.0046          $ & $0.2954                   $ & $0.2970\pm 0.0048          $\\

$\sigma_8                  $ & $0.8067                    $ & $0.8054\pm 0.0057          $ & $0.8263                    $ & $0.823\pm 0.010            $\\

$S_8$                        & $0.8115                    $ & $0.8051\pm 0.0087          $ & $0.8199                    $ & $0.819\pm 0.011            $\\

$\mathrm{log}_{10}(f/{\mathrm{eV}})$ & $-$ & $-$ & $26.47                    $ & $26.57^{+0.11}_{-0.16}     $\\

$\mathrm{log}_{10}(m/{\mathrm{eV}})$ & $-$ & $-$ & $-26.76                   $ & $-27.03\pm 0.38            $\\

    \hline
  \end{tabular} 
    \caption{The best-fit and mean $\pm1\sigma$ constraints on the cosmological parameters in $\Lambda$CDM and in the EDE scenario with $n=3$, as inferred from the combination of \emph{Planck} 2018 primary CMB data (TT+TE+EE); \emph{Planck} 2018 CMB lensing data; BAO data from 6dF, SDSS DR7, and SDSS DR12; Pantheon SNIa data; the latest SH0ES $H_0$ constraint; SDSS DR12 RSD data; and DES-Y1 3x2pt data.  The inclusion of the DES data decreases the evidence for EDE to $2.6 \sigma$ significance (using the credible-interval-derived error bar here). These constraints are computed with the credible-interval approach discussed near the beginning of Sec.~\ref{sec:constraints}, and can be compared with the equal-tail limits presented in Table~\ref{table:params-uberlikelihood} (the best-fit and mean values are, of course, identical in the two tables).}
  \label{table:params-uberlikelihood-margestats}
\end{table*}

\begin{table*}[htb!]
Constraints from \emph{Planck} 2018 TT+TE+EE + CMB Lensing, BAO, SNIa, SH0ES, RSD, DES-Y1, KiDS-$S_8$, and HSC-$S_8$ \vspace{2pt} \\
  \centering
  \begin{tabular}{|l|c|c|}
    \hline\hline Parameter &$\Lambda$CDM Marg.~~&~~~EDE ($n=3$) Marg.\\ \hline \hline

{\boldmath$\ln(10^{10} A_\mathrm{s})$} & $3.046\pm 0.014            $ & $3.053\pm 0.015$\\

{\boldmath$n_\mathrm{s}$} & $0.9710\pm 0.0035          $ & $0.9813\pm 0.0074          $\\

{\boldmath$100\theta_\mathrm{s}$} & $1.04209\pm 0.00028        $ & $1.04169^{+0.00037}_{-0.00034}$\\

{\boldmath$\Omega_\mathrm{b} h^2$} & $0.02260\pm 0.00013        $ & $0.02285^{+0.00019}_{-0.00022}$\\

{\boldmath$\Omega_\mathrm{c} h^2$} & $0.11718\pm 0.00076        $ & $0.1230^{+0.0029}_{-0.0039}$\\

{\boldmath$\tau_\mathrm{reio}$} & $0.0581\pm 0.0072$ & $0.0574\pm 0.0073$\\

{\boldmath$\mathrm{log}_{10}(z_c)$} & $-$ & $3.73^{+0.18}_{-0.23}           $\\

{\boldmath$f_\mathrm{EDE} $} & $-$ & $0.062\pm 0.030  $\\

{\boldmath$\theta_i$} & $-$ & $2.49^{+0.50}_{+0.048}    $\\

    \hline

$H_0 \, [\mathrm{km/s/Mpc}] $ & $68.67\pm 0.35             $ & $70.45^{+0.94}_{-1.2}$\\

$\Omega_\mathrm{m}         $ & $0.2978\pm 0.0044          $ & $0.2952\pm 0.0046$\\

$\sigma_8                  $ & $0.8032\pm 0.0055          $ & $0.8157\pm 0.0094$\\

$S_8$                        & $0.8002\pm 0.0082          $ & $0.809\pm 0.010$\\

$\mathrm{log}_{10}(f/{\mathrm{eV}})$ & $-$ & $26.55^{+0.12}_{-0.19}     $\\

$\mathrm{log}_{10}(m/{\mathrm{eV}})$ & $-$ & $-26.94^{+0.33}_{-0.50}            $\\

    \hline
  \end{tabular} 
    \caption{The mean $\pm1\sigma$ constraints on the cosmological parameters in $\Lambda$CDM and in the EDE scenario with $n=3$, as inferred from the combination of \emph{Planck} 2018 primary CMB data (TT+TE+EE); \emph{Planck} 2018 CMB lensing data; BAO data from 6dF, SDSS DR7, and SDSS DR12; Pantheon SNIa data; the latest SH0ES $H_0$ constraint; SDSS DR12 RSD data; DES-Y1 3x2pt data; and priors on $S_8$ derived from KiDS and HSC data.  The inclusion of the KiDS and HSC data decreases the evidence for EDE to $2.1 \sigma$ significance (using the credible-interval-derived error bar here). These constraints are computed with the credible-interval approach discussed near the beginning of Sec.~\ref{sec:constraints}, and can be compared with the equal-tail limits presented in Table~\ref{table:params-uberlikelihoodKiDSHSC} (the mean values are, of course, identical in the two tables).}
  \label{table:params-uberlikelihoodKiDSHSC-margestats}
\end{table*}

\begin{table*}[htb!]
Constraints from \emph{Planck} 2018 TT+TE+EE + CMB Lensing, BAO, SNIa, RSD, DES-Y1, KiDS-$S_8$, and HSC-$S_8$ (No-SH0ES) \vspace{2pt} \\
  \centering
  \begin{tabular}{|l|c|c|}
    \hline\hline Parameter &$\Lambda$CDM Marg.~~&~~~EDE ($n=3$) Marg.\\ \hline \hline

{\boldmath$\ln(10^{10} A_\mathrm{s})$} & $3.041\pm 0.014            $ & $3.044\pm 0.014$\\

{\boldmath$n_\mathrm{s}$} & $0.9691\pm 0.0035          $ & $0.9718^{+0.0041}_{-0.0055}          $\\

{\boldmath$100\theta_\mathrm{s}$} & $1.04200\pm 0.00028       $ & $1.04177^{+0.00038}_{-0.00030}$\\

{\boldmath$\Omega_\mathrm{b} h^2$} & $0.02253\pm 0.00013        $ & $0.02264^{+0.00015}_{-0.00018}$\\

{\boldmath$\Omega_\mathrm{c} h^2$} & $0.11785\pm 0.00077        $ & $0.11956^{+0.00088}_{-0.0020}$\\

{\boldmath$\tau_\mathrm{reio}$} & $0.0552\pm 0.0070$ & $0.0558\pm 0.0070$\\

{\boldmath$\mathrm{log}_{10}(z_c)$} & $-$ & $> 3.28$\\

{\boldmath$f_\mathrm{EDE} $} & $-$ & $ < 0.060 $\\

{\boldmath$\theta_i$} & $-$ & $> 0.35$\\

    \hline

$H_0 \, [\mathrm{km/s/Mpc}] $ & $68.33\pm 0.36            $ & $68.92^{+0.40}_{-0.72}$\\

$\Omega_\mathrm{m}         $ & $0.3021\pm 0.0045          $ & $0.3008\pm 0.0047$\\

$\sigma_8                  $ & $0.8032\pm 0.0053         $ & $0.8064^{+0.0057}_{-0.0073}$\\

$S_8$                        & $0.8060\pm 0.0082         $ & $0.8074\pm 0.0089$\\

$\mathrm{log}_{10}(f/{\mathrm{eV}})$ & $-$ & $26.52^{+0.28}_{-0.44}     $\\

$\mathrm{log}_{10}(m/{\mathrm{eV}})$ & $-$ & $-26.67\pm 0.69            $\\

    \hline
  \end{tabular} 
    \caption{The mean $\pm1\sigma$ constraints on the cosmological parameters in $\Lambda$CDM and in the EDE scenario with $n=3$, as inferred from the combination of \emph{Planck} 2018 primary CMB data (TT+TE+EE); \emph{Planck} 2018 CMB lensing data; BAO data from 6dF, SDSS DR7, and SDSS DR12; Pantheon SNIa data; SDSS DR12 RSD data; DES-Y1 3x2pt data; and priors on $S_8$ derived from KiDS and HSC data.  The SH0ES $H_0$ measurement is not included here.  Upper and lower limits are quoted at 95\% CL.  With SH0ES excluded, there is no evidence for the EDE component.  These constraints are computed with the credible-interval approach discussed near the beginning of Sec.~\ref{sec:constraints}, and can be compared with the equal-tail limits presented in Table~\ref{table:params-uberlikelihoodKiDSHSCNoSH0ES} (the mean values are, of course, identical in the two tables).}
  \label{table:params-uberlikelihoodKiDSHSCNoSH0ES-margestats}
\end{table*}

\end{document}